\documentclass[prb,aps,amssymb,shownopacs,twocolumn,10pt]{revtex4-1}
\usepackage{amsmath}
\usepackage{amssymb}
\usepackage{amsthm}
\usepackage{amsfonts}
\usepackage{listings}
\usepackage{easybmat}
\usepackage{enumerate}
\usepackage{latexsym}
\usepackage{epstopdf}
\usepackage{psfrag}
\usepackage{color}
\usepackage{bm}
\usepackage[pdftex]{graphicx}
\usepackage{subfigure}
\usepackage{hyperref}
\usepackage{ulem} 
\newcommand{\beq}{\begin{equation}}
\newcommand{\eneq}{\end{equation}}
\newcommand{\bs}[1]{\boldsymbol{#1}}

\newcommand{\ket}[1]{\left|#1\right\rangle}

\def\be{\begin{equation}}
\def\ee{\end{equation}}
\def\ba{\begin{eqnarray}}
\def\ea{\end{eqnarray}}

\input{epsf}

\begin{document}

\tolerance 10000
\newcommand{\vk}{{\bf k}}
\title{
Projective construction of the $\mathbb{Z}_k$ Read-Rezayi fractional quantum Hall states 
and their excitations on the torus geometry}

\author{C\'ecile Repellin$^1$}
\author{Titus Neupert$^2$} 
\author{B. Andrei Bernevig$^{3}$}
\author{Nicolas Regnault$^{1,3}$}

\affiliation{$^1$
Laboratoire Pierre Aigrain, Ecole Normale Sup\'erieure-PSL Research
University, CNRS, Universit\'e Pierre et Marie Curie-Sorbonne Universit\'es,
Universit\'e Paris Diderot-Sorbonne Paris Cit\'e, 24 rue Lhomond, 75231
Paris Cedex 05, France\\
$^2$ Princeton Center for Theoretical Science, Princeton University, Princeton, NJ 08544, USA\\
$^3$ Department of Physics, Princeton University, Princeton, NJ 08544, USA} 

\begin{abstract}

Multilayer fractional quantum Hall  wave functions can be used to construct the non-Abelian states of the $\mathbb{Z}_k$ Read-Rezayi series upon symmetrization over the layer index. 
Unfortunately, this construction does not yield the complete set of $\mathbb{Z}_k$ ground states on the torus.
We develop an alternative projective construction of $\mathbb{Z}_k$ Read-Rezayi states that complements the existing one.
On the multi-layer torus geometry, our construction consists of introducing twisted boundary conditions connecting the layers before performing the symmetrization. 
We give a comprehensive account of this construction for bosonic states, and numerically show that the full ground state and quasihole manifolds are recovered for all computationally accessible system sizes. Furthermore, we analyze the neutral excitation modes above the Moore-Read on the torus through an extensive exact diagonalization study. We show numerically that 
our construction can be used to obtain excellent approximations to these modes. Finally, we extend the new symmetrization scheme to the plane and sphere geometries.
\end{abstract}

\date{\today}

\maketitle

\section{Introduction}
 Exotic correlated and topologically ordered quantum states can often be constructed from less exotic parent states via the action of specific many-body projection operators. Projective constructions can yield topological order by starting with entirely uncorrelated single Slater determinant parent states of (topological) band insulators. Well-studied examples are so-called parton constructions of fractional quantum Hall (FQH) states~\cite{PhysRevB.40.8079,doi:10.1142/S0217984991000058,PhysRevLett.66.802,PhysRevB.60.8827,doi:10.1142/S0217979292000840,Blok1992615} and more generally fractional topological insulators~\cite{PhysRevLett.105.246809,PhysRevB.83.195139,PhysRevB.85.125105}. 
Other projective constructions can change the form of topological order in a parent state to a more exotic one, for example by turning an Abelian FQH state into a non-Abelian one. This has been exploited in topological orders described by conformal field theories via the so-called coset projections~\cite{Cappelli-1999CMaPh.205..657C, Froehlich-2000cond.mat..2330F, cappelli-2001NuPhB.599..499C,Barkeshli-PhysRevB.81.155302}.

The $\mathbb{Z}_k$ Read-Rezayi series~\cite{Read-1999PhRvB..59.8084R} is a well known sequence of FQH states with non-Abelian topological order. The $k=1$ member of this series and parent state for our construction is the (Abelian) Laughlin state~\cite{Laughlin-PhysRevLett.50.1395}. The $k=2$ member of this series is the Moore-Read~\cite{Moore1991362} state supporting Majorana excitations and a well studied candidate state for the FQH plateau of electrons at $\nu=5/2$. The $k=3$ member of this series supports Fibonacci anyons, which can in principle be used to perform the operations of a universal topological quantum computer~\cite{Slingerland2001229,Nayak08}. For bosons, the $\mathbb{Z}_k$ Read-Rezayi series can be obtained using projective constructions, by starting from the Laughlin state as a parent state. Reference~\onlinecite{cappelli-2001NuPhB.599..499C} considered $k$ independent layers of Laughlin states. The projective construction then consists of symmetrizing over the layer degree of freedom, yielding a single layered $\mathbb{Z}_k$ Read-Rezayi state. The approach of Ref.~\onlinecite{cappelli-2001NuPhB.599..499C} was based on conformal field theory arguments, and was subsequently tested by numerical simulations on the sphere geometry~\cite{Regnault-PhysRevLett.101.066803}. Besides the exact zero-energy ground states and quasihole states, this procedure also provides accurate trial wave functions for the low energy neutral and quasielectron excitations of the $\mathbb{Z}_k$ Read-Rezayi states on the sphere geometry, as numerically shown in Refs.~\onlinecite{Rodriguez-PhysRevB.85.035128, Sreejith-PhysRevLett.107.136802, Sreejith-PhysRevLett.107.086806, Sreejith-PhysRevB.87.245125}. More than a useful mathematical trick, the projective constructions may also be of physical importance: several proposals~\cite{Read-PhysRevB.61.10267,Nomura-JPSJ.73.2612,Rezayi-2010arXiv1007.2022R,Barkeshli-2010PhRvL.105u6804B,Papic-PhysRevB.82.075302,Vaezi-2014PhRvL.113w6804V,Zhu-2015arXiv150205076Z,Geraedts-2015arXiv150201340G,Liu-2015arXiv150205391L,Peterson-2015arXiv150202671P} to build non-Abelian order from Abelian systems are indeed based on this construction, using either a tunneling term or an interaction between layers.  

However, this established multi-layer symmetrization procedure is not capable of constructing the $\mathbb{Z}_k$ Read-Rezayi states for all system sizes on the torus geometry. For instance, the Moore-Read ground state appearing for an odd number of particles on the torus cannot be written using this procedure, and similarly, other states in the $\mathbb{Z}_k$ series fail to have a multilayer symmetrization description. In this paper, we present an alternative projective construction that allows us to also obtain those missing ground states. 
We consider a $k$-layered torus with twisted instead of periodic boundary conditions, which can be realized through an extended topological defect that connects the layers. 
In addition, the construction can be used to obtain the
quasihole states of the $\mathbb{Z}_k$ Read-Rezayi series for all system sizes on the torus. 
We show that the multilayer torus with twisted boundary conditions is equivalent to an enlarged, single-layer torus. Our construction thus enables us to extract any bosonic Read-Rezayi state from a single Laughlin wave function.
We generalize this novel scheme to the plane and sphere geometries as an alternative to the symmetrization over $k$ independent layers.

A second part of our study is dedicated to the neutral excitation modes above $\mathbb{Z}_k$ Read-Rezayi states. These modes are the higher-$k$ cousins of the magnetoroton excitation~\cite{GMP-PhysRevLett.54.581, GMP-PhysRevB.33.2481} above the $\mathbb{Z}_1$ Laughlin state. We detail their structure on the torus geometry for the simplest non-Abelian case, the Moore-Read state. Furthermore, we show that symmetrization of the magnetoroton excitations above the Laughlin state yields excellent approximations to the excitation modes above the Moore-Read state both on the torus and the sphere. While these trial wave functions are different from those obtained using the multilayer approach~\cite{Rodriguez-PhysRevB.85.035128, Sreejith-PhysRevLett.107.136802, Sreejith-PhysRevLett.107.086806, Sreejith-PhysRevB.87.245125}, they describe the neutral mode with the same accuracy. 

This paper is structured into three parts. Section~\ref{sec: main results} defines the projective construction on the torus with twisted boundary conditions, interprets it both geometrically and through its action on the orbitals, and presents the exact numerical results on zero energy states. The subsequent Sec.~\ref{sec: torus neutral modes ED} is concerned with the neutral excitation modes above the Moore-Read state on the torus, that are accessed both with exact diagonalization and with the projective construction. Finally, Sec.~\ref{sec: sphere} contains a detailed discussion of the results on the plane and sphere geometries, including both the zero-energy states and the neutral excitation modes.

\section{Read-Rezayi states on the torus}
\label{sec: main results}

\subsection{Ground states and their degeneracy}

The bosonic $\mathbb{Z}_k$ Read-Rezayi~\cite{Read-1999PhRvB..59.8084R} state $\ket{\Psi_k}$ is the densest zero-energy ground state of the $(k+1)$-body contact interaction
\be
H_{\mathrm{int},k}
=\int\mathrm{d}^2\bs{r}:[\rho(\bs{r})]^{k+1}:,
\label{eq: model Hamiltonian}
\ee
when projected into the lowest Landau level. 
Here,
\be
\rho(\bs{r})=\psi^\dagger (\bs{r})\psi (\bs{r}),
\ee
is the density operator written in terms of the field operator
$\psi^\dagger (\bs{r})$,  that creates a boson at position $\bs{r}$, and $::$ represents normal ordering.

On the torus, the $\mathbb{Z}_k$ Read-Rezayi ground state appears at filling $\nu=k/2$, that is, if the number of particles $N$ and the number of flux quanta $N_\phi$ obey
\begin{equation}
kN_\phi=2N.
\label{eq: filling factor relation torus}
\end{equation}
This relation cannot be satisfied for all system sizes, since both $N$  and $N_\phi$ have to be integer. For example, if $k$ is odd and $N_\phi$ is odd, there exists no ground state. If a ground state exists, it can exhibit a topological degeneracy on the torus (but not on the sphere). This degeneracy is a hallmark of the specific topological order and is in one-to-one correspondence with the number of topological anyon excitations of the phase. The topological order realized by a bosonic $\mathbb{Z}_k$ Read-Rezayi state is labelled by the affine Lie algebra SU$(2)_k$ and contains $(k+1)$ irreducible representations, corresponding to $(k+1)$ topological anyon excitations. Thus, on the torus with an even number of flux quanta $N_\phi$, any $\mathbb{Z}_k$ Read-Rezayi state is $(k+1)$-fold degenerate. The momentum quantum numbers of the degenerate ground states are given in Tab.~\ref{tab: GS momentum sectors} and Appendix~\ref{app: translation symmetry}. Note that there is a fundamental difference between $k$ even and $k$ odd. For $k$ odd, $N$ must be a multiple of $k$ for the filling fraction to be $\nu = k/2$. When $k$ is even, there are two alternatives: $N$ is either an integer multiple of $k$ (and $N_{\phi}$ is even) or a half-integer multiple of $k$ (and $N_{\phi}$ is odd). In the latter case, there is a unique $\mathbb{Z}_k$ state. Note that this feature is specific to the torus geometry, and absent on the sphere.

If one deviates from the filling factor $\nu=k/2$ by increasing  the number  of flux quanta, quasihole  excitations are nucleated. Their  wave functions correspond to the zero energy states of the model Hamiltonian~\eqref{eq: model Hamiltonian}. The number of zero energy states (and their quantum numbers) for a given number of particles and a given number of flux quanta can be deduced from the clustering properties of these states~\cite{Bernevig-PhysRevLett.100.246802,Bernevig-PhysRevB.77.184502} (for the sphere geometry, closed formulas are known~\cite{Gurarie1997685, Ardonne-JPhysA2002, Read-PhysRevB.73.245334}).

\begin{table}[t]
\begin{center}
\begin{equation}\nonumber
\begin{array}{| c | c c | c c |}
\hline
 & k \ \mathrm{odd}&  & k \ \mathrm{even}& \\
 & (\mathsf{k}_x, \mathsf{k}_y) & \mathrm{deg.} & (\mathsf{k}_x, \mathsf{k}_y) & \mathrm{deg.}  \\
\hline
N_{\phi} \ \mathrm{odd} & - &  & (0,0) & 1\\
\hline
N_{\phi} \ \mathrm{even} & (0,0) & \frac{k + 1}{2} & (0,0) & \frac{k}{2} \\
                         &  (0,N_\phi/2)     &   \frac{k + 1}{2}              &  (0,N_{\phi}/2) & 1 \\
                         &       &                 &  (N_{\phi}/2, 0) & 1 \\
                         &       &                 &  (N_{\phi}/2,N_{\phi}/2) & \frac{k}{2}-1 \\
\hline
\end{array}
\end{equation}
\end{center}
\caption{Ground state degeneracies and momenta of the $\mathbb{Z}_k$ Read-Rezayi state on the torus. 
For $k$ odd (respectively even), the momentum quantum numbers $\mathsf{k}_x, \mathsf{k}_y$ thus belong to a  $N_{\phi}/2 \times N_{\phi}$ (respectively $N_{\phi} \times N_{\phi}$) Brillouin zone.
These momentum quantum numbers are defined in the Landau $x$-gauge with vector potential $\bs{A}(\bs{r})=(0,-Bx)$. 
See Appendix~\ref{app: translation symmetry} for the definition of the momentum quantum numbers.
}
\label{tab: GS momentum sectors}
\end{table}%

\subsection{$\mathbb{Z}_k$ Read-Rezayi states from symmetrization}
\label{sec: intro to symmetrization}

In this work, we exploit the fact that a $\mathbb{Z}_{gk}$ Read-Rezayi  state 
$\ket{\Psi_{gk}}$ can be obtained by symmetrizing
 over $g$ independent copies of a $\mathbb{Z}_{k}$ Read-Rezayi  state $\ket{\Psi_k}$.
This procedure was first introduced for the Moore-Read state, which can be written as a bilayer Laughlin state in the strongly paired regime~\cite{greiter1992paired, greiter1991paired}. In the conformal field theory language, the coset construction described in Refs.~\onlinecite{Cappelli-1999CMaPh.205..657C, Froehlich-2000cond.mat..2330F} (and later generalized to the Read-Rezayi series in Ref.~\onlinecite{cappelli-2001NuPhB.599..499C}) uses Abelian theories to build the parafermion Hall states.

Introducing a symmetrization operator ${\mathcal S}_{g \rightarrow 1}$, we write
\be
\ket{\Psi_{gk}}=
{\mathcal S}_{g \rightarrow 1}\bigl(
\underset{g\ \text{times}}{\underbrace{
\ket{\Psi_k}
\otimes
\cdots 
\otimes
\ket{\Psi_k}
}}\bigr).
\label{eq: symmatrization}
\ee
Here, we can picture the direct product of $\ket{\Psi_k}$ states as being defined on $g$ ``layers" of the same manifold. For example, we can construct a $\mathbb{Z}_g$ Read-Rezayi state for any $g$ by symmetrizing over $g$ layers of $\nu=1/2$ bosonic Laughlin states.

 Let $\phi^\dagger_j$ be any basis of single-particle bosonic creation operators acting on the single particle Hilbert space $\mathcal{H}_T$ in the position basis on the torus and denote by $\phi^\dagger_{j,l}$, $l=0,\cdots, g-1,$ the corresponding operators belonging the $l$-th layer of $\mathcal{H}_T$ in the direct sum $\mathcal{H}_T\oplus\cdots\oplus\mathcal{H}_T$ that constitutes the single-particle Hilbert space of the $g$-layered torus. Then, the symmetrization
${\mathcal S}_{g \rightarrow 1}$ is a map from 
$\mathcal{H}_T\oplus\cdots\oplus\mathcal{H}_T$ to $\mathcal{H}_T$ defined by
\be
\phi^\dagger_{j,l}\stackrel{{\mathcal S}_{g \rightarrow 1}}{\rightarrow}\phi^\dagger_j.
\label{eq: def S}
\ee 

We note that there is a fundamental problem with the symmetrization construction~\eqref{eq: symmatrization} for systems of finite size. If a densest ground state $\ket{\Psi_{gk}}$ has a particle numbers $N$ that is not divisible by $g$, this particle number cannot be spread equally over $g$ densest states $\ket{\Psi_k}$. To give a simple example, consider the bosonic Moore-Read state $\ket{\Psi_2}$ with $N=7$ particles, which is a zero-energy ground state of the Hamiltonian~\eqref{eq: model Hamiltonian} on a torus with $N_\phi=7$ flux quanta (filling $\nu=1$). On the same torus, the bosonic $\nu=1/2$ Laughlin state $\ket{\Psi_1}$ does not exist as a zero energy ground state, since $N_\phi=7$ is not divisible by two. One could consider symmetrizing over a Laughlin quasihole state with $N=3$ and a Laughlin quasielectron state with $N=4$ (the latter being at finite energy). However, this construction can at best yield an approximation of the desired $\ket{\Psi_2}$ ground state (we checked numerically that the corresponding overlap was of the order of $0.998$ for $N=17$). It might be possible to obtain the exact state by using the composite fermion~\cite{jain-PhysRevLett.63.199} expression of the quasielectron states, but these are very hard to obtain on the torus~\cite{Hermanns-PhysRevB.87.235128,Fremling-PhysRevB.89.125303}.

A similar obstruction appears for all $\mathbb{Z}_k$ Read-Rezayi states with $k$ even on a torus with odd $N_\phi$. Adding flux quanta to the original $\nu=k/2$ system is a way to nucleate quasihole excitations of the $\mathbb{Z}_k$ Read-Rezayi states. While the obstruction disappears for a sufficient number of added flux quanta, not all quasihole states are immune to this issue. The system with $7$ particles on a torus pierced by $N_\phi=5$ flux quanta has a filling fraction strictly lower than $3/2$, and thus admits zero energy states of the $4$-body interaction of the $\mathbb{Z}_3$ Read-Rezayi state. However, the $7$ particles can at best be spread into two layers with $2$ particles (Laughlin quasihole state) and one layer with $3$ particles (Laughlin quasielectron), resulting in a non-zero energy state after symmetrization. In general, we see that for odd $N_\phi$ and the largest $N$ such that $N/N_{\phi} \leq k/2$, the $\mathbb{Z}_k$ states cannot be constructed using the symmetrization technique.

\begin{figure}[t]
\includegraphics[width = 0.99\linewidth]{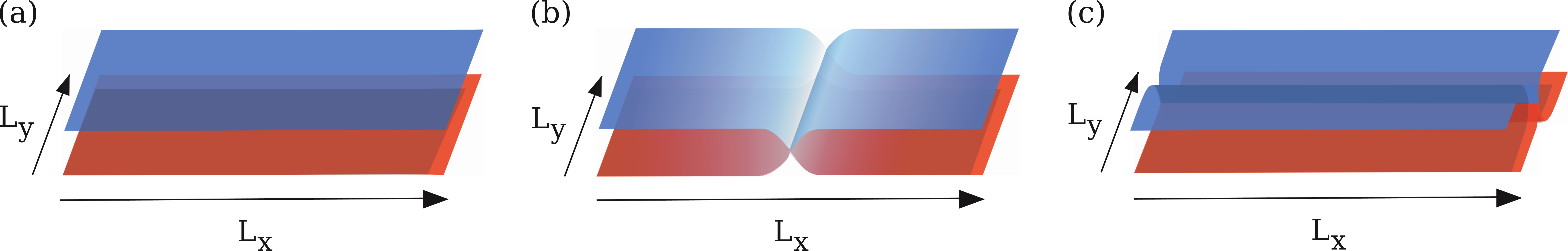}
\caption{Bilayer system with different types of topological defects. a) Bilayer system with no defect. b) Bilayer system with a twist defect along the $y$ direction. If the system has periodic boundary conditions, it is transformed into a monolayer torus with a doubled length in the $x$ direction. c) Bilayer system with a defect along the $x$ direction. If the system has periodic boundary conditions, it is transformed into a monolayer torus with a doubled length in the $y$ direction.}
\label{fig:twistedBilayer}
\end{figure}

A main result of this paper is an alternative symmetrization scheme on the torus that remedies this obstruction. We show that this scheme can also be used on the sphere as an equally powerful alternative to the multi-layer symmetrization. 
On the torus, the key idea is to change the boundary conditions between the layers from periodic to twisted, as depicted in Fig.~\ref{fig:twistedBilayer}. Equivalently, this can be seen as introducing a topological defect line that permutes the layer indices~\cite{Barkeshli-PhysRevB.87.045130, barkeshli-PhysRevB.88.241103, barkeshli-PhysRevB.88.235103, barkeshli-2014arXiv1410.4540B, Teo-2015arXiv150306812T}. Similarly to the symmetrization over multiple independent layers, our construction with twisted boundary conditions in one direction between the layers alone does not yield the complete $\mathbb{Z}_k$ manifold. To obtain the complete manifold, two of the three symmetrization schemes (untwisted multilayer, twist in $x$ direction, twist in $y$ direction) have to be combined.

\begin{figure}[t]
\includegraphics[width = 0.99\linewidth]{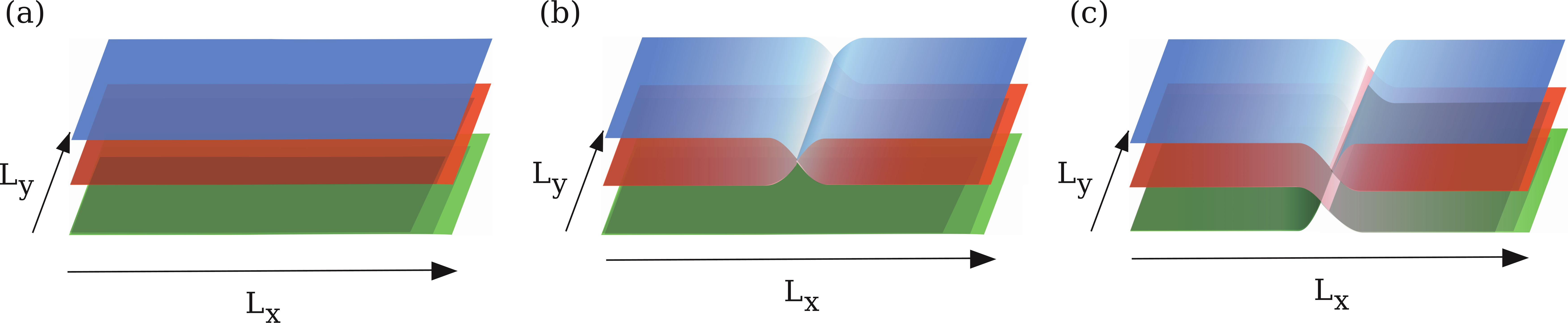}
\caption{Trilayer system with different types of topological defects. a) Trilayer system with no defect. b) Trilayer system with a defect along the $y$ direction connecting the two upper layers. With periodic boundary conditions, this system is equivalent to a bilayer torus, with the first layer having aspect ratio 2, and the second layer having aspect ratio 1. c) Trilayer system with a defect along the $y$ direction connecting respectively layers 1 (blue) and 2 (red), 2 and 3 (green), 3 and 1. With periodic boundary conditions, this system is equivalent to a torus with aspect ratio 3.}
\label{fig:twistedTrilayer}
\end{figure}

As a manifold, a double layer torus of size $L_x\times L_y$ with twisted boundary conditions in the $x$-direction (respectively $y$-direction) is equivalent to a single layer torus of size $2L_x\times L_y$ (respectively $L_x\times 2L_y$). The symmetrization is then taken over the particle coordinates with $0\leq x<L_x$ and $L_x\leq x<2L_x$ (respectively with $0\leq y<L_y$ and $L_y\leq y<2L_y$). This carries over to a $g$ layered torus, for boundary conditions that fully permute the layer coordinates. (See Fig.~\ref{fig:twistedTrilayer} for an illustration of different possible boundary conditions of three layers.)
In the next section, we will show that this equivalence between manifolds is also respected by the model Hamiltonian~\eqref{eq: model Hamiltonian} for the  $\mathbb{Z}_k$ Read-Rezayi states. Moreover, we will work out the action of the symmetrization operator ${\mathcal S}_{g \rightarrow 1}$ on the basis states of the $g$ times larger tori: $gL_x\times L_y$ (the $(gT)_x$ torus) and $L_x\times gL_y$ (the $(gT)_y$ torus).

One may wonder whether yet another relevant symmetrization scheme is obtained by imposing twisted boundary conditions in both $x$ and $y$ directions in a double layer system. As we illustrate in Appendix~\ref{app: twisted}, the resulting surface is equivalent to a single-layer torus whose spanning vectors form a given relative angle and norm ratio that depend on the symmetrization scheme that will be applied. The  symmetrization schemes are then the same as for twisted boundary conditions in one direction, with the difference that the symmetrized state is also defined on a torus with that different angle between its spanning vectors.

\subsection{Equivalence between twisted boundary conditions and enlarged torus}
\label{sec: geometrical equivalence on Hamiltonian level}

We want to show that the model Hamiltonian~\eqref{eq: model Hamiltonian} for $\mathbb{Z}_k$ Read-Rezayi states on the $g$-layered torus with twisted boundary conditions is equivalent to the same Hamiltonian on a $g$ times larger torus.
For this, we momentarily neglect the projection on the lowest Landau level, which will later be used to obtain all numerical results. We will see that the equivalence holds for the entire system and thus also for the lowest Landau level.

Let us for concreteness fix the gauge to be the Landau $x$-gauge with vector potential $\bs{A}(\bs{r})=(0,-xB)$ and consider separately the cases of twisted boundary conditions in the $y$-direction and twisted boundary conditions in the $x$-direction.
A complete set of single-particle operators on the $g$-layer torus $T=[0,L_x)\times[0,L_y)$ with (fully) twisted boundary conditions is given by
the operators $\widetilde{\psi}^\dagger_l (\widetilde{\bs{r}})$, that create a boson at position $\widetilde{\bs{r}}\in T$ in layer $l=0,\cdots,g-1$. 

\begin{subequations}
If twisted boundary conditions are applied along the $x$-direction, 
they obey 
\be
\widetilde{\psi}^\dagger_l (\widetilde{\bs{r}}+L_x\bs{e}_x)=e^{2\pi \mathrm{i}N_\phi y/L_y}\widetilde{\psi}^\dagger_{(l+1)\,\mathrm{mod}\,g} (\widetilde{\bs{r}}),
\label{eq: boundary conditions 1}
\ee
where $N_\phi=BL_xL_y$.
A complete set of single-particle operators on the $g$ times larger torus $(gT)_x=[0,gL_x)\times[0,L_y)$ is given by
the operators $\psi^\dagger (\bs{r})$,  that create a boson at position $\bs{r}\in (gT)_x$. They obey periodic boundary conditions 
\be
\psi^\dagger (\bs{r}+gL_x\bs{e}_x)=e^{2\pi \mathrm{i}gN_\phi y/L_y}\psi^\dagger (\bs{r}).
\label{eq: boundary conditions 1 b}
\ee
\end{subequations}

\begin{subequations}
If twisted boundary conditions are applied along the $y$-direction, 
they obey 
\be
\widetilde{\psi}^\dagger_l (\widetilde{\bs{r}}+L_y\bs{e}_y)=\widetilde{\psi}^\dagger_{(l+1)\,\mathrm{mod}\,g} (\widetilde{\bs{r}}).
\label{eq: boundary conditions 2}
\ee
A complete set of single-particle operators on the $g$ times larger torus $(gT)_y=[0,L_x)\times[0,gL_y)$ is given by
the operators $\psi^\dagger (\bs{r})$,  that create a boson at position $\bs{r}\in (gT)_y$. They obey periodic boundary conditions 
\be
\psi^\dagger (\bs{r}+gL_y\bs{e}_y)=\psi^\dagger (\bs{r}).
\label{eq: boundary conditions 2 b}
\ee
\end{subequations}

The model Hamiltonian for the $\mathbb{Z}_k$ Read-Rezayi state on either manifold decomposes into a noninteracting Landau level Hamiltonian and an interaction part
\be
H_k=H_0+H_{\mathrm{int},k},\qquad
\widetilde{H}_k=\widetilde{H}_0+\widetilde{H}_{\mathrm{int},k},
\ee
that act on the Fock space build from the single-particle operators in Eq.~\eqref{eq: boundary conditions 1 b}/Eq.~\eqref{eq: boundary conditions 2 b} and Eq.~\eqref{eq: boundary conditions 1}/Eq.~\eqref{eq: boundary conditions 2}, respectively. 
Here, $H_0$ and $\widetilde{H}_0$ are the single-particle operators for the $T$ and $(gT)_i,\ i=x,y,$ tori, respectively.
The respective interacting parts $H_{\mathrm{int},k}$ and $\widetilde{H}_{\mathrm{int},k}$ are given by Eq.~\eqref{eq: model Hamiltonian} and 
\be
\widetilde{H}_{\mathrm{int},k}
=\sum_{l=0}^{g-1}\int\limits_{T}\mathrm{d}^2\widetilde{\bs{r}}[\widetilde{\rho}_l(\widetilde{\bs{r}})]^{k+1},
\ee
 with the latter expressed in terms of the density operator
$
\widetilde{\rho}_l(\widetilde{\bs{r}})
=
\widetilde{\psi}^\dagger_l (\widetilde{\bs{r}})
\widetilde{\psi}_l (\widetilde{\bs{r}})
$
in the $l$-th layer. 

\begin{subequations}
\label{eq: identification sp operators}
If the boundary conditions are twisted in the $x$-direction, the identification
\be
\widetilde{\psi}^\dagger_l (\widetilde{\bs{r}})\equiv\psi^\dagger (\widetilde{\bs{r}}+lL_x\bs{e}_x),
\quad l=0,\cdots, g-1, \quad \widetilde{\bs{r}}\in T,
\label{eq: identification sp operators a}
\ee
provides a mapping between the two single-particle Hilbert spaces (and thus also the Fock spaces) under which 
$\widetilde{H}_{\mathrm{int},k}$ is exactly mapped into $H_{\mathrm{int},k}$.
Analogously, if the boundary conditions are twisted in the $y$-direction, the identification is
\be
\widetilde{\psi}^\dagger_l (\widetilde{\bs{r}})\equiv\psi^\dagger (\widetilde{\bs{r}}+lL_y\bs{e}_y),
\quad l=0,\cdots, g-1, \quad \widetilde{\bs{r}}\in T.
\label{eq: identification sp operators b}
\ee
\end{subequations}
The ultra-locality of the contact two-body interaction is crucial to make this connection to a $g$ times as long torus for the interacting system.  
Under this same mapping, the single-particle Hamiltonian
\be
\widetilde{H}_0=
\sum_{l=0}^{g-1}\int\limits_{T}\mathrm{d}^2\widetilde{\bs{r}}\,
\widetilde{\psi}^\dagger_l (\widetilde{\bs{r}})
\left[\mathrm{i}\bs{\nabla}_{\widetilde{\bs{r}}}-e \widetilde{\bs{A}}_l(\widetilde{\bs{r}})\right]^2
\widetilde{\psi}_l (\widetilde{\bs{r}})
\ee
with 
\be
\bs{\nabla}_{\widetilde{\bs{r}}}\wedge\widetilde{\bs{A}}_l(\widetilde{\bs{r}})=B,
\quad
l=0,\cdots, g-1,
\quad
\widetilde{\bs{r}}\in T
\ee
maps into
\be
H_0=
\int_{(gT)_i}\mathrm{d}^2\bs{r}\,
\psi^\dagger(\bs{r})
\left[\mathrm{i}\bs{\nabla}_{\bs{r}}-e \bs{A}(\bs{r})\right]^2
\psi (\bs{r}),
\ee
with $i=x,y$, and 
\be
\widetilde{\bs{A}}_l(\widetilde{\bs{r}})=\bs{A}(\widetilde{\bs{r}}+lL_i\bs{e}_i)
\label{eq: gauge field on gT}
\ee
defining the gauge potential of $(gT)_i$ that thus also obeys $\bs{\nabla}_{\bs{r}}\wedge\bs{A}(\bs{r})=B$
for every $\bs{r}\in (gT)_i$.

With this, we have shown the equivalence of the interacting model Read-Rezayi systems on the $g$-layered torus with twisted boundary conditions and on the $g$ times larger torus.
Since the magnetic field $B$ is preserved under this mapping, the $g$ times larger torus is pierced by $g$ times as many flux quanta, i.e., $gN_\phi$, as the initial torus.

\subsection{Symmetrization on the torus with twisted boundary conditions}
\label{sec: symm with twisted Bc theory}

To implement the symmetrization operation, we want to compute the action of the operator $\mathcal{S}_{(gT)_i \rightarrow T}$ on a many-body wave function defined on the $g$ times larger torus $(gT)_i$, $i=x,y$ [$(gT)_i$ is the torus enlarged $g$ times in the $i$ direction]. The action of ${\mathcal S}_{(gT)_i\rightarrow T}$ can be written in first quantized notation. Let us denote the translation operator ${\mathcal T}_{L_i\bs{e}_i}$ that translates single-particle operators by $L_i\bs{e}_i$ on $(gT)_i$, $i=x,y$. The symmetrization identifies positions that are related by any magnetic translation ${\mathcal T}_{\delta L_i\bs{e}_i}$, where $0 \le \delta < g$, $\delta\in\mathbb{Z}$. Starting from a many-body wave function $\Psi$ on a $(gT)_i$ torus, the symmetrized wave function reads
\begin{equation}
\begin{split}
&{\mathcal S}_{(gT)_i\rightarrow T}\Psi\left(\bs{r}_1,\cdots,\bs{r}_N\right)=
\\
&\qquad
\sum_{\{0 \le \delta_j < g\}} 
\left(\prod_{j=1}^N {\mathcal T}_{j,\delta_j L_i\bs{e}_i}\right) \Psi\left(\bs{r}_1,\cdots,\bs{r}_N\right).
\label{eq:firstquantizedsymmetrization}
\end{split}
\end{equation}
where $\bs{r}_j$ denotes the coordinates of the $j$-th particle restricted to the $L_x \times L_y$ region and the $j$ label on the translation operator indicates it acts on the $j$-th particle. With our gauge choice, Eq.~\eqref{eq:firstquantizedsymmetrization} has an even more explicit expression for ${\mathcal S}_{(gT)_y\rightarrow T}$
\begin{equation}
\begin{split}
&{\mathcal S}_{(gT)_y\rightarrow T}\Psi\left(\bs{r}_1,\cdots,\bs{r}_N\right)=
\\
&\qquad
\sum_{\{0 \le \delta_j < g\}}  \Psi\left(\bs{r}_1+\delta_1 L_y\bs{e}_y,\cdots,\bs{r}_N+\delta_N L_y\bs{e}_y\right)
\label{eq:firstquantizedsymmetrization2}
\end{split}
\end{equation}
The vanishing properties of the symmetrized wave function can immediately be deduced from those of $\Psi$. Indeed, if $\Psi$ is a Laughlin wave function (or one of its quasihole excitations), it vanishes when two particles are at the same point. There are terms in the sum of Eq.~\eqref{eq:firstquantizedsymmetrization2} where $\delta_1,\cdots,\delta_g$ are all distinct. If we put the $g$ first particles at the same position, these terms do not need to vanish. We now bring the $(g+1)$-th first particles at the same position. In each term of the sum, at least two $\delta_i$ among $\delta_1,\cdots,\delta_{g+1}$ are equal, ensuring that $\Psi$ is evaluated at a position including two equal coordinates. Thus ${\mathcal S}_{(gT)_y\rightarrow T} \Psi$ vanishes when $g+1$ particles are at the same position, proving it is a zero energy eigenstate of the $(g+1)$-body contact interaction~\eqref{eq: model Hamiltonian}, i.e. a $\mathbb{Z}_g$ Read-Rezayi wave function (or one of its quasihole excitations). A similar argument holds true for the symmetrization on the $(gT)_x$ torus, including Eq.~\eqref{eq:firstquantizedsymmetrization}.

A more practical expression of these symmetrization operators can be obtained by considering only their action on the single-particle operators or basis states. Using the definition of ${\mathcal S}_{g \rightarrow 1}$ from Eq.~\eqref{eq: def S}, and the boundary conditions~\eqref{eq: boundary conditions 1} and~\eqref{eq: boundary conditions 2}, we see that ${\mathcal S}_{(gT)_i\rightarrow T}$ indeed maps into operators $\phi^\dagger (\widetilde{\bs{r}})$ that obey the correct boundary conditions on $T$, irrespective of whether the initial torus is $(gT)_x$ or $(gT)_y$, namely
\begin{subequations}
\be
\begin{split}
\widetilde{\psi}^\dagger_l (\widetilde{\bs{r}})
\stackrel{{\mathcal S}_{(gT)_i\rightarrow T}}{\longrightarrow}
\phi^\dagger (\widetilde{\bs{r}}),
\qquad\ \forall l=0,\cdots, g-1,
\end{split}
\ee
with
\be
\begin{split}
\phi^\dagger (\widetilde{\bs{r}})=&\,
e^{2\pi\mathrm{i}N_\phi y/L_y}
\phi^\dagger (\widetilde{\bs{r}}+L_x\bs{e}_x),
\\
\phi^\dagger (\widetilde{\bs{r}})=&\,
\phi^\dagger (\widetilde{\bs{r}}+L_y\bs{e}_y).
\end{split}
\ee
\end{subequations}
Thus, these single particle operators can be defined on a $g$ times smaller torus. 
Using the identification~\eqref{eq: identification sp operators}, we obtain the identity
\be
\begin{split}
{\mathcal S}_{(gT)_i\rightarrow T}\left[\psi^\dagger (\bs{r})\right]=
{\mathcal S}_{(gT)_i\rightarrow T}\left[\psi^\dagger (\bs{r}+lL_i\bs{e}_i)\right],
\\
\qquad \forall\ l=0,\cdots, g-1,
\end{split}
\ee
for the operators $\psi^\dagger (\bs{r})$ on $(gT)_i$, $i=x,y$.
Let us interpret this result in terms of the eigenspaces of the translation operator ${\mathcal T}_{L_i\bs{e}_i}$. It has eigenvalues $e^{2\pi \mathrm{i}  s/g}$, $s=0,\cdots, g-1$. We observe that ${\mathcal S}_{(gT)_i\rightarrow T}$ is precisely the projector on the single-particle eigenstates of ${\mathcal T}_{L_i\bs{e}_i}$ with  eigenvalue $1$ on $(gT)_i$. (Note that the projection on any eigenvalue $e^{2\pi\mathrm{i} s/g}$ is an equally good option for the symmetrization as we will discuss below). While this result has been derived for the entire Hilbert space, it is also true in any ${\mathcal T}_{L_i\bs{e}_i}$-invariant subspace, such as the lowest Landau level.

In the $x$-Landau gauge, a basis $\phi^{T}_{\tilde{j}}(\bs{r})$ of the lowest Landau level on the torus $T$ with $N_\phi$ flux quanta is given by
\begin{equation}
\begin{split}
\phi^{T}_{\tilde{j}}(\bs{r})
=&\,\frac{1}{\sqrt{L_y\sqrt{\pi}}}
\sum_{m=-\infty}^\infty
e^{\mathrm{i}y\left(2\pi\, {\tilde{j}}/L_y+mL_x\right)}\\
&\quad\times
e^{-\left(x+2\pi\,{\tilde{j}}/L_y+mL_x\right)^2/2},
\label{eq: Landau gauge orbitals}
\end{split}
\end{equation}
with the orbital quantum number ${\tilde{j}}=0,\cdots, N_\phi-1$. (We will consider single-particle states instead of second quantized operators from here on.)
The basis $\phi^{(gT)_x}_{j}(\bs{r})$ of the lowest Landau level on $(gT)_x$ with $gN_\phi$ flux quanta is obtained from Eq.~\eqref{eq: Landau gauge orbitals} by replacing $L_x$ with $gL_x$ and thus $j$ may take values $j=0,\cdots, gN_\phi-1$.
The action of ${\mathcal T}_{L_x\bs{e}_x}$ on the basis states of $(gT)_x$ is given by 
\begin{equation}
{\mathcal T}_{L_x\bs{e}_x}
\ket{\phi^{(gT)_x}_j}=\ket{\phi^{(gT)_x}_{(j+N_\phi)\mathrm{mod}\,gN_\phi}}.
\end{equation}
A basis for the eigenspace of ${\mathcal T}_{L_x\bs{e}_x}$ of eigenvalue $1$ in the lowest Landau level is given by the equal-amplitude superposition of basis states with momenta $j$ that are $N_\phi$ apart. With the help of Eq.~\eqref{eq: Landau gauge orbitals} one can explicitly check that
\begin{equation}
\ket{\phi^{T}_{\tilde{j}}}
=\frac{1}{\sqrt{g}}
\sum_{s=0}^{g-1}\ket{\phi^{(gT)_x}_{\tilde{j}+sN_\phi}},
\quad \tilde{j}=0,\cdots, N_\phi-1.
\end{equation}
The interpretation of this result is that symmetrization of a many-body state over the orbitals that are spaced $N_\phi$ on the $(gT)_x$ torus is equivalent to symmetrizing over the layers of the $g$-layered $T$ torus with twisted boundary conditions in the $x$ direction.

Next, we want to study the case where the twist in the boundary conditions is along the $y$ direction, while keeping the Landau gauge fixed in the $x$-direction. We are now seeking the eigenspace of ${\mathcal T}_{L_y\bs{e}_y}$ on the   $(gT)_y$ torus. From Eq.~\eqref{eq: Landau gauge orbitals} we find
\begin{equation}
{\mathcal T}_{L_y\bs{e}_y}
\ket{\phi^{(gT)_y}_j}=
e^{2\pi \mathrm{i} j/g}
\ket{\phi^{(gT)_y}_{j}}.
\end{equation}
Thus, the eigenvalue $1$ subspace of ${\mathcal T}_{L_y\bs{e}_y}$ is spanned by the basis states with $j=\tilde{j}g$, where $\tilde{j}=0,\cdots, N_\phi-1$. In other words, the states
\begin{equation}
\ket{\phi^{T}_{\tilde{j}}}
=\sqrt{g}
\ket{\phi^{(gT)_y}_{g\tilde{j}}},
\quad \tilde{j}=0,\cdots, N_\phi-1
\end{equation}
form a basis for the $T$ torus. The interpretation of this result is that the projection of a many-body state on the orbitals that are spaced $g$ on the $(gT)_y$ torus is equivalent to symmetrizing over the layers of the $g$-layered $T$ torus with twisted boundary conditions in the $y$ direction.

\begin{figure}
\includegraphics[width = 0.99\linewidth]{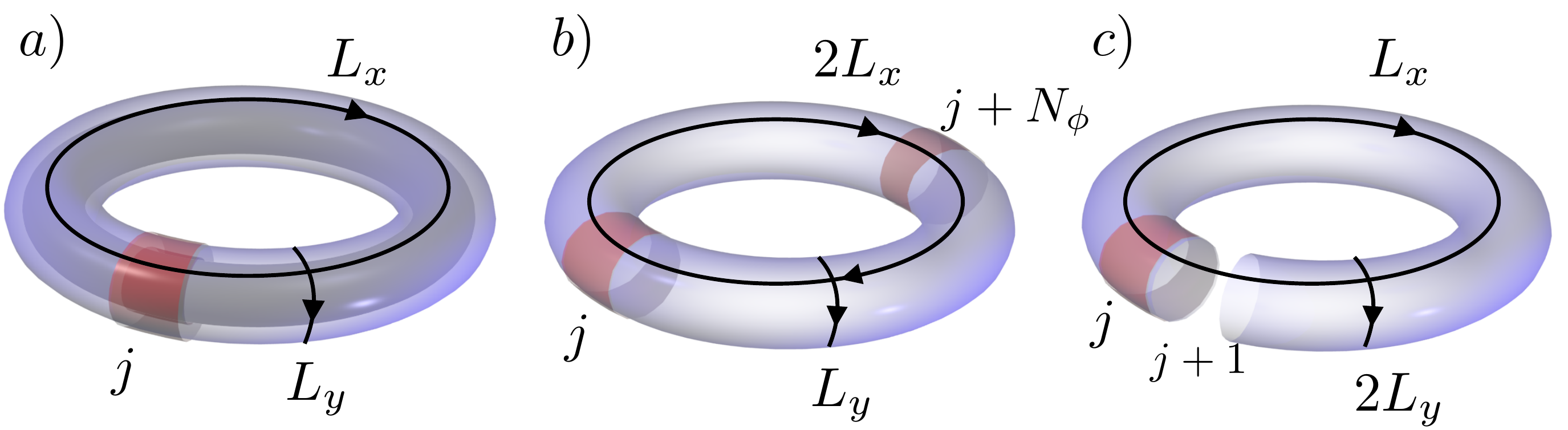}
\caption{Schematic representation of the symmetrization procedure to obtain the ground state on the $T$ torus of dimensions $L_x\times L_y$ pierced by $N_{\phi}$ flux quanta. The orbitals colored in red have become equivalent after symmetrization. a)~The $T$ torus bilayer of lengths $L_x\times L_y$, pierced by $N_{\phi}$ flux quanta. 
b)~The $(2T)_x$ torus of lengths $2L_x\times L_y$, pierced by $2N_{\phi}$ flux quanta. 
c)~The $(2T)_y$ torus of lengths $L_x\times 2L_y$, pierced by $2N_{\phi}$ flux quanta.
}
\label{fig:symmetrization procedure}
\end{figure}

Returning to second quantized formulation, we can 
summarize the action of $\mathcal{S}_{(gT)_x \rightarrow T}$ as follows.
On the torus $(gT)_x=[0,gL_x)\times[0,L_y)$
\be
\phi^{(gT)_x}_j
\stackrel{\mathcal{S}_{(gT)_x \rightarrow T}}{\longrightarrow}
\phi^{T}_{j\,\mathrm{mod}\,N_\phi},
\label{eq: symm Nphi apart orbitals}
\ee
i.e., a many-body state has to be symmetrized over all orbitals with quantum numbers $j$
differing by $N_\phi$.
On the torus $(gT)_y=[0,L_x)\times[0,gL_y)$
\be
\phi^{(gT)_y}_j
\stackrel{\mathcal{S}_{(gT)_y \rightarrow T}}{\longrightarrow}
\begin{cases}
\phi^{T}_{(j-s)/g},\qquad &j\,\mathrm{mod}\,g=s,
\\
0,\ & \mathrm{else},
\end{cases}
\label{eq: symm every goth orbital}
\ee
i.e., a many-body state is projected to the orbitals with quantum numbers $j$
that are multiples of $g$ (for $s=0$).
In choosing different $s=0,\cdots, g-1$, we obtain a family of equivalent symmetrization operators which only differ by a global translation between the origin of the coordinate systems on $T$ and $(gT)_y$.
These symmetrization schemes are schematically illustrated in Fig.~\ref{fig:symmetrization procedure}. Note that this freedom in the choice of $s$ has an analogue in first quantized notation. Indeed, we can rewrite Eq.~\eqref{eq:firstquantizedsymmetrization} to include this optional parameter
\begin{equation}
\begin{split}
&{\mathcal S}_{(gT)_i\rightarrow T}\Psi\left(\bs{r}_1,\cdots,\bs{r}_N\right)=
\\
&\qquad
\sum_{\{0 \le \delta_j < g\}} 
\left(\prod_{j=1}^N e^{-\frac{2\pi \mathrm{i} \delta_j s}{g}} {\mathcal T}_{j,\delta_j L_i\bs{e}_i}\right) \Psi\left(\bs{r}_1,\cdots,\bs{r}_N\right).
\label{eq:firstquantizedsymmetrizationwiths}
\end{split}
\end{equation}
We note that setting $s$ is akin to projecting on the eigenspace of ${\mathcal T}_{L_i\bs{e}_i}$ with ${\mathcal T}_{L_i\bs{e}_i}$ eigenvalue $e^{\frac{2\pi \mathrm{i} s}{g}}$ on $(gT)_i$.

\subsection{
Numerical results on zero-energy states}
\label{sec: torus}

We will now discuss the exact numerical results on the symmetrization construction in detail for the ground states and quasihole states of the Read-Rezayi series on the torus, i.e., for all zero energy states of the model interaction~\eqref{eq: model Hamiltonian}. While we have shown in the previous section that the symmetrization leads to Read-Rezayi states, we still have to investigate in which cases the procedure allows to recover all the states.

\subsubsection{Symmetrization with periodic boundary conditions}
We start by testing the previously known~\cite{cappelli-2001NuPhB.599..499C} multi-layer symmetrization construction for the case of two layers of Laughlin ground states on the torus. We choose a system with an even number of particles, so that they can be evenly distributed among the two layers.
\be
\begin{split}
\ket{\Psi^{(\mathsf{k}_y+\mathsf{k}'_y)\mathrm{mod}\,N_\phi}_{2}} 
= \,&
{\mathcal S}_{2\to1}\left(
\ket{\Psi^{\mathsf{k}_y}_{1}}  
\otimes
\ket{\Psi^{\mathsf{k}^\prime_y}_{1}}\right).
\end{split}
\ee

Notice that the states carry center of mass momentum quantum numbers $\mathsf{k}_y$ and $\mathsf{k}_y^\prime$, which can take the values $0$ and $N_\phi/2$ for the two degenerate Laughlin states. As the symmetrization makes the layers indistinguishable, the choices $(\mathsf{k}_y,\mathsf{k}_y^\prime)=(0,N_\phi/2)$ and $(\mathsf{k}_y,\mathsf{k}_y^\prime)=(N_\phi/2, 0)$ deliver the identical Moore-Read state.  
In contrast, the choices $(\mathsf{k}_y,\mathsf{k}_y^\prime)=(0,0)$ and $(\mathsf{k}_y,\mathsf{k}_y^\prime)=(N_\phi/2, N_\phi/2)$ yield two different Moore-Read states in the same $\mathsf{k}_y$ sector. 
The latter two states are not eigenstates of the relative translation operator $\mathcal{T}_x^\mathrm{rel}$, even if the Laughlin states were (see Appendix~\ref{app: translation symmetry}). The diagonalization of $\mathcal{T}_x^\mathrm{rel}$ in the subspace defined by the two degenerate Moore-Read states yields one state with $\mathsf{k}_x = 0$ and one with $\mathsf{k}_x = N_\phi/2$.


Going beyond the ground states, we numerically checked that one can obtain a Moore-Read quasihole state by symmetrizing two decoupled systems of Laughlin $1/2$ quasiholes (i.e. one flux quantum added compared to the ground state filling fraction). As this process can sometimes be redundant, we have to extract a linearly independent basis from all states obtained after the symmetrization. We compare the number of linearly independent symmetrized products of Laughlin quasiholes states with the number of Moore-Read quasihole states for different values of $N \leq N_{\phi}$. For each $N_{\phi} \leq 12$, we considered the systems with $4 \leq N \leq N_{\phi}$. For all even $N\leq N_{\phi}$, and all odd $N < N_{\phi}$, we found that all Moore-Read quasihole states can be constructed as a symmetrized product of Laughlin quasihole states.

As a result, the only zero-energy state of the three-body contact interaction that \emph{cannot} be reproduced using the symmetrization construction over two independent layers is the Moore-Read ground state that lies in the $N = N_{\phi}$ odd sector that we discussed in Sec.~\ref{sec: intro to symmetrization}.

\subsubsection{Symmetrization with twisted boundary conditions}
\label{sec: symm with twisted}

We will now numerically test the symmetrization constructions on the $g$-layer torus with twisted boundary conditions, or equivalently the $g$ times as long torus (see Sec.~\ref{sec: symm with twisted Bc theory}). We will start with a $\mathbb{Z}_k$ state on the $g$-layered  torus to construct a $\mathbb{Z}_{gk}$ state on the single-layer torus by symmetrization. 
Not only will this construction allow us to access the missing $\mathbb{Z}_{gk}$ states for $N_\phi$ odd, it also yields the entire quasihole manifold of $\mathbb{Z}_k$ states. This construction comes with a trade-off: some of the $\mathbb{Z}_{gk}$ ground states accessible using the multilayer (with periodic boundary conditions) representation cannot be reached using only one type of twisted boundary conditions. It is simply the consequence of the symmetrization operator yielding one state per vector it acts on. For instance, the Moore-Read state is constructed by choosing $g=2$, $k=1$. Acting with ${\mathcal S}_{2L_x\rightarrow L_x}$ on the twofold degenerate Laughlin state defined on the $(gT)_x$ torus will yield both Moore-Read states in the $\mathsf{k}_y = 0$ sector,  but not the Moore-Read state in the $\mathsf{k}_y = N/2$ sector. Conversely, only the Moore-Read states in the $\mathsf{k}_x = 0$ sector can be constructed by acting with ${\mathcal S}_{2L_y\rightarrow L_y}$ on the Laughlin state defined on the $(gT)_y$ torus. In summary, all of the $\mathbb{Z}_{gk}$ states can be constructed using the symmetrization method, as long as at least two out of the three boundary conditions (no twist, twist in $x$ direction, twist in $y$ direction) are used.

\emph{Enlarging the torus length in the $x$-direction} --- 
\label{sec:symmetrizeRatio2}
The numerical implementaion of ${\mathcal S}_{(gT)_x \rightarrow T}$ is described in Appendix~\ref{app: numerical symmetrization}. This symmetrization scheme conserves the center of mass momentum $\mathsf{k}_y$ modulo $N_\phi$, but does not conserve the relative momentum $\mathsf{k}_x$  of a many body state (see Appendix~\ref{app: quantum numbers} for details).

We checked numerically that the application of ${\mathcal S}_{(gT)_x \rightarrow T}$ to the degenerate $\mathbb{Z}_k$ state defined on the $(gT)_x$ torus yields only zero energy states of the $(gk+1)$-body model interaction. Focusing first on the densest $\mathbb{Z}_k$ state, we apply ${\mathcal S}_{(gT)_x \rightarrow T}$ to the Laughlin state. We numerically checked the above property for $N \leq 14$ ($g=2$), $N \leq 12$ ($g=3$), $N \leq 14$ ($g=4$). This implies that applying ${\mathcal S}_{2L_x\rightarrow L_x}$ to the Moore-Read state defined on the $(2T)_x$ torus yields part (respectively all) of the $\mathbb{Z}_4$ ground state manifold when $N_{\phi}$ is even (respectively odd). 

Decreasing the filling fraction, we have checked that the above property also holds true for quasihole states. In this latter case, we observe an additional property: the completeness of the symmetrized quasihole manifold. In other words, any zero energy state of the $(gk+1)$-body interaction with $N$ and $N_{\phi}$ such that $N/N_{\phi} < gk/2$ can be reached by applying ${\mathcal S}_{(gT)_x \rightarrow T}$ to the subspace of zero energy states of the $(k+1)$-body interaction, with the same number of particles and $gN_{\phi}$ flux quanta. We numerically verified this statement in a number of cases for $k=1$ and $g = 2,\ 3,\ 4$, including all cases with $N\leq 7$ and up to $3$ added flux quanta (compared to the number of flux quanta for the densest $\mathbb{Z}_{kg}$ state). For $gk = 3,\ 4$, we also numerically verified this property for $N \leq 10$ and the smallest fraction of flux added to obtain a quasihole state.

\emph{Enlarging the torus length in the $y$-direction} --- 
\label{sec:symmetrizeRatio0.5}
The numerical implementaion of ${\mathcal S}_{(gT)_y \rightarrow T}$ is described in Appendix~\ref{app: numerical symmetrization}. This symmetrization scheme does not conserve the center of mass momentum $\mathsf{k}_y$, but conserves the relative momentum $\mathsf{k}_x$ of a many body state (see Appendix~\ref{app: quantum numbers} for details). 

We confirmed numerically that the application of ${\mathcal S}_{(gT)_y \rightarrow T}$ to the degenerate $\mathbb{Z}_k$ state defined on the $(gT)_y$ torus yielded only zero energy states of the $(gk+1)$-body model interaction.  Focusing first on the densest $\mathbb{Z}_k$ states, we numerically verified this property by applying ${\mathcal S}_{(gT)_y \rightarrow T}$ to the Laughlin ($k=1$) and the Moore-Read ($k=2$) states. We checked that the property was true at least up to $N \leq 13$ ($g=2$), $N \leq 12$ ($g=3$), $N \leq 14$ ($g=4$) when $k=1$, and for $N \leq 14$ when $k=2$ and $g=2$.

Decreasing the filling fraction, we confirmed that the above property also held true for quasihole states. Again, we observe the completeness of the symmetrized quasihole subspace. We numerically verified this statement in a number of cases including all cases with $k=1$, $g=2,\ 3, \ 4$, $N\leq 6$ and up to $3$ added flux quanta (compared to the number of flux quanta for the densest $\mathbb{Z}_{kg}$ state) (for $N=7$, we checked all cases with one added flux quanta). For $gk = 3,\ 4$, we numerically verified this property for $N \leq 10$ and the smallest fraction of flux added to obtain a quasihole state.



\section{Neutral excitation modes on the torus}
\label{sec: torus neutral modes ED}

All properties discussed so far regard the construction of zero energy states of the $(k+1)$-body interaction. 
While these $\mathbb{Z}_k$ Read-Rezayi ground states are relatively well understood, little is known about their neutral low energy excitation modes on the torus beyond the Laughlin case $k=1$.~\cite{GMP-PhysRevLett.54.581, GMP-PhysRevB.33.2481, repellin-PhysRevB.90.045114}

Most of the existing literature concentrates on the sphere geometry, 
where the neutral excitations above the Moore-Read and $\mathbb{Z}_3$ Read-Rezayi states have been studied in Refs.~\onlinecite{moller-PhysRevLett.107.036803, bonderson-PhysRevLett.106.186802, Rodriguez-PhysRevB.85.035128, Sreejith-PhysRevLett.107.086806, Sreejith-PhysRevLett.107.136802} and in Ref.~\onlinecite{Sreejith-PhysRevB.87.245125}, respectively.
In contrast, we are not aware of any related study on the torus geometry for the $\mathbb{Z}_2$, or for any $\mathbb{Z}_k$ states with $k>2$.

\begin{figure}
\includegraphics[width = 0.99\linewidth]{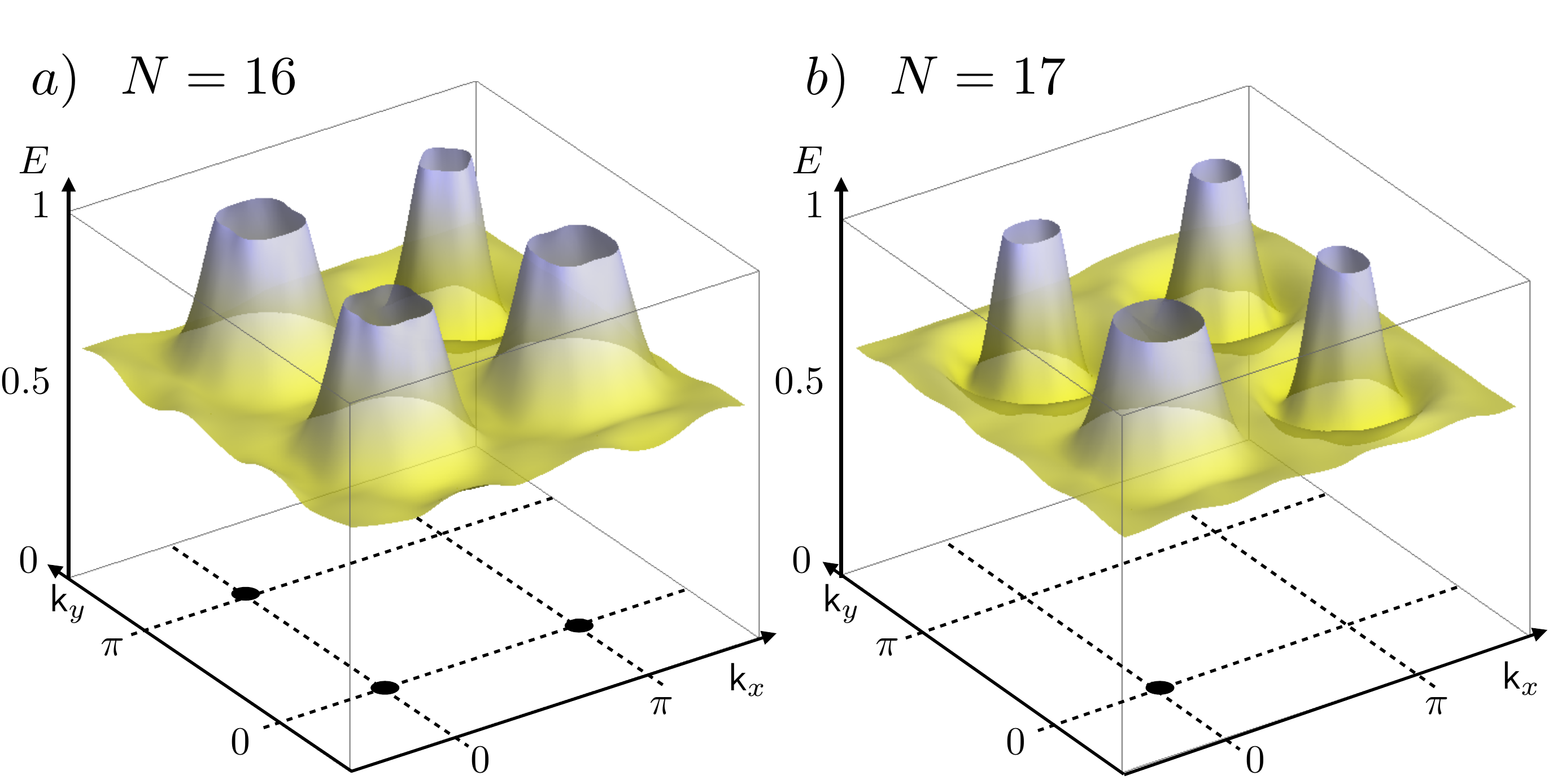}
\caption{
Dispersion of the neutral excitation mode above the Moore-Read ground states on the torus
 as a function of the center of mass and relative momenta of the many-body states in the FQH Brillouin zone. Interpolations of the dispersions are obtained from energy eigenvalues of finite size systems for a) $N=N_\phi=16$ and  b) $N=N_\phi=17$. In accordance with the momentum sectors Tab.~\ref{tab: GS momentum sectors}, these systems feature three ground states and one ground state, respectively (black dots). For energies larger than $E\sim1$, a continuum of excited states appears (not shown). The dispersions are centered around the inversion symmetric momenta. The slight anisotropy of the dispersive part of each mode near these momenta is a finite size effect. We clearly observe two types of neutral excitations depending on the presence or absence of a zero energy state at the center of the dispersive part: in the absence of a ground state, the neutral mode has a much more pronounced minimum.}
\label{fig: 3D modes}
\end{figure}

In this section, we start by analyzing the neutral modes above the Moore-Read state using exact diagonalization. We identify several dispersive modes (see Fig.~\ref{fig: 3D modes} for the neutral excitation mode above the Moore-Read state), and make a connection with the neutral modes on the sphere. Further, we test the validity of the symmetrization procedure beyond zero energy states. We show that the symmetrization construction captures the physics of inherent excitations of the system correctly. Symmetrization of excited states above the $g$-layer Laughlin state will yield good trial states for the excitations above the $\mathbb{Z}_g$. Instead of yielding exact states, 
the symmetrization provides a variational scheme to approximate the dispersion and eigenstates of the neutral excitation modes. We will quantitatively benchmark this construction.

\subsection{Neutral excitation modes from exact diagonalization}


The neutral excitation mode above the Laughlin state, the magnetoroton mode, can be understood as the dispersing bound state of a quasiparticle and quasihole. 
Trial states belonging to the neutral mode can be obtained by acting on the ground state with a (lowest Landau level projected) density operator. This constitutes the so-called single-mode approximation, which has been verified numerically both on the sphere~\cite{GMP-PhysRevLett.54.581, GMP-PhysRevB.33.2481, Yang-PhysRevLett.108.256807} and on the torus geometry~\cite{repellin-PhysRevB.90.045114}.

On the sphere, the Moore-Read densest ground states exists only for an even number of particles. There is a neutral mode above this state, which is also called magnetoroton mode, because it bears strong similarities to the magnetoroton mode above the Laughlin state. Indeed, it is well described by the single-mode approximation~\cite{Yang-PhysRevLett.108.256807}. 
For an odd number of particles, no zero energy ground state of the three-body contact Hamiltonian is found at filling fraction $\nu=1$ on the sphere. In spite of the absence of zero energy state, there is a neutral low energy mode, dubbed neutral fermion mode, as numerically shown in Refs.~\onlinecite{moller-PhysRevLett.107.036803, bonderson-PhysRevLett.106.186802}.
Both the magnetoroton mode and the neutral fermion mode can be viewed as states that minimally violate the generalized exclusion principle, as was shown in Ref.~\onlinecite{Yang-PhysRevLett.108.256807}. 
In the magnetoroton mode, the violation appears as a quasielectron-quasihole pair, similar to the Laughlin case. 
In the neutral fermion mode, the violation consists of a single unpaired particle in a background of paired particles. (See Appendix~\ref{app: thin torus} for details.)

\begin{figure*}
\includegraphics[width = 0.28\linewidth]{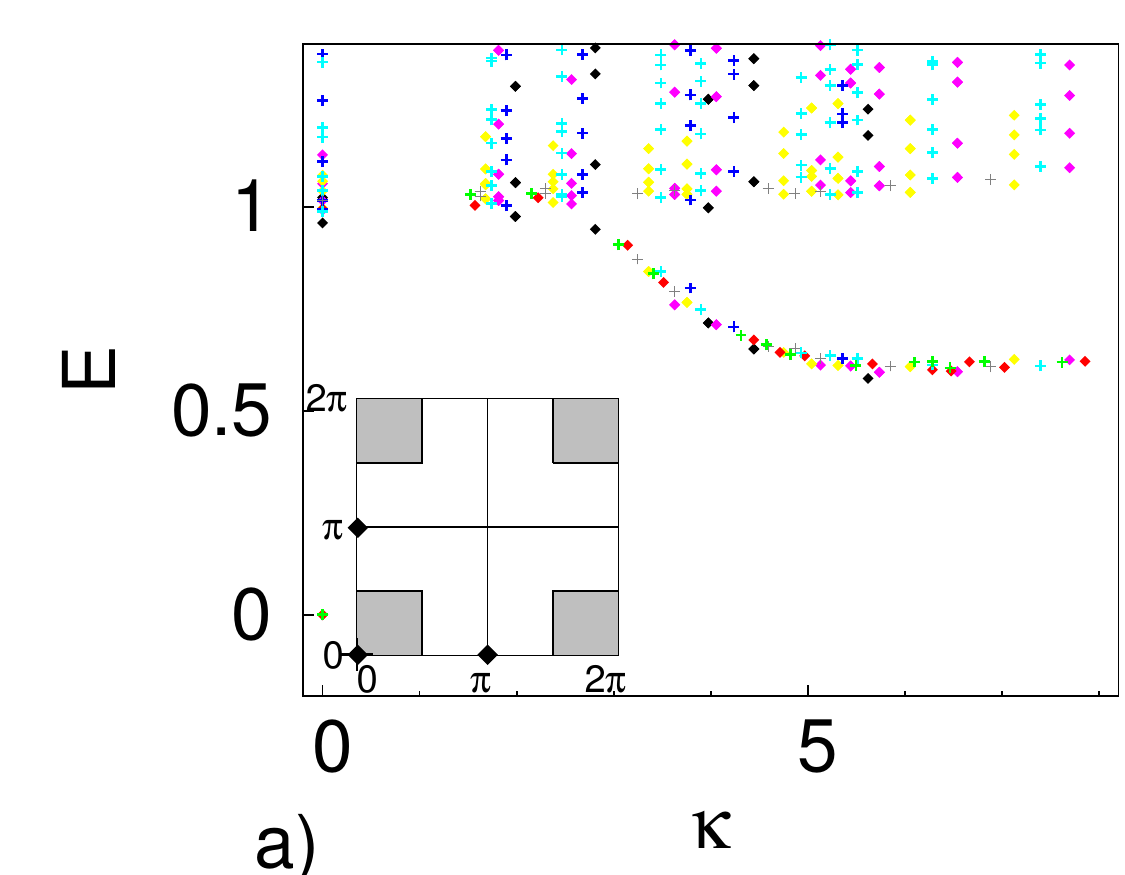}
\includegraphics[width = 0.28\linewidth]{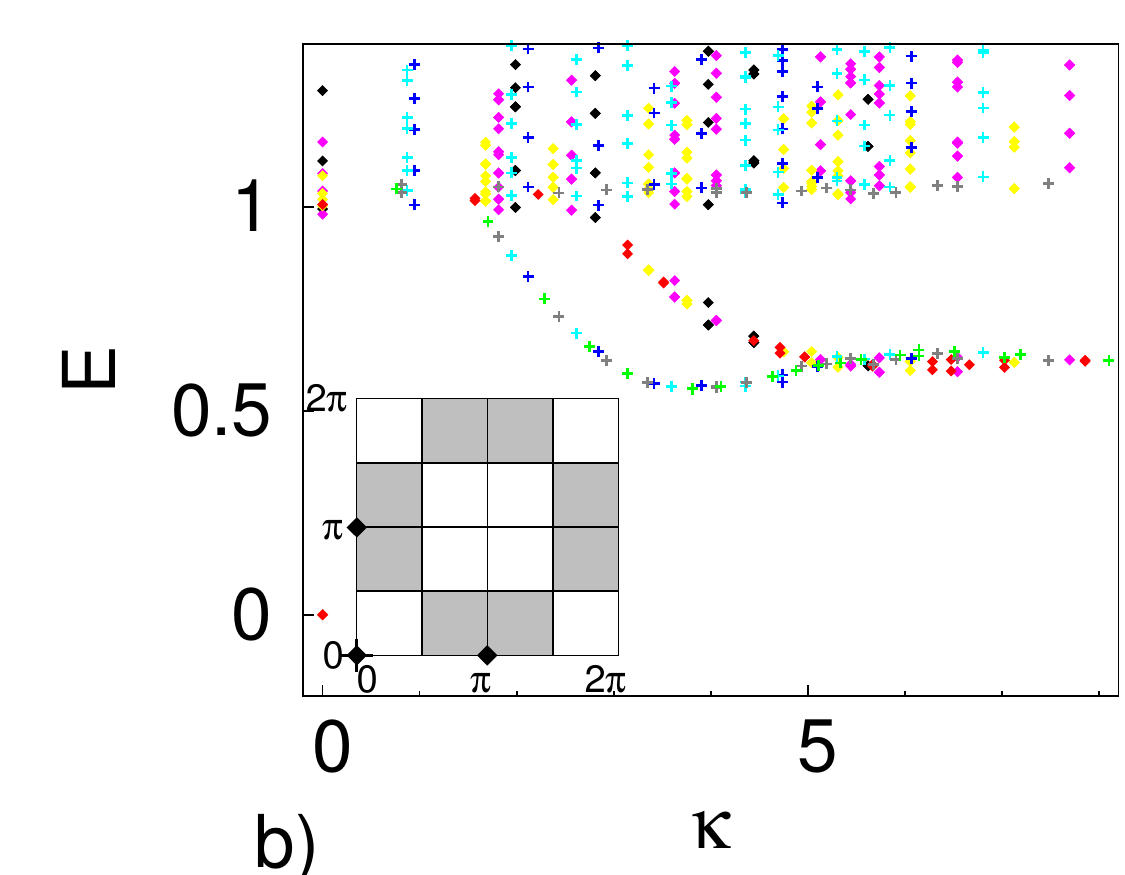}
\includegraphics[width = 0.28\linewidth]{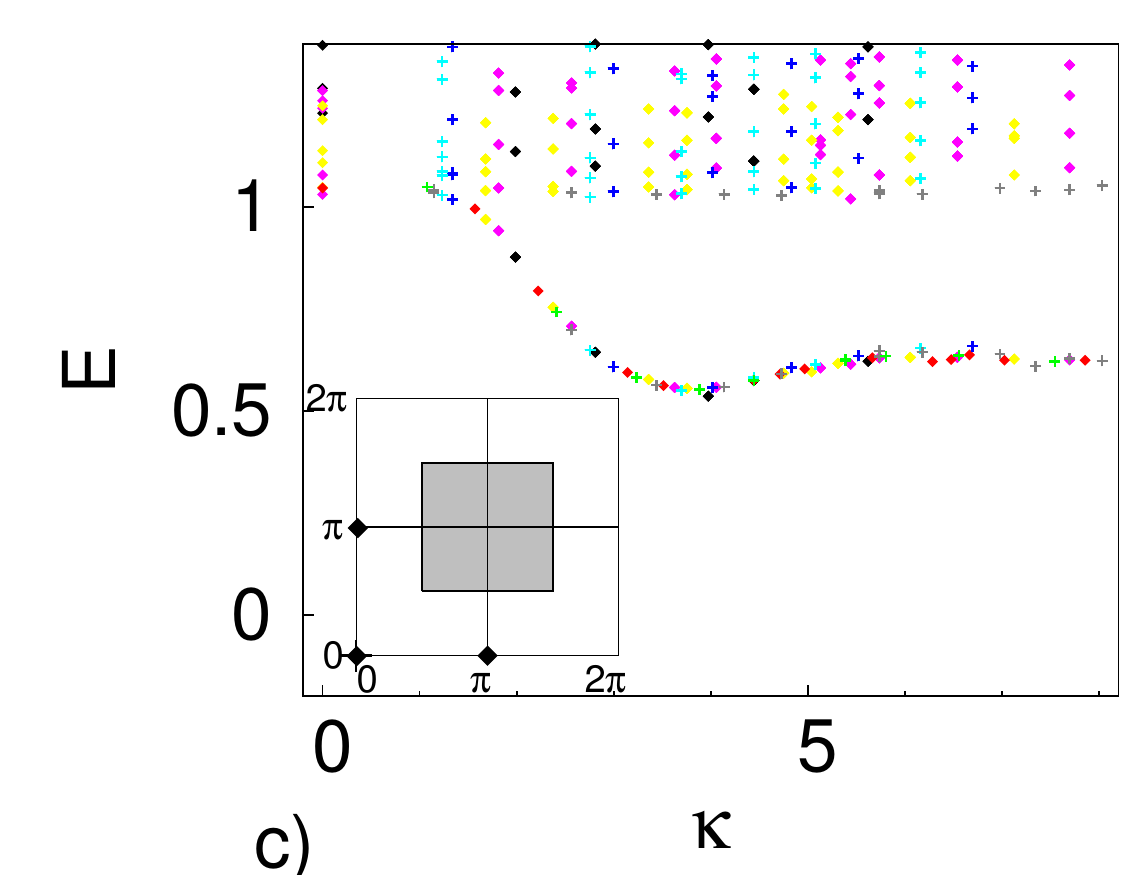}
\includegraphics[width = 0.1\linewidth]{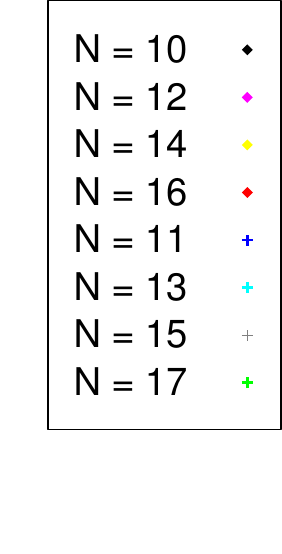}
\caption{Low energy spectrum of the model three-body contact interaction (Moore-Read or $\mathbb{Z}_2$ Read-Rezayi state) on the torus for as a function of the one-dimensional momentum $\kappa$ defined in Eq.~\eqref{eq:kappa}. In each graph, we show a different domain of the Brillouin zone, which is highlighted in grey in the inset. The black dots (respectively crosses) in the inset represent the Moore-Read ground states for $N$ even (respectively odd). In each domain, $\kappa$ is defined as the distance to the center of the domain (see also Eq.~\eqref{eq:kappa}). a) For both $N$ even (dots) and $N$ odd (crosses), there is a zero energy eigenstate lying at $\bs{ k}_0 = 0$. All points fall onto the same curve, defining a neutral excitation (the magnetoroton mode). b) A zero energy state is found at each inversion symmetric momentum in this domain for $N$ even only. The neutral excitation energies fall onto two distinct curves, with the repartition determined by the parity of $N$. c) No zero energy ground state is found at $\bs{ k}_3 = (\pi,\pi)$. All neutral excitation energies fall onto the same curve.}
\label{fig:magnetoroton}
\end{figure*}

On the torus, we start out by characterizing the neutral modes above the Moore-Read state via the exact diagonalization of the three-body contact interaction. We find that both types of neutral excitation modes above the Moore-Read state 
appear irrespective of the parity of $N$ (see Fig.~\ref{fig: 3D modes}) and 
are actually different parts of one common neutral excitation mode. This neutral mode
 can be accessed in a unified way within the symmetrization construction, as we will show in Sec.~\ref{sec:symmetrizedBilayer} by starting from the low-lying excitations above the Laughlin state.

Generically, in the low energy spectrum of the $(k+1)$-body contact interactions, we observe patterns of low energy modes, with up to four distinct dispersing branches that merge with the continuum of states near the four inversion symmetric momenta of the FQH Brillouin zone $\mathrm{BZ}_{\mathrm{FQH}}$
\be
\label{eq:inv_sym}
\bs{k}_0  =  0,\quad
\bs{k}_1  =  \pi \bs{\tilde{e}}_x, \quad
\bs{k}_2  =  \pi \bs{\tilde{e}}_y, \quad
\bs{k}_3  =  \pi \bs{\tilde{e}}_x + \pi \bs{\tilde{e}}_y.
\ee
Note that $\bs{k}_1$, $\bs{k}_2$ and $\bs{k}_3$ are not valid accessible momenta when $N$ is odd, but rather are defined as reference points in the Brillouin zone. 

On the torus, the continuous rotational symmetry of the quantum Hall problem is only broken by the boundary conditions. Still, the physics governing the neutral excitation modes should be dominated by shorter length scales comparable to the magnetic length. We can thus expect that the dispersion of the neutral excitation modes is almost rotationally symmetric and can be plotted as a function of a one-dimensional momentum $\kappa$. To unveil this dispersion, we have to account for the momentum shifts with respect to the inversion symmetric momenta~\eqref{eq:inv_sym} and include a geometric factor $\sqrt{L_x L_y/N}$ in order to obtain the data collapse with all system sizes.~\cite{repellin-PhysRevB.90.045114} We thus define the linearized momentum as the appropriately rescaled minimal distance to any of the inversion symmetric momenta
\be
\kappa := \sqrt{\frac{L_x L_y}{N}} \times \mathrm{min}\left\{\left.|\bs{k} - \bs{k}_i| \ \right| \ i = 0, \cdots,3\right\}.
\label{eq:kappa}
\ee

We show the low energy spectrum of the three-body contact interaction in Fig.~\ref{fig:magnetoroton} for systems with up to $N = 17$ bosons. For clarity, we represent the three inequivalent regions of the Brillouin zone centered around $\bs{k}_0$, $\bs{k}_1 \ / \ \bs{k}_2$, and $\bs{k}_3$ in separately panels a), b) and c), respectively.
We can clearly identify two modes with different dispersion relations. Above the four ground states, there is a mode flattening out at large momentum without forming an energy minimum [seen in Fig.~\ref{fig:magnetoroton}~a) for all $N$ and in Fig.~\ref{fig:magnetoroton}~b) for $N$ even]. This mode is similar to the magnetoroton mode observed above the Laughlin state~\cite{GMP-PhysRevLett.54.581, GMP-PhysRevB.33.2481}. 
Secondly, around inversion symmetric momenta that do not harbor a ground state for a given $N$, there is another type of neutral mode that features a clear but soft energy minimum before flattening out [seen in Fig.~\ref{fig:magnetoroton}~c) for all $N$ and in Fig.~\ref{fig:magnetoroton}~b) for $N$ odd]. 

These characteristics of a mode without minimum and a mode with minimum are similar to those of the magnetoroton and the neutral fermion modes as observed for the three-body contact interaction on the sphere geometry, for an even and odd number of particles, respectively. Further, we observe that the first mode shows a noticeable finite size effect, while the second one is far better defined. This is consistent with the interpretation given to either mode in terms of weakly interacting elementary excitations on the sphere. In a background of paired quasiparticles, the magnetoroton mode is interpreted as the dispersion relation of an interacting quasiparticle-quasihole pair (i.e., two $\sigma$ quasiparticles of the underlying Ising field theory), while the neutral fermion mode corresponds to the energy of an unpaired quasiparticle (i.e., one $\psi$ Ising quasiparticle). Since two quasiparticles induce more finite size effect in a system than one, the magnetoroton mode shows more finite size effects than the neutral fermion mode, as was observed for example in Ref.~\onlinecite{moller-PhysRevLett.107.036803}. This interpretation is relatively natural on the sphere, where the magnetoroton mode (respectively the neutral fermion mode) only appears at an even (respectively odd) number of particles. It is however less obvious that this holds on the torus geometry where both the magnetoroton and the neutral fermion mode are observed at the same filling.

\subsection{Approximating the neutral modes using symmetrization}
\label{sec:symmetrizedBilayer}

\label{sec:bilayerSymmMagn}

In Refs.~\onlinecite{Rodriguez-PhysRevB.85.035128, Sreejith-PhysRevLett.107.086806, Sreejith-PhysRevLett.107.136802}, trial wave functions for the neutral modes above the Moore-Read state were obtained by symmetrizing over excitations of the Laughlin state on the sphere. If one layer is in a Laughlin ground state, while the other is in a magnetoroton state, one obtains an approximation to the Moore-Read magnetoroton state~\cite{Rodriguez-PhysRevB.85.035128, Sreejith-PhysRevLett.107.086806}. On the other hand, if one layer is in a Laughlin quasihole state and the other is in a Laughlin quasielectron state, the symmetrized state approximates a neutral fermion state~\cite{Sreejith-PhysRevLett.107.136802}. A thin torus picture provides an intuitive understanding of this approach, as discussed in App.~\ref{app: thin torus}. 

To test this bilayer construction on the torus geometry, we have performed an extensive numerical study. We found that either periodic boundary conditions or twisted boundary conditions can be used to obtain a complete set of trial states for the entire neutral excitation branch of the Moore-Read state on the torus. We obtained good quantitative agreement both regarding the dispersion of the neutral excitation modes as well as the wave function overlap. 

We now compare the symmetrization construction of the neutral modes with the results from exact diagonalization.

\subsubsection{Neutral excitations from the bilayer torus with periodic conditions}
We start with the bosonic Moore-Read state on the torus with an even number of particles $N$. As a trial wave function for its neutral excitations we use
\be
\ket{\Psi^{\mathrm{ex}}_{2}}  = {\mathcal S}_{2\to 1}\left(
\ket{\Psi_{1}} \otimes 
\ket{\Psi^{\mathrm{ex}}_{1}}
\right),
\label{eq:SbilayerEx}
\ee
i.e., the symmetrized product of one layer with the Laughlin ground state $\ket{\Psi_{1}}$  and one layer with its neutral excitation $\ket{\Psi^{\mathrm{ex}}_{1}}$, the magnetoroton mode.

In Refs.~\onlinecite{Rodriguez-PhysRevB.85.035128,Sreejith-PhysRevLett.107.136802}, $\ket{\Psi_{1}}$ and $\ket{\Psi^{\mathrm{ex}}_{1}}$ were obtained using the composite fermion construction~\cite{jain-PhysRevLett.63.199}. Here, we use the states resulting from exact diagonalization of the model two-body contact interaction, since the composite fermion construction on the torus is at best a tedious task~\cite{Hermanns-PhysRevB.87.235128}.
Note that the symmetrization procedure does not preserve the $\mathsf{k}_x$ quantum number. Specific linear combinations of the resulting states have to be formed to obtain eigenstates of the corresponding translation operator $\mathcal{T}_x^{\mathrm{rel}}$. 
The overlap of the states constructed this way with the Moore-Read neutral excitation mode is of the order of $0.99$ for well-defined magnetoroton states, i.e., states below the continuum of excitations that are found above a critical $\kappa$. This is the same order of magnitude as the overlaps obtained on the sphere in Refs.~\onlinecite{Rodriguez-PhysRevB.85.035128,Sreejith-PhysRevLett.107.136802}. 

\begin{figure*}
\includegraphics[width = 0.3\linewidth]{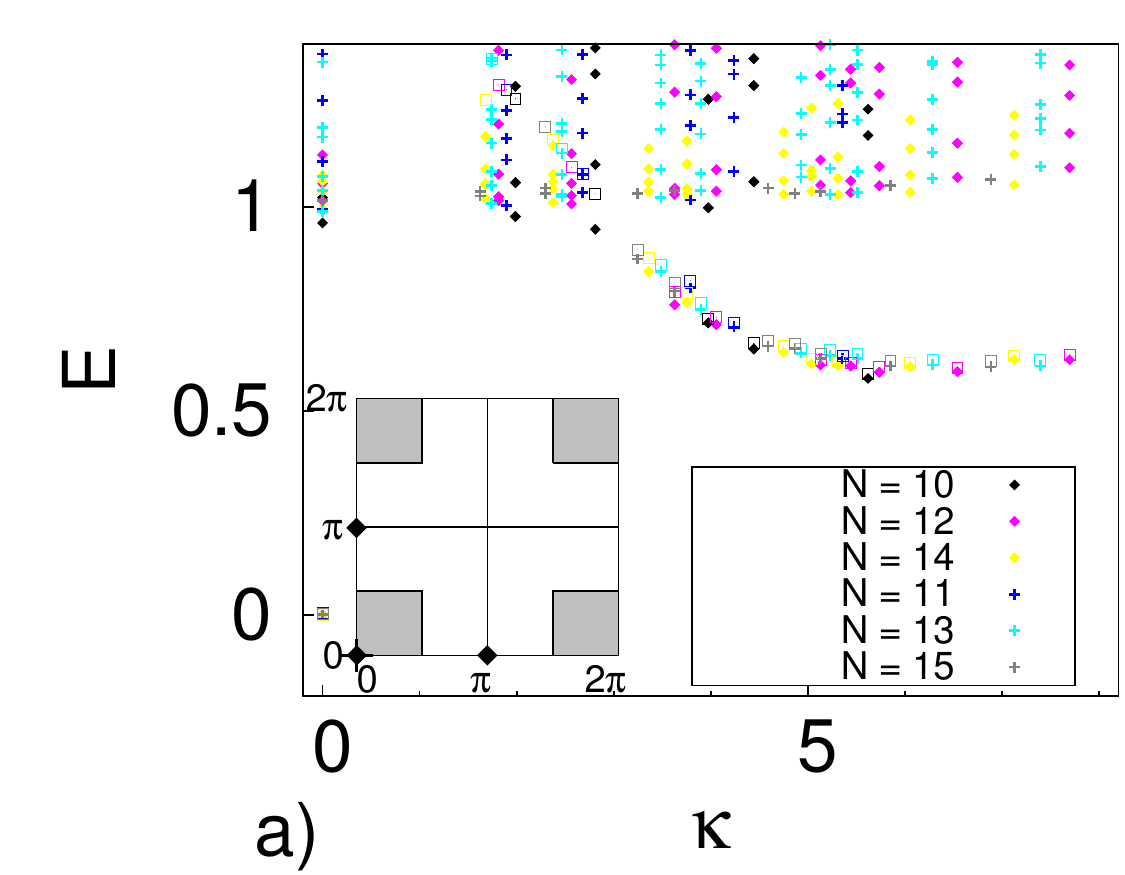}
\includegraphics[width = 0.3\linewidth]{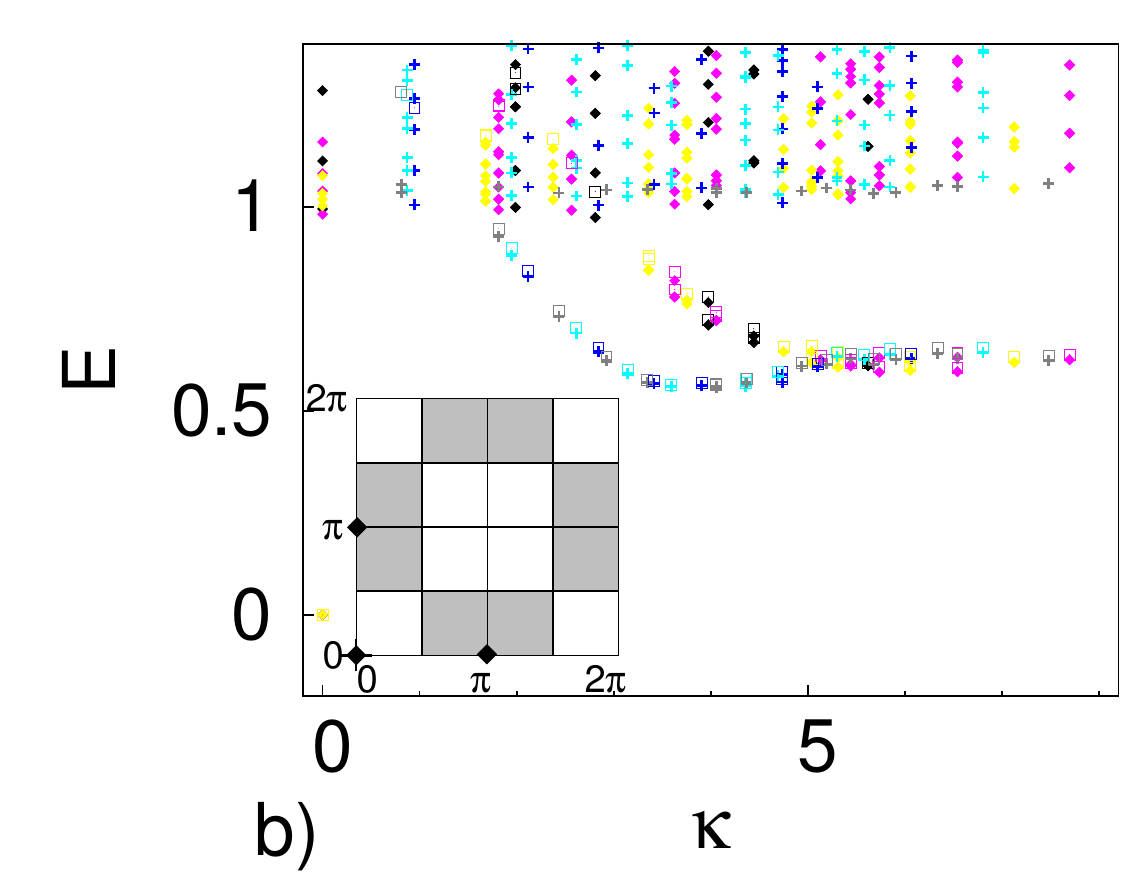}
\includegraphics[width = 0.3\linewidth]{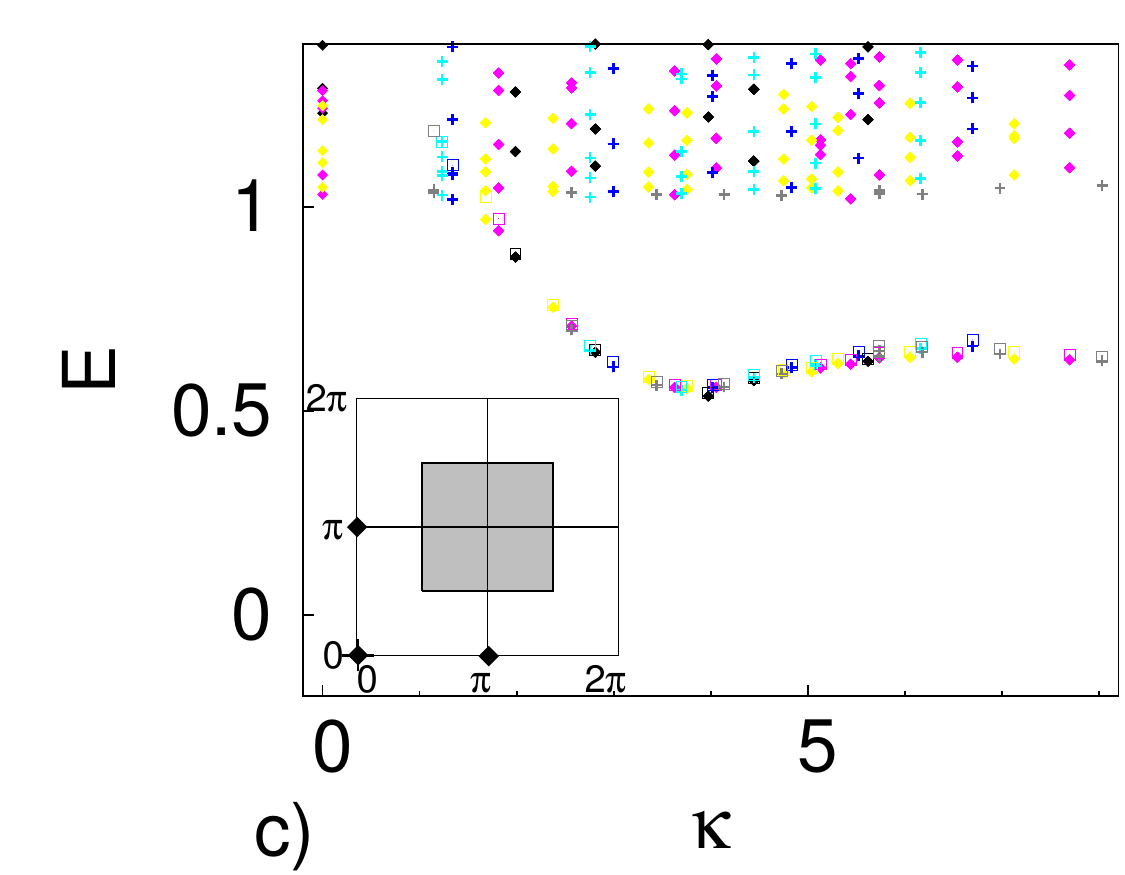}
\includegraphics[width = 0.3\linewidth]{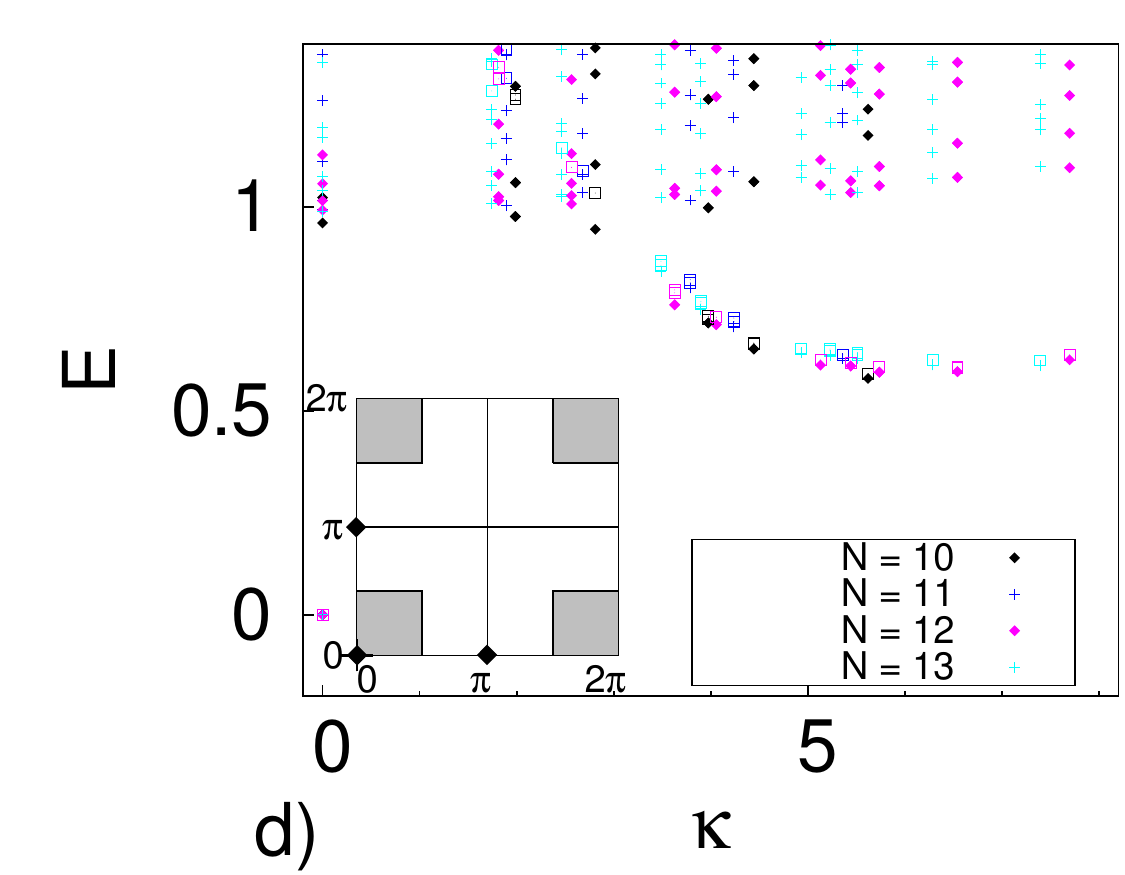}
\includegraphics[width = 0.3\linewidth]{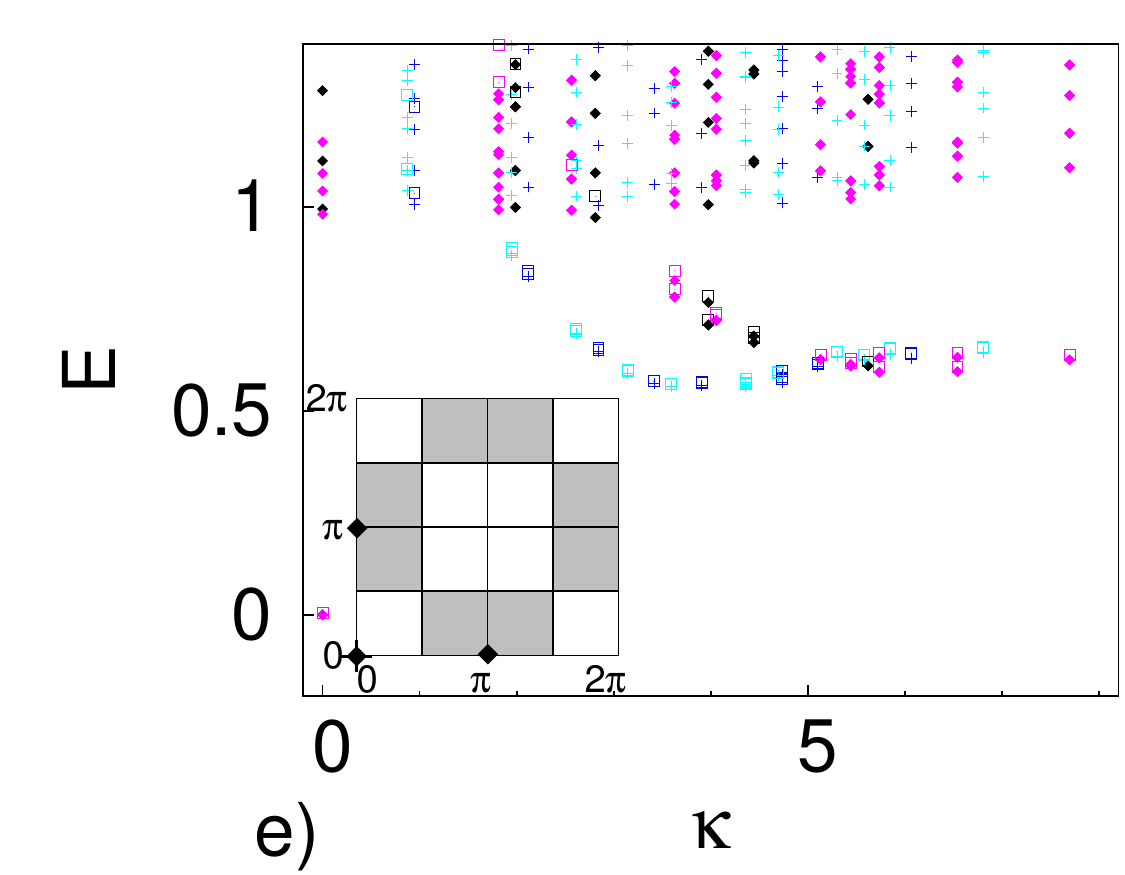}
\includegraphics[width = 0.3\linewidth]{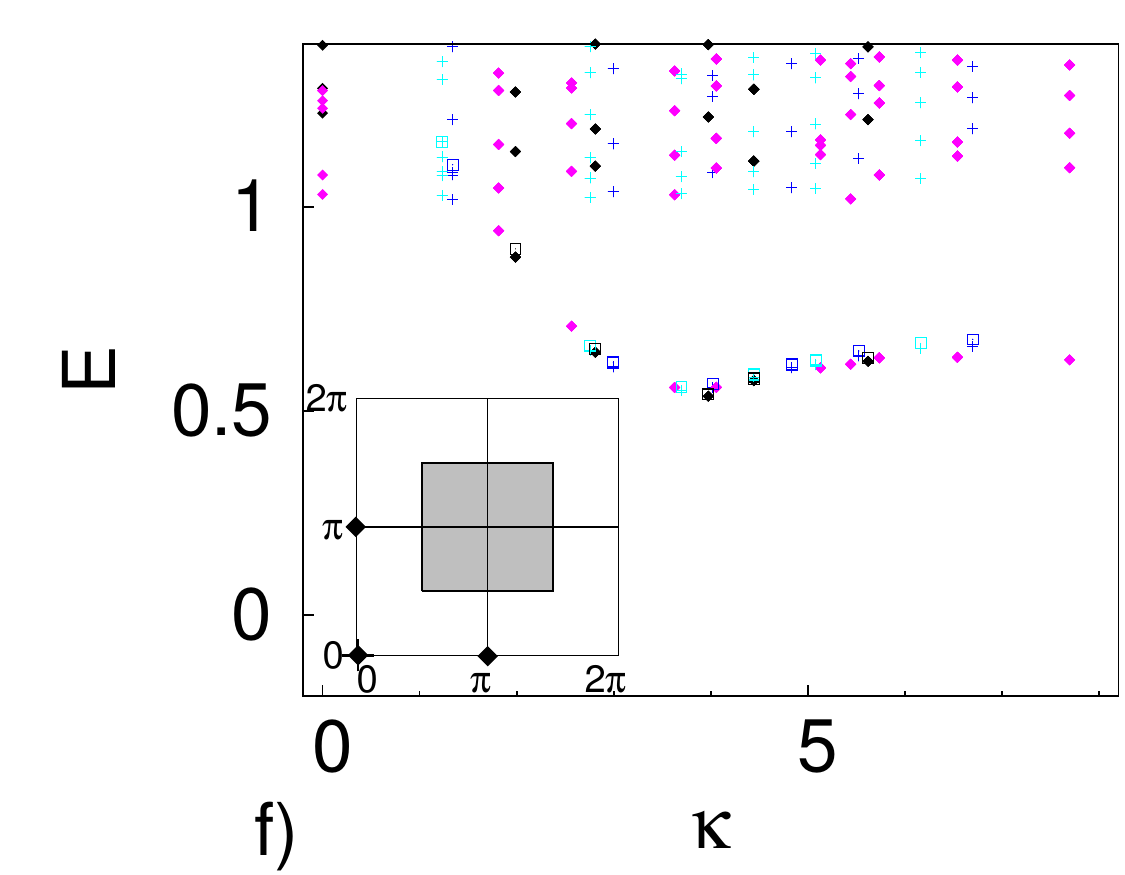}
\caption{
Low energy spectrum of the FQH on the torus as a function of the modulus of the momentum $\kappa$, for the model three-body contact interaction. We compare the energy of the trial states obtained by symmetrization (hollow squares) to the result of exact diagonalization (filled squares and crosses). The variational states are obtained by symmetrizing a decoupled bilayer system (a, b, c) or a single layer system defined on the $(2T)_x$ torus (c, d, e). The energies of these last states differ by less than a percent from the energies of the states obtained by the symmetrization of a system defined on the $(gT)_y$ system. We thus only show the former. See Sec.~\ref{sec:bilayerSymmMagn} for details on the construction of the trials states. In each graph, we show a different domain of the Brillouin zone, which is highlighted in grey in the inset. The black dots (respectively crosses) in the inset represent the Moore-Read ground states for $N$ even (respectively odd). In each domain, $\kappa$ is defined as the distance to the center of the domain (see also Eq.~\eqref{eq:kappa}).}
\label{fig:symBilayer}
\end{figure*}

When the number of particles $N$ is odd, we use the following trial state
\be
\ket{\Psi^{\mathrm{ex}}_{2}}  = {\mathcal S}_{2\to 1}\left(
\ket{\Psi^{\mathrm{qh}}_{1}} \otimes 
\ket{\Psi^{\mathrm{qe}}_{1}}
\right),
\label{eq:SbilayerQHQE}
\ee
i.e., the symmetrized product of one layer with one Laughlin quasihole state $\ket{\Psi^{\mathrm{qh}}_{1}}$ and one layer with a Laughlin quasielectron state $\ket{\Psi^{\mathrm{qe}}_{1}}$ (obtained using exact diagonalization). 
Equation~\eqref{eq:SbilayerQHQE} can be used to generate trial states for the Moore-Read ground state $\ket{\Psi^{\,}_{2}}$ as well as its neutral excitations $\ket{\Psi^{\mathrm{ex}}_{2}}$.
Although this method does not yield any exact zero energy state of the three-body contact interaction, the approximation it provides for the Moore-Read ground state is fairly good (with an overlap of $0.998$ with the Moore-Read ground state for $N=17$). We expect this trial wave function to become the exact Moore-Read ground state should we use the exact composite fermion wave functions for $\Psi^{\mathrm{qh}}_{1}$ and $\Psi^{\mathrm{qe}}_{1}$. This conjecture is motivated by results on the sphere that will be discussed in Sec.~\ref{sec: sphere}. Likewise, $\Psi^{\mathrm{ex}}_{2}$ is a good approximation of the neutral excitation states (with overlaps of the order of $0.99$ for states well into the neutral mode). 
The energies of the trial states for both $N$ even and odd are represented in Fig.~\ref{fig:symBilayer} along with the exact energies.

Hence, this method yields very satisfactory trial states for the neutral excitations above the Moore-Read state on the torus.

In the following, we will see that the alternative symmetrization with twisted boundary conditions developed in Sec.~\ref{sec: symm with twisted} also yields a good approximation of the neutral modes above the Moore-Read state.

\subsubsection{Neutral excitations from the bilayer torus with twisted boundary conditions}
As a trial wave function for the neutral excitation states above the Moore-Read state, we now use the following wave function
\be
\ket{\Psi^{\mathrm{ex}}_{2, x}}  = {\mathcal S}_{(2T)_x \rightarrow T}
\ket{\Psi^{\mathrm{ex}}_{1}}.
\label{eq:SxMagnetoroton}
\ee
This is the result of the action of the symmetrization operator on a Laughlin magnetoroton state defined on a twice enlarged torus $(2T)_x$. Similarly, applying the symmetrization operator ${\mathcal S}_{(2T)_y \rightarrow T}$ on a Laughlin magnetoroton state defined on the $(2T)_y$ torus yields another trial state for the Moore-Read neutral mode.

These states have high overlap with the exact neutral mode states (of the order of $0.99$ for $N = 13$). For a given system size, the overlaps with the trial states provided by the different symmetrization constructions [see Eqs.~\eqref{eq:SbilayerEx}, \eqref{eq:SbilayerQHQE}, and \eqref{eq:SxMagnetoroton}] have very similar values. We plot the energies of $\ket{\Psi^{\mathrm{ex}}_{2, x}}$ in Fig.~\ref{fig:symBilayer} ($\ket{\Psi^{\mathrm{ex}}_{2, y}}$ yields the same spectrum up to one percent accuracy), and compare them to the exact spectrum of the three-body Hamiltonian. We note that the two subspaces created by acting on  the Laughlin magneto-roton states with either ${\mathcal S}_{(2T)_x \rightarrow T}$ or ${\mathcal S}_{(2T)_y \rightarrow T}$ are distinct, and individually not invariant upon a $\pi/2$ rotation (as already noticeable for the ground state). However, we numerically verified that these two subspaces are related by a rotation of $\pi/2$.

We have thus verified that the symmetrization methods provided satisfactory approximations to the neutral modes above the Moore-Read state on the torus. As expected, the symmetrization scheme previously known on the sphere is equally valid on the torus, but can also be extended to a torus with twisted boundary conditions. It is however harder to reach large system sizes with this new scheme, because it would imply calculating the Laughlin magnetoroton states for larger number of particles, a great numerical challenge. The same limitation arises when approximating the neutral mode for larger $k$: we expect the symmetrized Laughlin magnetoroton states on a $g$ times enlarged torus to yield good trial states for the $\mathbb{Z}_g$ neutral mode. However, it is numerically very difficult to compute the Laughlin neutral mode for more than $N=13$ particles. For this number of particles, the neutral modes above the $\mathbb{Z}_g$ states with $g>2$ still show large finite size effects.

\section{Symmetrization on the sphere}
\label{sec: sphere}

The main result from the previous section are two alternative symmetrization schemes to construct exact ground states and trial excited states of the $\mathbb{Z}_{gk}$ Read-Rezayi type from $\mathbb{Z}_{k}$ parent states on the torus geometry. These symmetrization schemes were motivated by the geometrical equivalence between the double-layered torus with twisted boundary conditions and a twice as large torus. 

Fractional quantum Hall states have been studied on a variety of other manifolds, aside from the torus. For example, model wave functions are most commonly written in a planar geometry, while finite-size numerical studies are often also carried out on the sphere. In particular, the multilayer construction of $\mathbb{Z}_{k}$ Read-Rezayi states was first numerically verified~\cite{Regnault-PhysRevLett.101.066803} on the sphere. While there is no obstruction to use the multilayer construction for any of the Read-Rezayi states on the sphere or on the plane, as there was on the torus, one can ask whether these geometries also allow for an alternative symmetrization scheme. 
In this section, we will show that this is indeed the case, i.e., the symmetrization scheme developed in this article for the torus can be adapted to the sphere geometry or to any genus zero manifold (including the disk and the cylinder). Moreover, this will provide us with another elegant way to generate trial states for the neutral excitation modes in these geometries. In particular, we will give an alternative symmetrization construction on the sphere that consists in mapping a single-layer sphere to a smaller single-layer sphere in this section.

\subsection{$\mathbb{Z}_k$ Read-Reazayi states from symmetrization on the sphere}

\begin{figure}
\includegraphics[width = 0.9\linewidth]{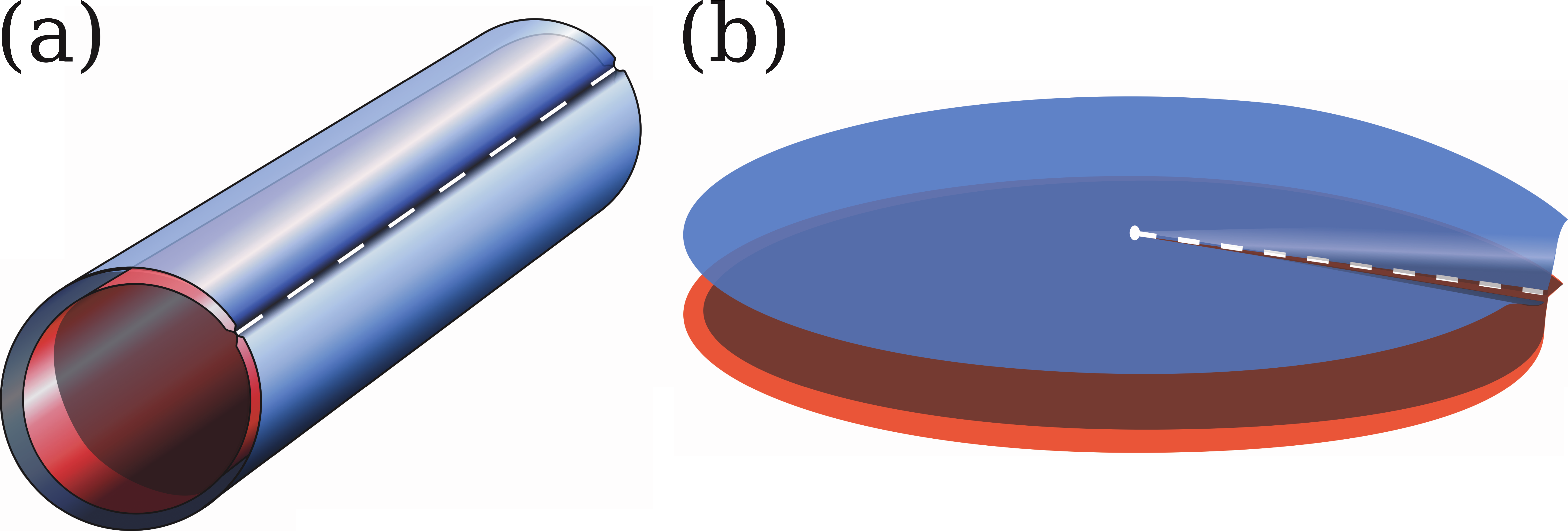}
\caption{
Geometrical interpretation of the alternative symmetrization prescription on rotationally symmetric manifolds, such as a) the cylinder or b) the disk. In either case, the double layer is joined into a singly connected surface via a topological twist defect.  
}
\label{fig:planesphere}
\end{figure}

The construction of Sec.~\ref{sec: intro to symmetrization} is motivated by the geometrical equivalence between a $g$-layered torus with twisted boundary conditions and a $g$ times longer torus. 
The symmetrization is a projection on single-particle eigenstates with a fixed eigenvalue $e^{2\pi\mathrm{i}s/g}$ under a translation of a $g$-th fraction of the length of the long torus. 
While the sphere or plane does not admit an analogous geometrical manipulation, we can nevertheless define the equivalent operation to the translation on the torus. It is the rotation by $2\pi/g$, that respects the symmetry of the cylinder, the sphere, and the plane (disk). We are lead to define a symmetrization operation on rotationally symmetric manifolds as the projection on the single-particle eigenstates with eigenvalue $e^{2\pi\mathrm{i}s/g}$ under the $g$-th fraction of a full rotation (see Fig.~\ref{fig:planesphere}).
We anticipate that this projection has a particularly simple representation in the basis of the single particle orbitals, if the chosen gauge is invariant under the same rotation.
On such rotationally symmetric geometry, the single-particle orbitals are eigenstates of the angular momentum $L_z$. Implementing the symmetrization operation amounts to projecting on single-particle orbitals whose angular momentum $j$ is a multiple of $g$.
On the polynomial part of the wave function, the symmetrization can be written in first quantization as
\begin{equation}
\begin{split}
&{\mathcal S}_{gS\rightarrow S}\Psi\left(z_1,...,z_N\right) =\\
&\quad \sum_{\{0 \le \delta_i < g\}} \Psi\left(e^{\frac{2\pi \mathrm{i} \delta_1}{g}}z_1^{\frac{1}{g}},...,e^{\frac{2\pi \mathrm{i} \delta_N}{g}}z_N^{\frac{1}{g}}\right) \prod_i \left[e^{-\frac{2\pi \mathrm{i} \delta_i s}{g}} z_i^{-\frac{s}{g}}\right],
\label{eq:firstquantizedsymmetrizationplane}
\end{split}
\end{equation}
The sum over all $0\leq\delta_i<g$ with $\delta_i\in \mathbb{Z}$ ensures that only integer powers of the coordinates $z_i$ survive in ${\mathcal S}_{gS\rightarrow S}\Psi\left(z_1,...,z_N\right) $, so that the wave function is single-valued and thus physical.
Using the stereographic projection, one can translate this procedure from the disk to the sphere.
The second-quantized representation of the symmetrization operation on the sphere or disk is exaclty the same as Eq.~\eqref{eq: symm every goth orbital} on the torus. From the same argument that the one developed for Eq.~\eqref{eq:firstquantizedsymmetrization} on the torus, we immediately deduce that Eq.~\eqref{eq:firstquantizedsymmetrizationplane} vanishes when we put $(g+1)$ particles at the same position if $\Psi$ is a Laughlin ground state or quasihole state. Moreover, the vanishing power of the symmetrized state is identical to  the one of $\Psi$. Notice that even though we focus on the $\mathbb{Z}_{k}$ Read-Rezayi series in this article, our conjecture holds true for any $(k,r)$ clustered states\cite{Bernevig-PhysRevLett.100.246802,Bernevig-PhysRevB.77.184502}. More specifically, Eq.~\eqref{eq:firstquantizedsymmetrizationplane} maps any $(k,r)$ clustered state onto a $(gk,r)$ clustered state.

We now focus on the sphere geometry. There, the symmetrization operation ${\mathcal S}_{gS\rightarrow S}$ precisely maps the Hilbert space of a sphere with $gS$ orbitals in the lowest Landau level to that of a sphere with $S$ orbitals, once the correct normalization factors are put in place. A complication that was not present on the torus is the shift of the sphere, i.e., the fact that the number of orbitals $S$ in the lowest Landau level is not equal to the number of flux quanta $N_\phi$.
We start from a system with a number of orbitals $gS=g(N_\phi+1)$ commensurate with $g$ on the sphere. Acting on a many-body state with the symmetrization operator ${\mathcal S}_{gS\rightarrow S}$ divides the number of one-body orbitals by $g$. The symmetrization thus maps a sphere with $N^{\mathrm{in}}_\phi$ flux quanta into a sphere with 
\be
N_\phi=\lfloor\frac{N^{\mathrm{in}}_\phi+1}{g} \rfloor-1
\ee
flux quanta, where $\lfloor \rfloor$ is the floor function
States on the sphere are characterized by a total angular momentum quantum number $L_z$ along the $z$-direction. 
For a system with $N^{\mathrm{in}}_\phi$ flux quanta, an eigenstate $\ket{\Psi\left(L_z\right)}$ is written as a superposition of Fock states $\ket{\lambda}$ such that
\be
\label{eq:state in configuration notation sphere}
\ket{\Psi\left(L_z\right)} = \sum_{\ket{\lambda} \in {\cal H}_{N^{\mathrm{in}}_\phi}\left(L_z\right)} b_{\lambda}\ket{\lambda},
\ee
where $b_{\lambda}\in\mathbb{R}$ and ${\cal H}_{N^{\mathrm{in}}_\phi}\left(L_z\right)$ is the Hilbert space restricted to the constraint
\be
\sum_{j = 0}^{N^{\mathrm{in}}_\phi}\left(j-N^{\mathrm{in}}_\phi/2\right) n_{j}\left(\lambda\right) = L_z.
\ee
As shown in Ref.~\onlinecite{Bernevig-PhysRevLett.100.246802}, the $\mathbb{Z}_k$ states are Jack polynomials~\cite{Stanley198976}. Up to geometrical and occupation factors, their components in the occupation number basis are thus rational numbers. We can write 
\be
b_{\lambda} = r_{\lambda}\prod_{j = 0}^{N^{\mathrm{in}}_\phi} \frac{{\cal N}_{j,N^{\mathrm{in}}_\phi}^{n_{j}\left(\lambda\right) }}{\sqrt{n_{j}\left(\lambda\right)!}},
\ee
where ${\cal N}_{j,N^{\mathrm{in}}_\phi}$ is the normalization factor of the orbital $j$, and $r_{\lambda}$ is a rational number.

Due to the curvature of the sphere, single particle orbitals on this geometry -- unlike single particle orbitals on the torus -- do not all have the same normalization factor for a given $N_{\phi}$, i.e., in general, ${\cal N}_{j,N^{\mathrm{in}}_\phi} \neq {\cal N}_{j',N^{\mathrm{in}}_\phi}$ if $j \neq j'$. To go from the large sphere to the small sphere using the symmetrization operator ${\mathcal S}_{gS\rightarrow S}$, one thus needs to adjust the normalization factors accordingly. The symmetrized state reads
\ba
{\mathcal S}_{gS\rightarrow S}  \ket{\Psi\left(L_z\right)} = 
\sum_{\ket{\lambda} \in {\cal H}_{N^{\mathrm{in}}_\phi}\left(L_z\right)} b_{s\left(\lambda\right)} \ket{s\left(\lambda\right)},
\ea
where $\ket{s\left(\lambda\right)}$ is defined by the occupation numbers
\be
n_j\left[s\left(\lambda\right)\right] = n_{gj+ s}\left(\lambda\right), \quad j = 0, \cdots, N_{\phi} 
\ee
and the coefficients $b_{\mu}$ of the symmetrized states are just renormalized by their partitions' respective amplitudes in the two states.
\be
b_{\mu} = \sum_{\lambda :\, s\left(\lambda\right) = \mu}b_{\lambda}\prod_{j=0}^{N_{\phi} }\left[\frac{{\cal N}_{gj + s,N_{\phi}}^{n_{gj + s}\left(\lambda\right)}}{\prod_{i = 0}^{g-1}{\cal N}_{gj + i,N^{\mathrm{in}}_{\phi}}^{n_{gj + i}\left(\lambda\right)}}\sqrt{\frac{n_{gj+ s}\left(\lambda\right)!}{\prod_{i = 0}^{g-1} n_{gj+ i}\left(\lambda\right)!}}\right],
\ee
where $0 \leq s < g$ is the sphere equivalent of the integer defined in Eq.~\eqref{eq: symm every goth orbital} on the torus geometry. In terms of the rational coefficients, this reads
\be
r_{\mu} = \sum_{\lambda :\, s\left(\lambda\right) = \mu}r_{\lambda} \prod_{j=0}^{N_{\phi} }\frac{n_{gj+ s}\left(\lambda\right)!}{\prod_{i = 0}^{g-1} n_{gj+ i}\left(\lambda\right)!}.
\ee
Starting from a state with fixed $L_z$, the symmetrized states generically have weight in different $L_z$ sectors.

\subsection{Numerical results}
We numerically checked that the application of the above described symmetrization operator ${\mathcal S}_{gS\rightarrow S}$ to a $\mathbb{Z}_k$ state yielded zero energy states of the $(gk + 1)$-body contact interaction. It is convenient to work in the rescaled basis where all state components are rational numbers $r_{\lambda}$. This allows for the exact comparison of states, despite the fact that we are using numerical tools. (In contrast, for the torus discussed in the previous section, to wording ``exactly equal" meant always ``equal up to numerical precision''.) We are thus effectively probing the properties of some Jack polynomials under symmetrization, and our conclusions will apply to the Jacks themselves.

As we are interested in finite-size systems, we have to account for an altered relation between the number of particles $N$ needed to obtain the $\mathbb{Z}_{k}$ Read-Rezayi ground state on a sphere with $N_\phi$ flux quanta
\begin{equation}
k(N_\phi-2)=2N
\label{eq: N Nphi sphere}
\end{equation}
as compared to the relation on the torus Eq.~\eqref{eq: filling factor relation torus}. Given this constraint, the number of single particle orbitals on the large sphere is rarely commensurate with $g$. The polynomial nature of the $\mathbb{Z}_{k}$ wave function allows us to easily solve this problem: adding an additional empty orbital at the end of an occupation number configuration is tantamount to multiplying each monomial by a factor of $1$, which leaves the wave function unaltered.

We numerically confirmed that the application of ${\mathcal S}_{gS\rightarrow S}$ to any zero energy state of the $(k+1)$-body contact interaction yields a zero energy state of the $(gk+1)$-body contact interaction. Focusing on the most compressible state with a given number of particles, this property was verified for $k=1$ and $g = 2,\ 3,\ 4$ with up to $12$ particles, and for $k=2$ and $g = 2$ with up to $12$ particles. Moreover, the quasihole space built by symmetrizing all $\mathbb{Z}_k$ quasihole states is complete (i.e. it spans the full $\mathbb{Z}_{gk}$ quasihole space for the same number of particles and a number of single particle orbitals reduced by one unit and divided by $g$). This property was verified for $k=1$ and $g = 2,\ 3,\ 4$ with up to $7$ particles and $3$ added flux quanta, and for $k=2$ and $g = 2$ with up to $7$ particles and $3$ added flux quanta. Note that in general, the symmetrized states are not Jacks, but some linear combination of Jacks.

Finally, we used this procedure to produce trial wave functions for the neutral mode above the Moore-Read state on the sphere. The initial states are the neutral low energy states above the Laughlin state. These states can be approximated using the composite fermion construction~\cite{jain-PhysRevLett.63.199}, by considering a Laughlin system with one quasielectron and one quasihole. Since composite fermion states have exact nontrivial vanishing properties,~\cite{PhysRevLett.103.016801} we expect that the symmetrized state will also possess similar features. Symmetrizing the magnetoroton states, with $g=2$, yields one state per value of $L$, except for $L=0$ and $L=1$ if $N$ is even (respectively $L = 1/2$ if $N$ is odd). Note that there is no neutral mode state in these sectors. For the accessible system sizes (up to $N = 14$ particles), we also verified that the trial states, although different from the ones obtained in Refs.~\onlinecite{Rodriguez-PhysRevB.85.035128, Sreejith-PhysRevLett.107.086806, Sreejith-PhysRevLett.107.136802}, provide an equally good approximation to the neutral modes. The symmetrization construction described in this section thus provides a unique method to approximate both the magnetoroton and the neutral fermion modes on the sphere. In contrast, the methods of Refs.~\onlinecite{Rodriguez-PhysRevB.85.035128, Sreejith-PhysRevLett.107.086806, Sreejith-PhysRevLett.107.136802} all used two different sets of initial states to obtain these modes.

We have thus provided an alternative symmetrization method on the sphere. $\mathbb{Z}_{gk}$ states can be obtained by symmetrizing a $\mathbb{Z}_k$ state defined on a $g$ times larger sphere. For a state written in the occupation number basis, the symmetrization procedure consists in selecting one in $g$ orbitals and discarding all of the other ones. This procedure was directly adapted from a similar symmetrization scheme on the torus, accounting for the sphere specificity in the normalization of the single particle orbitals. We numerically confirmed that all $\mathbb{Z}_{gk}$ quasihole states could be obtained using this method. This highlights a previously unknown property of some of the Jack polynomials. In contrast, the alternative symmetrization procedure 
that could be derived from the second quantized representation of $S_{(gT)_x\to T}$ (see Appendix~\ref{app:  numerical symmetrization}) destroys the squeezing hierarchy of the occupation number basis on the sphere, and therefore cannot be used.

\section{Conclusion}
\label{ref: Conclusion}

This study was motivated by the fact that the previously known multi-layer symmetrization construction of $\mathbb{Z}_k$ Read-Rezayi states misses several zero energy state of the model Hamiltonian on the torus. 
Our result is an alternative projective construction scheme that turns out to be as powerful as the multi-layer symmetrization construction and ideally complements it. 
This novel construction obtains a $\mathbb{Z}_{gk}$ state from a $\mathbb{Z}_{k}$ by reducing the size of the manifold  --  torus or sphere -- on which the state is defined by a factor $g$. 
The construction has a suggestive geometrical interpretation in terms of topological twist defects connecting the different layers.
On the sphere or the plane, it leads to a conjecture regarding the mathematical properties of some Jack polynomials, and more generically the clustered states.

Beyond exact statements on zero-energy states, we showed that the projective constructions also yield excellent approximations to the collective low energy neutral excitation modes above the $\mathbb{Z}_k$ Read-Rezayi states.

Our results open several natural directions for future work, such as the extension to other fractional quantum Hall states, in particular those of fermions.

\section*{Acknowledgement}
We acknowledge Z. Papi\ifmmode \acute{c}\else \'{c}\fi{}, B. Estienne and M. Barkeshli for discussions. We are especially grateful to Curt von Keyserlingk for helpful comments.
The authors acknowledge financial support from DARPA SPAWARSYSCEN Pacific N66001-11-1-4110. B.A.B and N.R. were supported by NSF CAREER DMR-095242, ONR-N00014-11-1-0635,MURI-130- 6082, MERSEC Grant, Packard Foundation, Keck grant and the Princeton Global Scholarship. NR and CR were supported by ANR-12-BS04-0002-02.

\appendix

\section{Translation operators and momentum quantum numbers}
\label{app: translation symmetry}

We consider a rectangular torus spanned by the vectors $L_x \bs{e}_x$ and $L_y \bs{e}_y$, where $\bs{e}_x$ and $\bs{e}_y$ are two unit vectors. 
Translation operators on the torus can be factorized into the product of a center of mass and a relative translation. The center of mass translation operator along the $y$ axis and the relative translation operator along the $x$ axis commute with each other and with the Hamiltonian. The eigenstates of the Hamiltonian thus carry the corresponding momentum quantum numbers $\bs{k}$ that belong to the FQH Brillouin zone
$\mathrm{BZ}_{\mathrm{FQH}}
\equiv\left\{
\bs{k}=\frac{2\pi}{L_x}\mathsf{k}_x \bs{\tilde{e}}_x + \frac{2\pi}{L_y}\mathsf{k}_y \bs{\tilde{e}}_y
\right\}$
with 
$\mathsf{k}_x=0,\cdots, \mathrm{GCD}(N,N_\phi)-1$ and 
$\mathsf{k}_y=0,\cdots, N_\phi-1$.
Here,  GCD stands for the greatest common divisor, and $\bs{\tilde{e}}_x, \bs{\tilde{e}}_y$ are such that $\bs{e}_i \cdot \bs{\tilde{e}}_j = \delta_{i,j}$.
Focusing on the densest $\mathbb{Z}_k$ states, for $k$ odd (respectively even), the momentum quantum numbers $\mathsf{k}_x, \mathsf{k}_y$ thus belong to a  $N_{\phi}/2 \times N_{\phi}$ (respectively $N_{\phi} \times N_{\phi}$) Brillouin zone.
These momentum quantum numbers are defined in the Landau $x$-gauge with vector potential $\bs{A}(\bs{r})=(0,-Bx)$. 

Due to the center of mass symmetry, there is a $q$-fold degeneracy of all states in the FQH Brillouin zone, where $q = N_\phi/\mathrm{GCD}(N,N_\phi)$. Without losing information, we can thus work in a reduced Brillouin zone~\cite{Haldane85-PhysRevLett.55.2095} of size $\mathrm{GCD}(N,N_\phi) \times \mathrm{GCD}(N,N_\phi)$
\be
\begin{split}
\mathrm{BZ}^{\mathrm{red}}_{\mathrm{FQH}}
\equiv&
\left\{
\bs{k}=\frac{2\pi}{L_x}\mathsf{k}_x \bs{\tilde{e}}_x + \frac{2\pi}{L_y}\mathsf{k}_y \bs{\tilde{e}}_y
\right|
\\
&
\left.
\mathsf{k}_x=0,\cdots, \mathrm{GCD}(N,N_\phi)-1 ;\right.
\\
&
\left.\vphantom{\frac{2\pi}{L_x}}
\mathsf{k}_y=0,\cdots, \mathrm{GCD}(N,N_\phi)-1
\right\}.
\end{split}
\ee
In the case of the $\mathbb{Z}_k$ Read-Rezayi densest state, the previous equation leads to a reduced Brillouin zone $\mathrm{BZ}^{\mathrm{red}}_{\mathrm{FQH}}$ of size $N_{\phi}/2 \times N_{\phi}/2$ (respectively $N_{\phi} \times N_{\phi}$) for $k$ odd (respectively $k$ even). Note that for $k$ even, the reduced Brillouin zone coincides with the full Brillouin zone, while for $k$ odd, $\mathrm{BZ}^{\mathrm{red}}_{\mathrm{FQH}}$ corresponds to half of the full Brillouin zone.

\section{Symmetrization on a double-layer torus with twisted boundary conditions in both directions} 
\label{app: twisted}

\begin{figure}
\includegraphics[width = 1.0\linewidth]{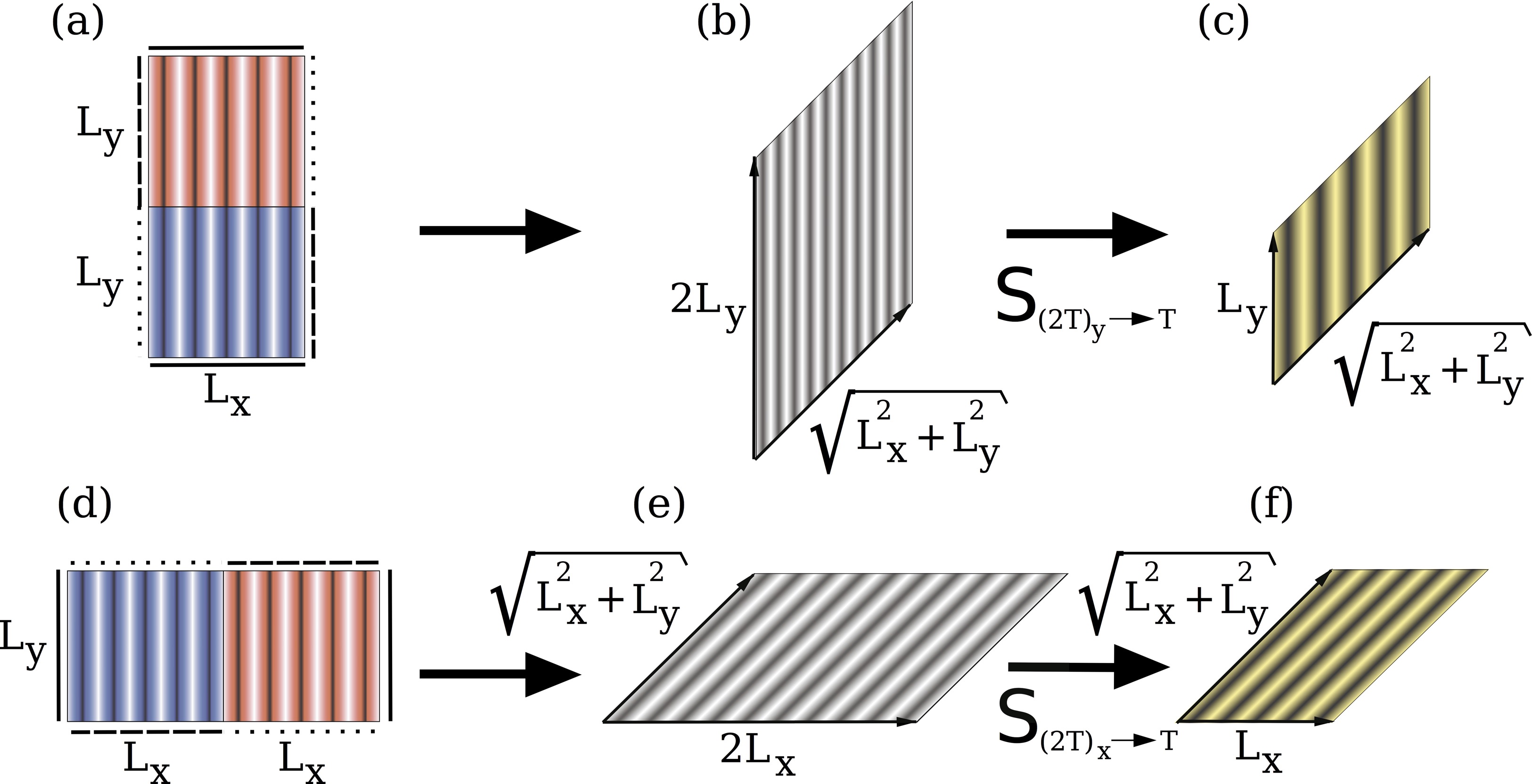}
\caption{
Sketch of the two possible symmetrizations for the double-layer torus with twisted boundary conditions in both directions. We look at the simple case where $L_x=L_y$. (a) and (d) show the initial system where the boundaries have to be glued according to the plain, dotted or dashed lines at the edge. (a) requires the equivalent single-layer system (b) to be defined by the spanning vectors $L_x \bs{e}_x+L_y \bs{e}_y$ of length $\sqrt{L_x^2 + L_y^2}$ and $2 L_y \bs{e}_y$. Analogously, the single-layer system (e) for (d) is set by the spanning vectors $2 L_x \bs{e}_x$ and $L_x \bs{e}_x + L_y \bs{e}_y$ of length $\sqrt{L_x^2 + L_y^2}$.  From (b) [resp. (e)], we apply the symmetrization procedure $\mathcal{S}_{(2T)_y \rightarrow T}$ (resp. $\mathcal{S}_{(2T)_x \rightarrow T}$) to obtain a Moore-Read state (c) [resp. (f)]. In both cases, one ends up with a Moore-Read state on a twisted torus. Notice that the orientation of the orbitals (along $\bs{L}_2$) depicted here is chosen accordingly to the gauge used when expending the wave functions on the many-body basis.
}
\label{fig:double twisted torus}
\end{figure}

In this Appendix, we discuss the result of a symmetrization over a multi-layered torus with twisted boundary conditions in both the $x$ and the $y$ directions. This manifold is geometrically equivalent to a single-layer torus as well. For simplicity, let us consider a double-layered torus spanned by two orthogonal vectors $L_x \bs{e}_x$ and $L_y \bs{e}_y$. 
It can be related to a single-layer torus in two ways (see Fig.~\ref{fig:double twisted torus}).

One option is to account for the twisted boundary conditions in $x$-direction by taking one spanning vector of the single layer torus as $\bs{L}_1=2L_x\bs{e}_x$. Then, to also account for the twisted boundary conditions in the $y$ direction, the other spanning vector must be chosen as $\bs{L}_2=L_x\bs{e}_x+L_y\bs{e}_y$. Thus, a rectangular double layer torus spanned by $L_x \bs{e}_x$ and $L_y \bs{e}_y$ with twisted boundary conditions in both directions is equivalent to a single layer torus spanned by $\bs{L}_1$ and $\bs{L}_2$.
By the arguments in Sec.~\ref{sec: geometrical equivalence on Hamiltonian level}, this geometrical equivalence also holds on the level on the Hamiltonian. 
Symmetrization reduces this $(\bs{L}_1,\bs{L}_2)$ torus to a torus spanned by $\bs{L}_1/2$ and $\bs{L}_2$. In the $x$-Landau gauge, the action of the symmetrization operator is given by Eq.~\eqref{eq: symm Nphi apart orbitals} in this case, despite the fact that the torus is nonrectangular. We numerically checked that this symmetrization of the $\nu=1/2$ Laughlin state on the  $(\bs{L}_1,\bs{L}_2)$ torus yielded the Moore-Read state on the $(\bs{L}_1/2,\bs{L}_2)$ torus for systems with up to $12$ particles.

Alternatively, to first account for the twisted boundary conditions in the $y$ direction, one may take one spanning vector of the single layer torus to be $\bs{L}_2=2L_y\bs{e}_y$. Then, to also account for the twisted boundary conditions in $x$ direction, the other spanning vector must be chosen as $\bs{L}_1=L_x\bs{e}_x+L_y\bs{e}_y$. Again, a rectangular double layer torus spanned by $L_x \bs{e}_x$ and $L_y \bs{e}_y$ with twisted boundary conditions in both directions is equivalent to a single layer torus spanned by these vectors $\bs{L}_1$ and $\bs{L}_2$.
By the arguments in Sec.~\ref{sec: geometrical equivalence on Hamiltonian level}, this geometrical equivalence also holds on the level on the Hamiltonian. 
Symmetrization reduces this $(\bs{L}_1,\bs{L}_2)$ torus to a torus spanned by $\bs{L}_1$ and $\bs{L}_2/2$. In the $x$-Landau gauge, the action of the symmetrization operator is given by Eq.~\eqref{eq: symm every goth orbital} in this case, despite the fact that the torus is nonrectangular. We numerically checked that this symmetrization of the $\nu=1/2$ Laughlin state on the  $(\bs{L}_1,\bs{L}_2)$ torus yielded the Moore-Read state on the $(\bs{L}_1,\bs{L}_2/2)$ torus for systems with up to $12$ particles.
\\

\section{Practical implementation of ${\mathcal S}_{(gT)_i\rightarrow T}$}
\label{app:  numerical symmetrization}

In this Appendix, we provide some technical information to implement both ${\mathcal S}_{(gT)_x \rightarrow T}$ and ${\mathcal S}_{(gT)_y \rightarrow T}$. As we will show, their action is simple to write once the wave function is decomposed on the occupation basis.

\emph{Enlarging the torus length in the $x$-direction} --- 
We detail this procedure sketched in Fig.~\ref{fig:symmetrization procedure}~c) and Eq.~\eqref{eq: symm Nphi apart orbitals} in the case, where we symmetrize a $\mathbb{Z}_k$ state on a torus of size $(gT)_x$, pierced by $gN_{\phi}$ flux quanta to obtain a candidate for a $\mathbb{Z}_{gk}$ state on a $T$ torus pierced by $N_{\phi}$ flux quanta, by projecting on  orbitals with quantum numbers $j$ that are integer multiples of $g$. 
For a system with $gN_{\phi}$ flux quanta, a state $\ket{\Psi\left(\mathsf{k}_y\right)}$ in a given $\mathsf{k}_y$ sector is written as a superposition of Fock states $\ket{\lambda}$ such that
\be
\label{eq:state in configuration notation}
\ket{\Psi\left(\mathsf{k}_y\right)} = \sum_{\ket{\lambda} \in {\cal H}_{gN_{\phi}}\left(\mathsf{k}_y\right)} b_{\lambda}\ket{\lambda},
\ee
where $b_{\lambda}\in\mathbb{C}$ and ${\cal H}_{gN_{\phi}}\left(\mathsf{k}_y\right)$ is the Hilbert space restricted to the total momentum constraint
\be
\mathsf{k}_y = \left(\sum_{j = 0}^{gN_{\phi} - 1}j n_{j}\left(\lambda\right)\right)\ \mathrm{mod} \left(gN_{\phi}\right),
\ee
with $\mathsf{k}_y\in\{ 0,\cdots, gN_\phi-1\}$.
Here and below, a Fock state $\ket{\lambda}$ corresponds to the occupation-number configuration $n\left(\lambda\right)$ with
\be
 n\left(\lambda\right) = \left\{n_j\left(\lambda\right) , \ j = 0, \cdots,N_{\phi} - 1 \right\},
\ee
where $n_j\left(\lambda\right)\in\mathbb{N}_0$ is the occupation number of the single-particle orbital with momentum $j$.

The symmetrized state ${\mathcal S}_{(gT)_x \rightarrow T}  \ket{\Psi\left(\mathsf{k}_y\right)} $ appears in the $\mathsf{k}_y \ \mathrm{mod} \ N_{\phi}$ momentum sector and reads
\ba
{\mathcal S}_{(gT)_x \rightarrow T}  \ket{\Psi\left(\mathsf{k_y}\right)} = 
\sum_{\ket{\lambda} \in {\cal H}_{gN_{\phi}}(\mathsf{k}_y)} b_{s_x\left(\lambda\right)} \ket{s_x\left(\lambda\right)},
\ea
where $\ket{s_x\left(\lambda\right)}$ is defined by the occupation numbers
\be
n_j\left[s_x\left(\lambda\right)\right] = \sum_{i = 0}^{g-1} n_{j + i N_{\phi}}\left(\lambda\right), \quad j = 0, \cdots, N_{\phi} - 1
\ee
and the coefficients $b_{\mu}$ of the symmetrized states are given by
\be
b_{\mu} = \sum_{\lambda \ :\, \ s_x\left(\lambda\right) = \mu}b_{\lambda}\prod_{j=0}^{N_{\phi} - 1}\sqrt{\frac{\left[\sum_{i = 0}^{g-1} n_{j + i N_{\phi}}\left(\lambda\right)\right]!}{\prod_{i = 0}^{g-1} n_{j + i N_{\phi}}\left(\lambda\right)!}}.
\label{eq:SxOrbital}
\ee

\emph{Enlarging the torus length in the $y$-direction} --- 
We now detail the procedure sketched in Fig.~\ref{fig:symmetrization procedure}~b) and Eq.~\eqref{eq: symm every goth orbital} in the case where we symmetrize a $\mathbb{Z}_k$ state on a torus of size $L_x\times gL_y$ pierced by $gN_\phi$ flux quanta. For each group of $g$ consecutive orbitals, the symmetrization operator selects the $s^{\mathrm{th}}$ orbital of them (where $s < g$ is a nonnegative integer), and discards all other $g - 1$ orbitals. The symmetrized state ${\mathcal S}_{(gT)_y \rightarrow T}  \ket{\Psi\left(\mathsf{k}_y\right)} $ is given by
\ba
{\mathcal S}_{(gT)_y \rightarrow T}  \ket{\Psi\left(\mathsf{k}_y\right)} = \sum_{\ket{\lambda} \in {\cal H}_{gN_{\phi}}} b_{s_y\left(\lambda\right)} \ket{s_y\left(\lambda\right)},
\ea
where $\ket{s_y\left(\lambda\right)}$ is defined by the occupation numbers
\be
n_j\left(s_y\left(\lambda\right)\right) = n_{gj + s}\left(\lambda\right),\quad \ j = 0,\cdots,N_{\phi} - 1
\ee
and the coefficients $b_{\mu}$ of the symmetrized states are
\be
b_{\mu} = \sum_{\lambda \ :\, \ s_y\left(\lambda\right) = \mu}b_{\lambda}\prod_{j=0}^{N_{\phi} - 1}\sqrt{\frac{n_{gj + s}\left(\lambda\right)!}{\prod_{i = 0}^{g - 1}n_{gj + i}\left(\lambda\right)!}}.
\label{eq:SyOrbital}
\ee
As a consequence of the second part of Eq.~\eqref{eq: symm every goth orbital}, all state components containing at least one $\tilde{j} \neq gj +s$ such that $n_{\tilde{j}}\left(\lambda\right) \neq 0$, are discarded. The number of particles $N$ is thus conserved by $\mathcal{S}(gT)_y\rightarrow T$.

\section{Symmetrization and many-body momentum quantum numbers}
\label{app: quantum numbers}
We denote by $\mathcal{T}^{\mathrm{rel}}_x$ the many-body operator that creates a relative translation by $L_x\bs{e}_x/N$ in position space. The properties of the momentum quantum numbers $\mathsf{k_x}$ and $\mathsf{k_y}$ under the symmetrization ${\mathcal S}_{(gT)_y \rightarrow T}$ are relatively simple. The former momentum is conserved, and the latter is mapped from $\mathsf{k_y}$ to 
\be
\mathsf{k_y'} = \frac{\mathsf{k}_{y} - sN}{g}
\ee
where $s$ is the integer defined in Eq.~\eqref{eq: symm every goth orbital} ($0 \leq s < g$).

The properties of the momentum quantum numbers under the action of ${\mathcal S}_{(gT)_x \rightarrow T}$ are slightly more complicated. We detail these properties in the following paragraph. 

When the symmetrization operation defined in Eq.~\eqref{eq: symm every goth orbital} is applied to a state on the $(gT)_x$ torus with center of mass momentum $\mathsf k_y \ \mathrm{mod} \ (gN_{\phi})$, it yields a state on the $T$ torus with the momentum $\mathsf k_y  \ \mathrm{mod} \ N_{\phi}$. The relative momentum quantum number $\mathsf{k}_x$ is not generally conserved during this transformtion. However, in some cases, the symmetrized state is still an eigenvector of the relative translation operator. We examplify this feature on the case of $k=1$ (i.e., the initial state is the Laughlin state), for a symmetrized state with filling fraction $\nu = g/2$, and compute the corresponding eigenvalue. We will treat the cases of even and odd $g$ separately. 

\emph{Case $g$ odd} ---
We start from a state on the $(gT)_x$ torus $\ket{\Psi\left(\mathsf{k_y}\right)}$. Since the Brillouin zones of both the original and the symmetrized tori are twice as large as their reduced Brillouin zone, a state on either of these systems can be an eigenstate of $\left(\mathcal{T}^{\mathrm{rel}}_x\right)^{g}$. On the $(gT)_x$ torus, an eigenstate of $\left(\mathcal{T}^{\mathrm{rel}}_x\right)^{g}$ with relative momentum $\mathsf{k}_x$ writes
\be
\ket{\phi(\mathsf{k}_x, \mathsf{k}_y)} = \sum_{p = 0}^{\frac{gN_{\phi}}{2} - 1}e^\frac{2\pi \mathrm{i} p\mathsf{k}_x}{gN_{\phi}/2} \ \left(\mathcal{T}^{\mathrm{rel}}_x\right)^{gp} \ \ket{\Psi(\mathsf{k}_y)}.
\label{eq: define initial kx state}
\ee
\begin{widetext}
Note that at filling fraction $g/2$ with $g$ odd, $N_{\phi}$ must be even. The action of the symmetrization operator on this momentum eigenstate is given by
\be
\begin{split}
{\mathcal S}_{(gT)_x \rightarrow T} \ket{\phi(\mathsf{k}_x, \mathsf{k}_y)} 
= & \sum_{p = 0}^{\frac{gN_{\phi}}{2} - 1}e^\frac{2\pi \mathrm{i} p\mathsf{k}_x}{gN_{\phi}/2}
\left(\mathcal{T}^{\mathrm{rel}}_x\right)^{gp} \ {\mathcal S}_{(gT)_x \rightarrow T}\ket{\Psi(\mathsf{k}_y)}
\\ 
= & \sum_{j = 0}^{g-1}\sum_{p = 0}^{N_{\phi}/2-1} e^\frac{2\pi \mathrm{i} (p + j N_{\phi}/2)\mathsf{k}_x}{gN_{\phi}/2}  \left(\mathcal{T}^{\mathrm{rel}}_x\right)^{g(p + jN_{\phi}/2)} {\mathcal S}_{(gT)_x \rightarrow T}\ket{\Psi(\mathsf{k}_y)}
\\ 
= & \left[ \sum_{j = 0}^{g-1} e^{\frac{2\pi \mathrm{i} j\mathsf{k}_x}{g}}\right] \sum_{p = 0}^{N_{\phi}/2-1} e^\frac{2\pi \mathrm{i} p(\mathsf{k}_x/g)}{N_{\phi}/2}\left(\mathcal{T}^{\mathrm{rel}}_x\right)^{gp}{\mathcal S}_{(gT)_x \rightarrow T}\ket{\Psi(\mathsf{k}_y)} ,
 \label{eq:Tx symmetry of symmetrized state}
\end{split}
\ee
where we have used the fact that $\left(\mathcal{T}^{\mathrm{rel}}_x\right)^{gN_{\phi}/2} = \left(\mathcal{T}^{\mathrm{rel}}_x\right)^N = \openone$. 
If $\mathsf{k}_x$ is not a multiple of $g$, the result is zero due to the factor $ \sum_{j = 0}^{g-1} e^{\frac{2\pi \mathrm{i} j\mathsf{k}_x}{g}}$, and there is no symmetrized state. If  $\mathsf{k}_x$ is a multiple of $g$, however, the symmetrized state is an eigenstate of $\left(\mathcal{T}^{\mathrm{rel}}_x\right)^g$ and the corresponding eigenvalue is
\be
\mathsf{k}_x' = \frac{\mathsf{k}_x}{g}.
\ee

\emph{Case $g$ even} ---
The filling fraction $\nu = g/2$ of the symmetrized state is an integer, therefore $N_{\phi}$ can have either parity. When $N_{\phi}$ is even (i.e., $N$ is a multiple of $g$), the symmetrized state is not an eigenstate of the relative translation operator. We focus on the case where $N_{\phi}$ is odd. Due to their respective filling fractions $1/2$ and $g/2 \in \mathbb{N}$, the original and symmetrized states may be eigenstates of $\left(\mathcal{T}^{\mathrm{rel}}_x\right)^g$ and $\left(\mathcal{T}^{\mathrm{rel}}_x\right)^{g/2}$, respectively. 

We start from a state on the $(gT)_x$ torus, with relative momentum $\mathsf{k}_x$, such as defined in Eq.~\eqref{eq: define initial kx state}. The symmetrized state writes
\be
\begin{split}
{\mathcal S}_{(gT)_x \rightarrow T} \ket{\phi(\mathsf{k}_x, \mathsf{k}_y)} 
= & \sum_{p = 0}^{
\frac{gN_{\phi}}{2} - 1}e^{\frac{2\pi \mathrm{i} p\mathsf{k}_x}{N_{\phi}(g/2)}}
\left(\mathcal{T}^{\mathrm{rel}}_x\right)^{gp} \ {\mathcal S}_{(gT)_x \rightarrow T}\ket{\Psi(\mathsf{k}_y)} 
\\ 
= & \sum_{j = 0}^{g/2-1}\sum_{p = 0}^{N_{\phi}-1} e^\frac{2\pi \mathrm{i} (p + j N_{\phi})\mathsf{k}_x}{N_{\phi}(g/2)}  \left(\mathcal{T}^{\mathrm{rel}}_x\right)^{g(p + jN_{\phi})} {\mathcal S}_{(gT)_x \rightarrow T}\ket{\Psi(\mathsf{k}_y)}
\\ 
= & \left[ \sum_{j = 0}^{g/2-1} e^{\frac{2\pi \mathrm{i} j \mathsf{k}_x}{g/2} }\right] \sum_{p = 0}^{N_{\phi} - 1} e^\frac{2\pi \mathrm{i} p(2\mathsf{k}_x/g)}{N_{\phi}}\left(\mathcal{T}^{\mathrm{rel}}_x\right)^{gp}{\mathcal S}_{(gT)_x \rightarrow T}\ket{\Psi(\mathsf{k}_y)} ,
\end{split}
\ee
where we have used the fact that $\left(\mathcal{T}^{\mathrm{rel}}_x\right)^{gN_{\phi}} = \left(\mathcal{T}^{\mathrm{rel}}_x\right)^{2N}  = \openone$. If $\mathsf{k}_x$ is not a multiple of $g/2$, the symmetrized state vanishes due to the Fourier sum. Otherwise, the above expression can be reexpressed in the following way (up to an overall factor $g/2$)
\be
\begin{split}
{\mathcal S}_{(gT)_x \rightarrow T} \ket{\phi(\mathsf{k}_x, \mathsf{k}_y)} 
= & \sum_{p = 0}^{(N_{\phi} - 1)/2} e^\frac{2\pi \mathrm{i} p(\mathsf{k}_x/g)}{N_{\phi}/2}\left(\mathcal{T}^{\mathrm{rel}}_x\right)^{gp}{\mathcal S}_{(gT)_x \rightarrow T}\ket{\Psi(\mathsf{k}_y)} 
+ 
\sum_{p =(N_{\phi} + 1)/2 }^{N_{\phi} - 1} e^\frac{2\pi \mathrm{i} p(\mathsf{k}_x/g)}{N_{\phi}/2}\left(\mathcal{T}^{\mathrm{rel}}_x\right)^{gp}{\mathcal S}_{(gT)_x \rightarrow T}\ket{\Psi(\mathsf{k}_y)} 
\\
= & \sum_{q = 0, \ even}^{N_{\phi} - 1} e^\frac{2\pi \mathrm{i} q(\mathsf{k}_x/g)}{N_{\phi}}\left(\mathcal{T}^{\mathrm{rel}}_x\right)^{\frac{g}{2}q}{\mathcal S}_{(gT)_x \rightarrow T}\ket{\Psi(\mathsf{k}_y)} 
\\
&+ 
\sum_{q = 1, \ odd}^{N_{\phi} - 1} e^\frac{2\pi \mathrm{i} (q + N_{\phi})(\mathsf{k}_x/g)}{N_{\phi}}\left(\mathcal{T}^{\mathrm{rel}}_x\right)^{\frac{g}{2}q + \frac{g}{2} N_{\phi}}{\mathcal S}_{(gT)_x \rightarrow T}\ket{\Psi(\mathsf{k}_y)} 
\\
= & \sum_{q = 0}^{N_{\phi} - 1} e^\frac{2\pi \mathrm{i} q (N_{\phi} + 1)(\mathsf{k}_x/g)}{N_{\phi}}\left(\mathcal{T}^{\mathrm{rel}}_x\right)^{\frac{g}{2}q}{\mathcal S}_{(gT)_x \rightarrow T}\ket{\Psi(\mathsf{k}_y)} .
\end{split}
\ee
The symmetrized state is thus an eigenstate of $\left(\mathcal{T}^{\mathrm{rel}}_x\right)^{g/2}$ with eigenvalue 
\be
\mathsf{k}_x' = \frac{(N_{\phi} + 1)\mathsf{k}_x}{g} \ \mathrm{mod} \ N_{\phi}.
\ee
\end{widetext}

\section{Thin torus perspective on symmetrization}
\label{app: thin torus}

\subsection{Thin torus and zero energy states}

To intuitively understand which states appear under the symmetrization operation, we can consider the action of symmetrization on the root configurations or on the thin torus limit configurations that represent  many-body states. Theses are single occupation number configurations in the single particle basis labelled by linearized momentum index $j$  on the sphere and torus geometry, respectively.

On the sphere, the approach of Ref.~\onlinecite{Bernevig-PhysRevLett.100.246802} makes use of the clustering properties of the states to rigorously derive fractional quantum Hall ground states~\cite{Greiter93} and quasihole states~\cite{Bernevig-PhysRevLett.100.246802} starting from a root configuration that obeys a generalized exclusion principle~\cite{Haldane-PhysRevLett.67.937}. Such a derivation of the state from a root configuration is not possible on the torus.
Nevertheless, for many FQH states (in particular for the Read-Rezayi series), the global counting of ground states and more generally of zero modes (i.e., quasihole states) is correctly predicted~\cite{Ardonne-2008JSMTE..04..016A} by counting single occupation number configurations that obey the generalized exclusion principle on the torus. The physical significance of the single occupation number configurations on the torus is that they correspond to a charge density wave pattern in the limit where the circumference $L_x$ is taken to zero~\cite{Bergholtz-PhysRevB.77.155308, Bergholtz-PhysRevLett.99.256803, Seidel-PhysRevLett.95.266405}. They are thus referred to as ``thin torus'' configurations.

Let us for concreteness consider the Laughlin ($k=1$) and Moore-Read ($k=2$) states at filling $\nu=1/2$ and $\nu=2/2$, respectively. Their allowed root configurations obey a $(k,2)$-exclusion principle\cite{Bernevig-PhysRevLett.100.246802,Bernevig-PhysRevB.77.184502}, meaning that no more than $k$ particles may be found in $2$ consecutive orbitals. Periodic boundary conditions on the torus imply that the $(k,2)$ exclusion principle has to be satisfied also between the first and the last orbital. Imposing this constraint on the thin torus configurations immediately delivers the $k+1$ degenerate ground states on the torus when the number of bosons $N$ is a multiple of $k$.
For example, the two topologically degenerate Laughlin states ($k=1$) on a torus with $N_\phi=6$ orbitals are represented by the unique two densest configurations that obey this principle
\begin{equation}
101010,\qquad
010101.
\label{eq: laughlin thin torus}
\end{equation}
(The notation for the thin torus configurations lists the sequence of occupation numbers $n_j$ of the consecutive orbitals $j=0,\cdots, N_\phi-1$. For the first example $n_0=1$, $n_1=1$, etc.)
The symmetrization of the root configuration of two layers of Laughlin states with respect to the layer quantum number of the orbitals, represented by the operator $\mathcal{S}_{2\to1}$, yields
\begin{equation}
\mathcal{S}_{2\to1}
\begin{pmatrix}
101010\\
101010
\end{pmatrix}
=202020,
\end{equation}
which is one of the correct thin torus configurations of the Moore-Read state ($k=2$). The other Moore-Read ground states
\begin{equation}
111111,\qquad
020202
\end{equation}
can be obtained analogously by symmetrizing over different combinations of the two Laughlin ground states~\eqref{eq: laughlin thin torus} in the two layers in accordance with what is listed in Tab.~\ref{tab: GS momentum sectors}.
That symmetrization of the entire Laughlin state (not only its thin torus configuration) exactly yields the Moore-Read state is a highly nontrivial fact which cannot be deduced from this observation on the thin torus configurations alone. However, if the symmetrization already fails on the level of thin torus configurations, it cannot work for the entire state either.

\begin{figure}
\includegraphics[width = 0.8\linewidth]{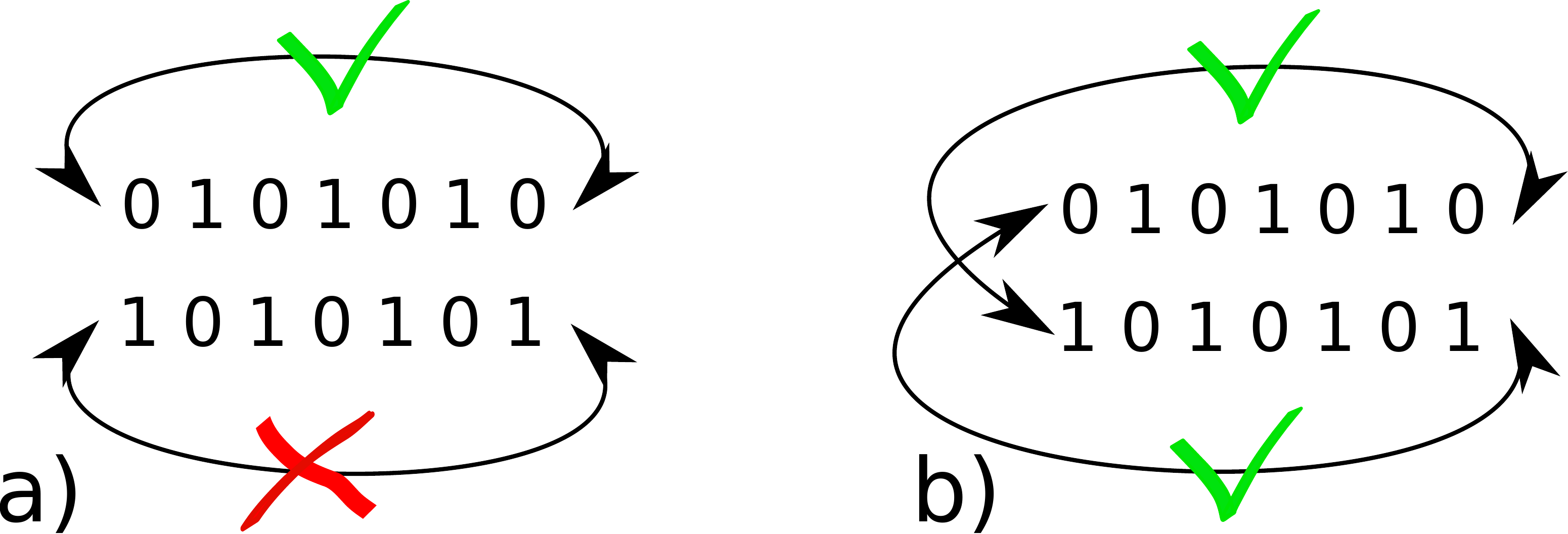}
\caption{a) Schematic representation of the impossibility to find a thin-torus configuration respecting the $(1,2)$ exclusion principle for a bilayer torus with $N = 7$ bosons, and $N_{\phi} = 7$ flux quanta. b)~Inserting a topological defect that connects the two tori relieves the frustration. A configuration respecting the exclusion principle is then possible.}
\label{fig:PBC}
\end{figure}

We can use this fact to illustrate why symmetrizing over several layers of the Laughlin state cannot yield all ground states of the Moore-Read type, or more generally of the $\mathbb{Z}_k$ Read-Rezayi type, on the torus. Turning to the torus with odd number of flux quanta, e.g. $N_\phi=7$, a single Moore-Read ground state exists with thin torus configuration
\begin{equation}
1111111
\end{equation}
that obeys the (2,2) exclusion principle.
Form the thin torus configurations, one can understand that this state cannot be obtained by symmetrization over two independent layers of Laughlin tori. The problem is that there exists no zero energy state that obeys the $(1,2)$ exclusion principle on a torus with odd $N_\phi$ and the desired number of particles.
One would need the thin torus configurations to be
\begin{equation}
\mathcal{S}_{2\to1}
\begin{pmatrix}
0101010\\
1010101
\end{pmatrix}
=1111111.
\label{eq: obstruction}
\end{equation}
However, due to the periodic boundary conditions, the lower of the two configurations, 1010101 is not allowed, for it violates the $(1,2)$ exclusion principle between its first and last orbital (see Fig.~\ref{fig:PBC}). 
The twisted boundary conditions are exactly what is needed to make the thin torus configuration on the lefthand side of Eq.~\eqref{eq: obstruction} allowed: With them in place, the (1,2) exclusion principle has to be satisfied between the last orbital in the upper layer and the first orbital in the lower layer as well as between the first orbital in the upper layer and the last orbital in the lower layer. 

Equivalently, we can apply the two symmetrization schemes for the $(gT)_y$ torus and for the $(gT)_x$ that we summarized in Eqs.~\eqref{eq: symm Nphi apart orbitals}
and~\eqref{eq: symm every goth orbital}, respectively,
\begin{equation}
\begin{split}
1111111=&\,\mathcal{S}_{(gT)_y \rightarrow T}
(10101010101010)
\\
=&\,
\mathcal{S}_{(gT)_x \rightarrow T}
(10101010101010).
\end{split}
\label{eq: obstruction2}
\end{equation}
Again, the fact that these equalities hold not only for the thin-torus configuration, but for the entire Laughlin states, is a highly non-trivial fact. 

\subsection{Thin torus perspective on neutral modes}

In terms of thin torus configurations, states belonging to the neutral modes appear as minimal violations of the generalized exclusion principle at fixed particle density. For example, a thin-torus configuration belonging to the magnetoroton mode of the Laughlin state is given by
\begin{equation}
10\underset{\bullet}{11} 0101\underset{\circ}{00}10,
\end{equation}
where dot ($\bullet$) and circle ($\circ$) mark the positions of the quasiparticle (violating the exclusion principle) and quasihole, respectively. 
Using this construction, explicit wave functions were derived in Ref.~\onlinecite{Yang-PhysRevLett.108.256807}, thus validating this interpretation of the neutral mode.

As discussed in Sec.~\ref{sec:bilayerSymmMagn}, neutral excitations above the Moore-Read state can be captured by a bilayer construction involving symmetrization over Laughlin charged or neutral excitations. Let us illustrate this scheme on the torus geometry with the help of thin torus configurations. For that, we consider the symmetrization over double layer configurations with a single violation of the $(1,2)$ exclusion principle to construct a single violation of the $(2,2)$ exclusion principle. We have two choices to obtain the same symmetrized root configuration: In the first case, one Laughlin layer is in a magnetoroton state and the other layer is in the Laughlin ground state, for example
\be
{\mathcal S}_{2\to 1}
\begin{pmatrix}
10\underset{\bullet}{11} 0101\underset{\circ}{00}10\\
010101010101
\end{pmatrix}
=
11\underset{\bullet}{12}0202\underset{\circ}{01}11.
\label{eq: thin torus MR 1}
\ee
In the second case, one layer is in a Laughlin quasielectron state and the other layer is in a Laughlin quasihole state
\be
{\mathcal S}_{2\to 1}
\begin{pmatrix}
10\underset{\bullet}{11} 01010101\\
01010101\underset{\circ}{00}10
\end{pmatrix}
=
11\underset{\bullet}{12}0202\underset{\circ}{01}11.
\label{eq: thin torus MR 2}
\ee
Observe that Eq.~\eqref{eq: thin torus MR 1} is an allowed (minimally violating) configuration for periodic  boundary conditions, while Eq.~\eqref{eq: thin torus MR 2} is allowed for twisted boundary conditions.

\bibliography{SymmetrizationDraft}

\begin{thebibliography}{10}%
\makeatletter
\providecommand \@ifxundefined [1]{%
 \ifx #1\undefined \expandafter \@firstoftwo
 \else \expandafter \@secondoftwo
\fi
}%
\providecommand \@ifnum [1]{%
 \ifnum #1\expandafter \@firstoftwo
 \else \expandafter \@secondoftwo
\fi
}%
\providecommand \enquote [1]{``#1''}%
\providecommand \bibnamefont  [1]{#1}%
\providecommand \bibfnamefont [1]{#1}%
\providecommand \citenamefont [1]{#1}%
\providecommand\href[0]{\@sanitize\@href}%
\providecommand\@href[1]{\endgroup\@@startlink{#1}\endgroup\@@href}%
\providecommand\@@href[1]{#1\@@endlink}%
\providecommand \@sanitize [0]{\begingroup\catcode`\&12\catcode`\#12\relax}%
\@ifxundefined \pdfoutput {\@firstoftwo}{%
 \@ifnum{\z@=\pdfoutput}{\@firstoftwo}{\@secondoftwo}%
}{%
 \providecommand\@@startlink[1]{\leavevmode\special{html:<a href="#1">}}%
 \providecommand\@@endlink[0]{\special{html:</a>}}%
}{%
 \providecommand\@@startlink[1]{%
  \leavevmode
  \pdfstartlink
   attr{/Border[0 0 1 ]/H/I/C[0 1 1]}%
   user{/Subtype/Link/A<</Type/Action/S/URI/URI(#1)>>}%
  \relax
 }%
 \providecommand\@@endlink[0]{\pdfendlink}%
}%
\providecommand \url  [0]{\begingroup\@sanitize \@url }%
\providecommand \@url [1]{\endgroup\@href {#1}{\urlprefix}}%
\providecommand \urlprefix [0]{URL }%
\providecommand \Eprint[0]{\href }%
\@ifxundefined \urlstyle {%
  \providecommand \doi [1]{doi:\discretionary{}{}{}#1}%
}{%
  \providecommand \doi [0]{doi:\discretionary{}{}{}\begingroup
  \urlstyle{rm}\Url }%
}%
\providecommand \doibase [0]{http://dx.doi.org/}%
\providecommand \Doi[1]{\href{\doibase#1}}%
\providecommand \bibAnnote [3]{%
  \BibitemShut{#1}%
  \begin{quotation}\noindent
    \textsc{Key:}\ #2\\\textsc{Annotation:}\ #3%
  \end{quotation}%
}%
\providecommand \bibAnnoteFile [2]{%
  \IfFileExists{#2}{\bibAnnote {#1} {#2} {\input{#2}}}{}%
}%
\providecommand \typeout [0]{\immediate \write \m@ne }%
\providecommand \selectlanguage [0]{\@gobble}%
\providecommand \bibinfo [0]{\@secondoftwo}%
\providecommand \bibfield [0]{\@secondoftwo}%
\providecommand \translation [1]{[#1]}%
\providecommand \BibitemOpen[0]{}%
\providecommand \bibitemStop [0]{}%
\providecommand \bibitemNoStop [0]{.\EOS\space}%
\providecommand \EOS [0]{\spacefactor3000\relax}%
\providecommand \BibitemShut [1]{\csname bibitem#1\endcsname}%
\bibitem{PhysRevB.40.8079}%
  \BibitemOpen
  \bibfield{author}{%
  \bibinfo {author} {\bibfnamefont{J.~K.}\ \bibnamefont{Jain}},\ }%
  \bibfield{journal}{%
  \Doi{10.1103/PhysRevB.40.8079}{\bibinfo {journal} {Phys. Rev. B}}\ }%
  \textbf{\bibinfo {volume} {40}},\ \bibinfo {pages} {8079} (\bibinfo {month}
  {Oct}\ \bibinfo {year} {1989})%
  \bibAnnoteFile{NoStop}{PhysRevB.40.8079}%
\bibitem{doi:10.1142/S0217984991000058}%
  \BibitemOpen
  \bibfield{author}{%
  \bibinfo {author} {\bibfnamefont{X.-G.}\ \bibnamefont{Wen}},\ }%
  \bibfield{journal}{%
  \Doi{10.1142/S0217984991000058}{\bibinfo {journal} {Modern Physics Letters
  B}}\ }%
  \textbf{\bibinfo {volume} {05}},\ \bibinfo {pages} {39} (\bibinfo {year}
  {1991})%
  \bibAnnoteFile{NoStop}{doi:10.1142/S0217984991000058}%
\bibitem{PhysRevLett.66.802}%
  \BibitemOpen
  \bibfield{author}{%
  \bibinfo {author} {\bibfnamefont{X.~G.}\ \bibnamefont{Wen}},\ }%
  \bibfield{journal}{%
  \Doi{10.1103/PhysRevLett.66.802}{\bibinfo {journal} {Phys. Rev. Lett.}}\ }%
  \textbf{\bibinfo {volume} {66}},\ \bibinfo {pages} {802} (\bibinfo {month}
  {Feb}\ \bibinfo {year} {1991})%
  \bibAnnoteFile{NoStop}{PhysRevLett.66.802}%
\bibitem{PhysRevB.60.8827}%
  \BibitemOpen
  \bibfield{author}{%
  \bibinfo {author} {\bibfnamefont{X.-G.}\ \bibnamefont{Wen}},\ }%
  \bibfield{journal}{%
  \Doi{10.1103/PhysRevB.60.8827}{\bibinfo {journal} {Phys. Rev. B}}\ }%
  \textbf{\bibinfo {volume} {60}},\ \bibinfo {pages} {8827} (\bibinfo {month}
  {Sep}\ \bibinfo {year} {1999})%
  \bibAnnoteFile{NoStop}{PhysRevB.60.8827}%
\bibitem{doi:10.1142/S0217979292000840}%
  \BibitemOpen
  \bibfield{author}{%
  \bibinfo {author} {\bibfnamefont{X.-G.}\ \bibnamefont{Wen}},\ }%
  \bibfield{journal}{%
  \Doi{10.1142/S0217979292000840}{\bibinfo {journal} {International Journal of
  Modern Physics B}}\ }%
  \textbf{\bibinfo {volume} {06}},\ \bibinfo {pages} {1711} (\bibinfo {year}
  {1992})%
  \bibAnnoteFile{NoStop}{doi:10.1142/S0217979292000840}%
\bibitem{Blok1992615}%
  \BibitemOpen
  \bibfield{author}{%
  \bibinfo {author} {\bibfnamefont{B.}~\bibnamefont{Blok}}\ and\ \bibinfo
  {author} {\bibfnamefont{X.-G.}\ \bibnamefont{Wen}},\ }%
  \bibfield{journal}{%
  \Doi{http://dx.doi.org/10.1016/0550-3213(92)90402-W}{\bibinfo {journal}
  {Nuclear Physics B}}\ }%
  \textbf{\bibinfo {volume} {374}},\ \bibinfo {pages} {615 } (\bibinfo {year}
  {1992})%
  \bibAnnoteFile{NoStop}{Blok1992615}%
\bibitem{PhysRevLett.105.246809}%
  \BibitemOpen
  \bibfield{author}{%
  \bibinfo {author} {\bibfnamefont{J.}~\bibnamefont{Maciejko}}, \bibinfo
  {author} {\bibfnamefont{X.-L.}\ \bibnamefont{Qi}}, \bibinfo {author}
  {\bibfnamefont{A.}~\bibnamefont{Karch}},\ and\ \bibinfo {author}
  {\bibfnamefont{S.-C.}\ \bibnamefont{Zhang}},\ }%
  \bibfield{journal}{%
  \Doi{10.1103/PhysRevLett.105.246809}{\bibinfo {journal} {Phys. Rev. Lett.}}\
  }%
  \textbf{\bibinfo {volume} {105}},\ \bibinfo {pages} {246809} (\bibinfo
  {month} {Dec}\ \bibinfo {year} {2010})%
  \bibAnnoteFile{NoStop}{PhysRevLett.105.246809}%
\bibitem{PhysRevB.83.195139}%
  \BibitemOpen
  \bibfield{author}{%
  \bibinfo {author} {\bibfnamefont{B.}~\bibnamefont{Swingle}}, \bibinfo
  {author} {\bibfnamefont{M.}~\bibnamefont{Barkeshli}}, \bibinfo {author}
  {\bibfnamefont{J.}~\bibnamefont{McGreevy}},\ and\ \bibinfo {author}
  {\bibfnamefont{T.}~\bibnamefont{Senthil}},\ }%
  \bibfield{journal}{%
  \Doi{10.1103/PhysRevB.83.195139}{\bibinfo {journal} {Phys. Rev. B}}\ }%
  \textbf{\bibinfo {volume} {83}},\ \bibinfo {pages} {195139} (\bibinfo {month}
  {May}\ \bibinfo {year} {2011})%
  \bibAnnoteFile{NoStop}{PhysRevB.83.195139}%
\bibitem{PhysRevB.85.125105}%
  \BibitemOpen
  \bibfield{author}{%
  \bibinfo {author} {\bibfnamefont{J.}~\bibnamefont{McGreevy}}, \bibinfo
  {author} {\bibfnamefont{B.}~\bibnamefont{Swingle}},\ and\ \bibinfo {author}
  {\bibfnamefont{K.-A.}\ \bibnamefont{Tran}},\ }%
  \bibfield{journal}{%
  \Doi{10.1103/PhysRevB.85.125105}{\bibinfo {journal} {Phys. Rev. B}}\ }%
  \textbf{\bibinfo {volume} {85}},\ \bibinfo {pages} {125105} (\bibinfo {month}
  {Mar}\ \bibinfo {year} {2012})%
  \bibAnnoteFile{NoStop}{PhysRevB.85.125105}%
\bibitem{Cappelli-1999CMaPh.205..657C}%
  \BibitemOpen
  \bibfield{author}{%
  \bibinfo {author} {\bibfnamefont{A.}~\bibnamefont{{Cappelli}}}, \bibinfo
  {author} {\bibfnamefont{L.~S.}\ \bibnamefont{{Georgiev}}},\ and\ \bibinfo
  {author} {\bibfnamefont{I.~T.}\ \bibnamefont{{Todorov}}},\ }%
  \bibfield{journal}{%
  \Doi{10.1007/s002200050693}{\bibinfo {journal} {Communications in
  Mathematical Physics}}\ }%
  \textbf{\bibinfo {volume} {205}},\ \bibinfo {pages} {657} (\bibinfo {year}
  {1999})%
  \bibAnnoteFile{NoStop}{Cappelli-1999CMaPh.205..657C}%
\bibitem{Froehlich-2000cond.mat..2330F}%
  \BibitemOpen
  \bibfield{author}{%
  \bibinfo {author} {\bibfnamefont{J.}~\bibnamefont{{Froehlich}}}, \bibinfo
  {author} {\bibfnamefont{B.}~\bibnamefont{{Pedrini}}}, \bibinfo {author}
  {\bibfnamefont{C.}~\bibnamefont{{Schweigert}}},\ and\ \bibinfo {author}
  {\bibfnamefont{J.}~\bibnamefont{{Walcher}}},\ }%
  \bibfield{journal}{%
  \bibinfo {journal} {eprint arXiv:cond-mat/0002330}}%
   (\bibinfo {month} {Feb.}\ \bibinfo {year} {2000}),\
  \Eprint{http://arxiv.org/abs/cond-mat/0002330}{cond-mat/0002330}%
  \bibAnnoteFile{NoStop}{Froehlich-2000cond.mat..2330F}%
\bibitem{cappelli-2001NuPhB.599..499C}%
  \BibitemOpen
  \bibfield{author}{%
  \bibinfo {author} {\bibfnamefont{A.}~\bibnamefont{{Cappelli}}}, \bibinfo
  {author} {\bibfnamefont{L.~S.}\ \bibnamefont{{Georgiev}}},\ and\ \bibinfo
  {author} {\bibfnamefont{I.~T.}\ \bibnamefont{{Todorov}}},\ }%
  \bibfield{journal}{%
  \Doi{10.1016/S0550-3213(00)00774-4}{\bibinfo {journal} {Nuclear Physics B}}\
  }%
  \textbf{\bibinfo {volume} {599}},\ \bibinfo {pages} {499} (\bibinfo {month}
  {Apr.}\ \bibinfo {year} {2001})%
  \bibAnnoteFile{NoStop}{cappelli-2001NuPhB.599..499C}%
\bibitem{Barkeshli-PhysRevB.81.155302}%
  \BibitemOpen
  \bibfield{author}{%
  \bibinfo {author} {\bibfnamefont{M.}~\bibnamefont{Barkeshli}}\ and\ \bibinfo
  {author} {\bibfnamefont{X.-G.}\ \bibnamefont{Wen}},\ }%
  \bibfield{journal}{%
  \Doi{10.1103/PhysRevB.81.155302}{\bibinfo {journal} {Phys. Rev. B}}\ }%
  \textbf{\bibinfo {volume} {81}},\ \bibinfo {pages} {155302} (\bibinfo {month}
  {Apr}\ \bibinfo {year} {2010})%
  \bibAnnoteFile{NoStop}{Barkeshli-PhysRevB.81.155302}%
\bibitem{Read-1999PhRvB..59.8084R}%
  \BibitemOpen
  \bibfield{author}{%
  \bibinfo {author} {\bibfnamefont{N.}~\bibnamefont{{Read}}}\ and\ \bibinfo
  {author} {\bibfnamefont{E.}~\bibnamefont{{Rezayi}}},\ }%
  \bibfield{journal}{%
  \Doi{10.1103/PhysRevB.59.8084}{\bibinfo {journal} {\prb}}\ }%
  \textbf{\bibinfo {volume} {59}},\ \bibinfo {pages} {8084} (\bibinfo {month}
  {Mar.}\ \bibinfo {year} {1999})%
  \bibAnnoteFile{NoStop}{Read-1999PhRvB..59.8084R}%
\bibitem{Laughlin-PhysRevLett.50.1395}%
  \BibitemOpen
  \bibfield{author}{%
  \bibinfo {author} {\bibfnamefont{R.~B.}\ \bibnamefont{Laughlin}},\ }%
  \bibfield{journal}{%
  \Doi{10.1103/PhysRevLett.50.1395}{\bibinfo {journal} {Phys. Rev. Lett.}}\ }%
  \textbf{\bibinfo {volume} {50}},\ \bibinfo {pages} {1395} (\bibinfo {month}
  {May}\ \bibinfo {year} {1983})%
  \bibAnnoteFile{NoStop}{Laughlin-PhysRevLett.50.1395}%
\bibitem{Moore1991362}%
  \BibitemOpen
  \bibfield{author}{%
  \bibinfo {author} {\bibfnamefont{G.}~\bibnamefont{Moore}}\ and\ \bibinfo
  {author} {\bibfnamefont{N.}~\bibnamefont{Read}},\ }%
  \bibfield{journal}{%
  \Doi{http://dx.doi.org/10.1016/0550-3213(91)90407-O}{\bibinfo {journal}
  {Nuclear Physics B}}\ }%
  \textbf{\bibinfo {volume} {360}},\ \bibinfo {pages} {362 } (\bibinfo {year}
  {1991}),\ ISSN \bibinfo {issn} {0550-3213}%
  \bibAnnoteFile{NoStop}{Moore1991362}%
\bibitem{Slingerland2001229}%
  \BibitemOpen
  \bibfield{author}{%
  \bibinfo {author} {\bibfnamefont{J.}~\bibnamefont{Slingerland}}\ and\
  \bibinfo {author} {\bibfnamefont{F.}~\bibnamefont{Bais}},\ }%
  \bibfield{journal}{%
  \Doi{http://dx.doi.org/10.1016/S0550-3213(01)00308-X}{\bibinfo {journal}
  {Nuclear Physics B}}\ }%
  \textbf{\bibinfo {volume} {612}},\ \bibinfo {pages} {229 } (\bibinfo {year}
  {2001}),\ ISSN \bibinfo {issn} {0550-3213}%
  \bibAnnoteFile{NoStop}{Slingerland2001229}%
\bibitem{Nayak08}%
  \BibitemOpen
  \bibfield{author}{%
  \bibinfo {author} {\bibfnamefont{C.}~\bibnamefont{Nayak}}, \bibinfo {author}
  {\bibfnamefont{S.~H.}\ \bibnamefont{Simon}}, \bibinfo {author}
  {\bibfnamefont{A.}~\bibnamefont{Stern}}, \bibinfo {author}
  {\bibfnamefont{M.}~\bibnamefont{Freedman}},\ and\ \bibinfo {author}
  {\bibfnamefont{S.}~\bibnamefont{Das~Sarma}},\ }%
  \bibfield{journal}{%
  \Doi{10.1103/RevModPhys.80.1083}{\bibinfo {journal} {Rev. Mod. Phys.}}\ }%
  \textbf{\bibinfo {volume} {80}},\ \bibinfo {pages} {1083} (\bibinfo {month}
  {Sep}\ \bibinfo {year} {2008})%
  \bibAnnoteFile{NoStop}{Nayak08}%
\bibitem{Regnault-PhysRevLett.101.066803}%
  \BibitemOpen
  \bibfield{author}{%
  \bibinfo {author} {\bibfnamefont{N.}~\bibnamefont{Regnault}}, \bibinfo
  {author} {\bibfnamefont{M.~O.}\ \bibnamefont{Goerbig}},\ and\ \bibinfo
  {author} {\bibfnamefont{T.}~\bibnamefont{Jolicoeur}},\ }%
  \bibfield{journal}{%
  \Doi{10.1103/PhysRevLett.101.066803}{\bibinfo {journal} {Phys. Rev. Lett.}}\
  }%
  \textbf{\bibinfo {volume} {101}},\ \bibinfo {pages} {066803} (\bibinfo
  {month} {Aug}\ \bibinfo {year} {2008})%
  \bibAnnoteFile{NoStop}{Regnault-PhysRevLett.101.066803}%
\bibitem{Rodriguez-PhysRevB.85.035128}%
  \BibitemOpen
  \bibfield{author}{%
  \bibinfo {author} {\bibfnamefont{I.~D.}\ \bibnamefont{Rodriguez}}, \bibinfo
  {author} {\bibfnamefont{A.}~\bibnamefont{Sterdyniak}}, \bibinfo {author}
  {\bibfnamefont{M.}~\bibnamefont{Hermanns}}, \bibinfo {author}
  {\bibfnamefont{J.~K.}\ \bibnamefont{Slingerland}},\ and\ \bibinfo {author}
  {\bibfnamefont{N.}~\bibnamefont{Regnault}},\ }%
  \bibfield{journal}{%
  \Doi{10.1103/PhysRevB.85.035128}{\bibinfo {journal} {Phys. Rev. B}}\ }%
  \textbf{\bibinfo {volume} {85}},\ \bibinfo {pages} {035128} (\bibinfo {month}
  {Jan}\ \bibinfo {year} {2012})%
  \bibAnnoteFile{NoStop}{Rodriguez-PhysRevB.85.035128}%
\bibitem{Sreejith-PhysRevLett.107.136802}%
  \BibitemOpen
  \bibfield{author}{%
  \bibinfo {author} {\bibfnamefont{G.~J.}\ \bibnamefont{Sreejith}}, \bibinfo
  {author} {\bibfnamefont{A.}~\bibnamefont{W\'ojs}},\ and\ \bibinfo {author}
  {\bibfnamefont{J.~K.}\ \bibnamefont{Jain}},\ }%
  \bibfield{journal}{%
  \Doi{10.1103/PhysRevLett.107.136802}{\bibinfo {journal} {Phys. Rev. Lett.}}\
  }%
  \textbf{\bibinfo {volume} {107}},\ \bibinfo {pages} {136802} (\bibinfo
  {month} {Sep}\ \bibinfo {year} {2011})%
  \bibAnnoteFile{NoStop}{Sreejith-PhysRevLett.107.136802}%
\bibitem{Sreejith-PhysRevLett.107.086806}%
  \BibitemOpen
  \bibfield{author}{%
  \bibinfo {author} {\bibfnamefont{G.~J.}\ \bibnamefont{Sreejith}}, \bibinfo
  {author} {\bibfnamefont{C.}~\bibnamefont{T\ifmmode~\mbox{\H{o}}\else
  \H{o}\fi{}ke}}, \bibinfo {author} {\bibfnamefont{A.}~\bibnamefont{W\'ojs}},\
  and\ \bibinfo {author} {\bibfnamefont{J.~K.}\ \bibnamefont{Jain}},\ }%
  \bibfield{journal}{%
  \Doi{10.1103/PhysRevLett.107.086806}{\bibinfo {journal} {Phys. Rev. Lett.}}\
  }%
  \textbf{\bibinfo {volume} {107}},\ \bibinfo {pages} {086806} (\bibinfo
  {month} {Aug}\ \bibinfo {year} {2011})%
  \bibAnnoteFile{NoStop}{Sreejith-PhysRevLett.107.086806}%
\bibitem{Sreejith-PhysRevB.87.245125}%
  \BibitemOpen
  \bibfield{author}{%
  \bibinfo {author} {\bibfnamefont{G.~J.}\ \bibnamefont{Sreejith}}, \bibinfo
  {author} {\bibfnamefont{Y.-H.}\ \bibnamefont{Wu}}, \bibinfo {author}
  {\bibfnamefont{A.}~\bibnamefont{W\'ojs}},\ and\ \bibinfo {author}
  {\bibfnamefont{J.~K.}\ \bibnamefont{Jain}},\ }%
  \bibfield{journal}{%
  \Doi{10.1103/PhysRevB.87.245125}{\bibinfo {journal} {Phys. Rev. B}}\ }%
  \textbf{\bibinfo {volume} {87}},\ \bibinfo {pages} {245125} (\bibinfo {month}
  {Jun}\ \bibinfo {year} {2013})%
  \bibAnnoteFile{NoStop}{Sreejith-PhysRevB.87.245125}%
\bibitem{Read-PhysRevB.61.10267}%
  \BibitemOpen
  \bibfield{author}{%
  \bibinfo {author} {\bibfnamefont{N.}~\bibnamefont{Read}}\ and\ \bibinfo
  {author} {\bibfnamefont{D.}~\bibnamefont{Green}},\ }%
  \bibfield{journal}{%
  \Doi{10.1103/PhysRevB.61.10267}{\bibinfo {journal} {Phys. Rev. B}}\ }%
  \textbf{\bibinfo {volume} {61}},\ \bibinfo {pages} {10267} (\bibinfo {month}
  {Apr}\ \bibinfo {year} {2000})%
  \bibAnnoteFile{NoStop}{Read-PhysRevB.61.10267}%
\bibitem{Nomura-JPSJ.73.2612}%
  \BibitemOpen
  \bibfield{author}{%
  \bibinfo {author} {\bibfnamefont{K.}~\bibnamefont{Nomura}}\ and\ \bibinfo
  {author} {\bibfnamefont{D.}~\bibnamefont{Yoshioka}},\ }%
  \bibfield{journal}{%
  \Doi{10.1143/JPSJ.73.2612}{\bibinfo {journal} {Journal of the Physical
  Society of Japan}}\ }%
  \textbf{\bibinfo {volume} {73}},\ \bibinfo {pages} {2612} (\bibinfo {year}
  {2004}),\
  \Eprint{http://arxiv.org/abs/http://dx.doi.org/10.1143/JPSJ.73.2612}{http://%
dx.doi.org/10.1143/JPSJ.73.2612}%
  \bibAnnoteFile{NoStop}{Nomura-JPSJ.73.2612}%
\bibitem{Rezayi-2010arXiv1007.2022R}%
  \BibitemOpen
  \bibfield{author}{%
  \bibinfo {author} {\bibfnamefont{E.}~\bibnamefont{{Rezayi}}}, \bibinfo
  {author} {\bibfnamefont{X.-G.}\ \bibnamefont{{Wen}}},\ and\ \bibinfo {author}
  {\bibfnamefont{N.}~\bibnamefont{{Read}}},\ }%
  \bibfield{journal}{%
  \bibinfo {journal} {ArXiv e-prints}}%
   (\bibinfo {month} {Jul.}\ \bibinfo {year} {2010}),\
  \Eprint{http://arxiv.org/abs/1007.2022}{arXiv:1007.2022 [cond-mat.mes-hall]}%
  \bibAnnoteFile{NoStop}{Rezayi-2010arXiv1007.2022R}%
\bibitem{Barkeshli-2010PhRvL.105u6804B}%
  \BibitemOpen
  \bibfield{author}{%
  \bibinfo {author} {\bibfnamefont{M.}~\bibnamefont{{Barkeshli}}}\ and\
  \bibinfo {author} {\bibfnamefont{X.-G.}\ \bibnamefont{{Wen}}},\ }%
  \bibfield{journal}{%
  \Doi{10.1103/PhysRevLett.105.216804}{\bibinfo {journal} {Physical Review
  Letters}}\ }%
  \textbf{\bibinfo {volume} {105}},\ \bibinfo {eid} {216804} (\bibinfo {month}
  {Nov.}\ \bibinfo {year} {2010})%
  \bibAnnoteFile{NoStop}{Barkeshli-2010PhRvL.105u6804B}%
\bibitem{Papic-PhysRevB.82.075302}%
  \BibitemOpen
  \bibfield{author}{%
  \bibinfo {author} {\bibfnamefont{Z.}~\bibnamefont{Papi\ifmmode~\acute{c}\else
  \'{c}\fi{}}}, \bibinfo {author} {\bibfnamefont{M.~O.}\
  \bibnamefont{Goerbig}}, \bibinfo {author}
  {\bibfnamefont{N.}~\bibnamefont{Regnault}},\ and\ \bibinfo {author}
  {\bibfnamefont{M.~V.}\ \bibnamefont{Milovanovi\ifmmode~\acute{c}\else
  \'{c}\fi{}}},\ }%
  \bibfield{journal}{%
  \Doi{10.1103/PhysRevB.82.075302}{\bibinfo {journal} {Phys. Rev. B}}\ }%
  \textbf{\bibinfo {volume} {82}},\ \bibinfo {pages} {075302} (\bibinfo {month}
  {Aug}\ \bibinfo {year} {2010})%
  \bibAnnoteFile{NoStop}{Papic-PhysRevB.82.075302}%
\bibitem{Vaezi-2014PhRvL.113w6804V}%
  \BibitemOpen
  \bibfield{author}{%
  \bibinfo {author} {\bibfnamefont{A.}~\bibnamefont{{Vaezi}}}\ and\ \bibinfo
  {author} {\bibfnamefont{M.}~\bibnamefont{{Barkeshli}}},\ }%
  \bibfield{journal}{%
  \Doi{10.1103/PhysRevLett.113.236804}{\bibinfo {journal} {Physical Review
  Letters}}\ }%
  \textbf{\bibinfo {volume} {113}},\ \bibinfo {eid} {236804} (\bibinfo {month}
  {Dec.}\ \bibinfo {year} {2014})%
  \bibAnnoteFile{NoStop}{Vaezi-2014PhRvL.113w6804V}%
\bibitem{Zhu-2015arXiv150205076Z}%
  \BibitemOpen
  \bibfield{author}{%
  \bibinfo {author} {\bibfnamefont{W.}~\bibnamefont{{Zhu}}}, \bibinfo {author}
  {\bibfnamefont{S.~S.}\ \bibnamefont{{Gong}}}, \bibinfo {author}
  {\bibfnamefont{D.~N.}\ \bibnamefont{{Sheng}}},\ and\ \bibinfo {author}
  {\bibfnamefont{L.}~\bibnamefont{{Sheng}}},\ }%
  \bibfield{journal}{%
  \bibinfo {journal} {ArXiv e-prints}}%
   (\bibinfo {month} {Feb.}\ \bibinfo {year} {2015}),\
  \Eprint{http://arxiv.org/abs/1502.05076}{arXiv:1502.05076 [cond-mat.str-el]}%
  \bibAnnoteFile{NoStop}{Zhu-2015arXiv150205076Z}%
\bibitem{Geraedts-2015arXiv150201340G}%
  \BibitemOpen
  \bibfield{author}{%
  \bibinfo {author} {\bibfnamefont{S.}~\bibnamefont{{Geraedts}}}, \bibinfo
  {author} {\bibfnamefont{M.~P.}\ \bibnamefont{{Zaletel}}}, \bibinfo {author}
  {\bibfnamefont{Z.}~\bibnamefont{{Papi{\'c}}}},\ and\ \bibinfo {author}
  {\bibfnamefont{R.~S.~K.}\ \bibnamefont{{Mong}}},\ }%
  \bibfield{journal}{%
  \bibinfo {journal} {ArXiv e-prints}}%
   (\bibinfo {month} {Feb.}\ \bibinfo {year} {2015}),\
  \Eprint{http://arxiv.org/abs/1502.01340}{arXiv:1502.01340 [cond-mat.str-el]}%
  \bibAnnoteFile{NoStop}{Geraedts-2015arXiv150201340G}%
\bibitem{Liu-2015arXiv150205391L}%
  \BibitemOpen
  \bibfield{author}{%
  \bibinfo {author} {\bibfnamefont{Z.}~\bibnamefont{{Liu}}}, \bibinfo {author}
  {\bibfnamefont{A.}~\bibnamefont{{Vaezi}}}, \bibinfo {author}
  {\bibfnamefont{K.}~\bibnamefont{{Lee}}},\ and\ \bibinfo {author}
  {\bibfnamefont{E.-A.}\ \bibnamefont{{Kim}}},\ }%
  \bibfield{journal}{%
  \bibinfo {journal} {ArXiv e-prints}}%
   (\bibinfo {month} {Feb.}\ \bibinfo {year} {2015}),\
  \Eprint{http://arxiv.org/abs/1502.05391}{arXiv:1502.05391 [cond-mat.str-el]}%
  \bibAnnoteFile{NoStop}{Liu-2015arXiv150205391L}%
\bibitem{Peterson-2015arXiv150202671P}%
  \BibitemOpen
  \bibfield{author}{%
  \bibinfo {author} {\bibfnamefont{M.~R.}\ \bibnamefont{{Peterson}}}, \bibinfo
  {author} {\bibfnamefont{Y.-L.}\ \bibnamefont{{Wu}}}, \bibinfo {author}
  {\bibfnamefont{M.}~\bibnamefont{{Cheng}}}, \bibinfo {author}
  {\bibfnamefont{M.}~\bibnamefont{{Barkeshli}}}, \bibinfo {author}
  {\bibfnamefont{Z.}~\bibnamefont{{Wang}}},\ and\ \bibinfo {author}
  {\bibfnamefont{S.}~\bibnamefont{{Das Sarma}}},\ }%
  \bibfield{journal}{%
  \bibinfo {journal} {ArXiv e-prints}}%
   (\bibinfo {month} {Feb.}\ \bibinfo {year} {2015}),\
  \Eprint{http://arxiv.org/abs/1502.02671}{arXiv:1502.02671 [cond-mat.str-el]}%
  \bibAnnoteFile{NoStop}{Peterson-2015arXiv150202671P}%
\bibitem{GMP-PhysRevLett.54.581}%
  \BibitemOpen
  \bibfield{author}{%
  \bibinfo {author} {\bibfnamefont{S.~M.}\ \bibnamefont{Girvin}}, \bibinfo
  {author} {\bibfnamefont{A.~H.}\ \bibnamefont{MacDonald}},\ and\ \bibinfo
  {author} {\bibfnamefont{P.~M.}\ \bibnamefont{Platzman}},\ }%
  \bibfield{journal}{%
  \Doi{10.1103/PhysRevLett.54.581}{\bibinfo {journal} {Phys. Rev. Lett.}}\ }%
  \textbf{\bibinfo {volume} {54}},\ \bibinfo {pages} {581} (\bibinfo {month}
  {Feb}\ \bibinfo {year} {1985})%
  \bibAnnoteFile{NoStop}{GMP-PhysRevLett.54.581}%
\bibitem{GMP-PhysRevB.33.2481}%
  \BibitemOpen
  \bibfield{author}{%
  \bibinfo {author} {\bibfnamefont{S.~M.}\ \bibnamefont{Girvin}}, \bibinfo
  {author} {\bibfnamefont{A.~H.}\ \bibnamefont{MacDonald}},\ and\ \bibinfo
  {author} {\bibfnamefont{P.~M.}\ \bibnamefont{Platzman}},\ }%
  \bibfield{journal}{%
  \Doi{10.1103/PhysRevB.33.2481}{\bibinfo {journal} {Phys. Rev. B}}\ }%
  \textbf{\bibinfo {volume} {33}},\ \bibinfo {pages} {2481} (\bibinfo {month}
  {Feb}\ \bibinfo {year} {1986})%
  \bibAnnoteFile{NoStop}{GMP-PhysRevB.33.2481}%
\bibitem{Bernevig-PhysRevLett.100.246802}%
  \BibitemOpen
  \bibfield{author}{%
  \bibinfo {author} {\bibfnamefont{B.~A.}\ \bibnamefont{Bernevig}}\ and\
  \bibinfo {author} {\bibfnamefont{F.~D.~M.}\ \bibnamefont{Haldane}},\ }%
  \bibfield{journal}{%
  \Doi{10.1103/PhysRevLett.100.246802}{\bibinfo {journal} {Phys. Rev. Lett.}}\
  }%
  \textbf{\bibinfo {volume} {100}},\ \bibinfo {pages} {246802} (\bibinfo
  {month} {Jun}\ \bibinfo {year} {2008})%
  \bibAnnoteFile{NoStop}{Bernevig-PhysRevLett.100.246802}%
\bibitem{Bernevig-PhysRevB.77.184502}%
  \BibitemOpen
  \bibfield{author}{%
  \bibinfo {author} {\bibfnamefont{B.~A.}\ \bibnamefont{Bernevig}}\ and\
  \bibinfo {author} {\bibfnamefont{F.~D.~M.}\ \bibnamefont{Haldane}},\ }%
  \bibfield{journal}{%
  \Doi{10.1103/PhysRevB.77.184502}{\bibinfo {journal} {Phys. Rev. B}}\ }%
  \textbf{\bibinfo {volume} {77}},\ \bibinfo {pages} {184502} (\bibinfo {month}
  {May}\ \bibinfo {year} {2008})%
  \bibAnnoteFile{NoStop}{Bernevig-PhysRevB.77.184502}%
\bibitem{Gurarie1997685}%
  \BibitemOpen
  \bibfield{author}{%
  \bibinfo {author} {\bibfnamefont{V.}~\bibnamefont{Gurarie}}\ and\ \bibinfo
  {author} {\bibfnamefont{C.}~\bibnamefont{Nayak}},\ }%
  \bibfield{journal}{%
  \Doi{http://dx.doi.org/10.1016/S0550-3213(97)00612-3}{\bibinfo {journal}
  {Nuclear Physics B}}\ }%
  \textbf{\bibinfo {volume} {506}},\ \bibinfo {pages} {685 } (\bibinfo {year}
  {1997}),\ ISSN \bibinfo {issn} {0550-3213}%
  \bibAnnoteFile{NoStop}{Gurarie1997685}%
\bibitem{Ardonne-JPhysA2002}%
  \BibitemOpen
  \bibfield{author}{%
  \bibinfo {author} {\bibfnamefont{E.}~\bibnamefont{Ardonne}},\ }%
  \bibfield{journal}{%
  \bibinfo {journal} {Journal of Physics A: Mathematical and General}\ }%
  \textbf{\bibinfo {volume} {35}},\ \bibinfo {pages} {447} (\bibinfo {year}
  {2002})%
  \bibAnnoteFile{NoStop}{Ardonne-JPhysA2002}%
\bibitem{Read-PhysRevB.73.245334}%
  \BibitemOpen
  \bibfield{author}{%
  \bibinfo {author} {\bibfnamefont{N.}~\bibnamefont{Read}},\ }%
  \bibfield{journal}{%
  \Doi{10.1103/PhysRevB.73.245334}{\bibinfo {journal} {Phys. Rev. B}}\ }%
  \textbf{\bibinfo {volume} {73}},\ \bibinfo {pages} {245334} (\bibinfo {month}
  {Jun}\ \bibinfo {year} {2006})%
  \bibAnnoteFile{NoStop}{Read-PhysRevB.73.245334}%
\bibitem{greiter1992paired}%
  \BibitemOpen
  \bibfield{author}{%
  \bibinfo {author} {\bibfnamefont{M.}~\bibnamefont{Greiter}}, \bibinfo
  {author} {\bibfnamefont{X.}~\bibnamefont{Wen}},\ and\ \bibinfo {author}
  {\bibfnamefont{F.}~\bibnamefont{Wilczek}},\ }%
  \bibfield{journal}{%
  \bibinfo {journal} {Nuclear Physics B}\ }%
  \textbf{\bibinfo {volume} {374}},\ \bibinfo {pages} {567} (\bibinfo {year}
  {1992})%
  \bibAnnoteFile{NoStop}{greiter1992paired}%
\bibitem{greiter1991paired}%
  \BibitemOpen
  \bibfield{author}{%
  \bibinfo {author} {\bibfnamefont{M.}~\bibnamefont{Greiter}}, \bibinfo
  {author} {\bibfnamefont{X.-G.}\ \bibnamefont{Wen}},\ and\ \bibinfo {author}
  {\bibfnamefont{F.}~\bibnamefont{Wilczek}},\ }%
  \bibfield{journal}{%
  \bibinfo {journal} {Physical review letters}\ }%
  \textbf{\bibinfo {volume} {66}},\ \bibinfo {pages} {3205} (\bibinfo {year}
  {1991})%
  \bibAnnoteFile{NoStop}{greiter1991paired}%
\bibitem{jain-PhysRevLett.63.199}%
  \BibitemOpen
  \bibfield{author}{%
  \bibinfo {author} {\bibfnamefont{J.~K.}\ \bibnamefont{Jain}},\ }%
  \bibfield{journal}{%
  \Doi{10.1103/PhysRevLett.63.199}{\bibinfo {journal} {Phys. Rev. Lett.}}\ }%
  \textbf{\bibinfo {volume} {63}},\ \bibinfo {pages} {199} (\bibinfo {month}
  {Jul}\ \bibinfo {year} {1989})%
  \bibAnnoteFile{NoStop}{jain-PhysRevLett.63.199}%
\bibitem{Hermanns-PhysRevB.87.235128}%
  \BibitemOpen
  \bibfield{author}{%
  \bibinfo {author} {\bibfnamefont{M.}~\bibnamefont{Hermanns}},\ }%
  \bibfield{journal}{%
  \Doi{10.1103/PhysRevB.87.235128}{\bibinfo {journal} {Phys. Rev. B}}\ }%
  \textbf{\bibinfo {volume} {87}},\ \bibinfo {pages} {235128} (\bibinfo {month}
  {Jun}\ \bibinfo {year} {2013})%
  \bibAnnoteFile{NoStop}{Hermanns-PhysRevB.87.235128}%
\bibitem{Fremling-PhysRevB.89.125303}%
  \BibitemOpen
  \bibfield{author}{%
  \bibinfo {author} {\bibfnamefont{M.}~\bibnamefont{Fremling}}, \bibinfo
  {author} {\bibfnamefont{T.~H.}\ \bibnamefont{Hansson}},\ and\ \bibinfo
  {author} {\bibfnamefont{J.}~\bibnamefont{Suorsa}},\ }%
  \bibfield{journal}{%
  \Doi{10.1103/PhysRevB.89.125303}{\bibinfo {journal} {Phys. Rev. B}}\ }%
  \textbf{\bibinfo {volume} {89}},\ \bibinfo {pages} {125303} (\bibinfo {month}
  {Mar}\ \bibinfo {year} {2014})%
  \bibAnnoteFile{NoStop}{Fremling-PhysRevB.89.125303}%
\bibitem{Barkeshli-PhysRevB.87.045130}%
  \BibitemOpen
  \bibfield{author}{%
  \bibinfo {author} {\bibfnamefont{M.}~\bibnamefont{Barkeshli}}, \bibinfo
  {author} {\bibfnamefont{C.-M.}\ \bibnamefont{Jian}},\ and\ \bibinfo {author}
  {\bibfnamefont{X.-L.}\ \bibnamefont{Qi}},\ }%
  \bibfield{journal}{%
  \Doi{10.1103/PhysRevB.87.045130}{\bibinfo {journal} {Phys. Rev. B}}\ }%
  \textbf{\bibinfo {volume} {87}},\ \bibinfo {pages} {045130} (\bibinfo {month}
  {Jan}\ \bibinfo {year} {2013})%
  \bibAnnoteFile{NoStop}{Barkeshli-PhysRevB.87.045130}%
\bibitem{barkeshli-PhysRevB.88.241103}%
  \BibitemOpen
  \bibfield{author}{%
  \bibinfo {author} {\bibfnamefont{M.}~\bibnamefont{Barkeshli}}, \bibinfo
  {author} {\bibfnamefont{C.-M.}\ \bibnamefont{Jian}},\ and\ \bibinfo {author}
  {\bibfnamefont{X.-L.}\ \bibnamefont{Qi}},\ }%
  \bibfield{journal}{%
  \Doi{10.1103/PhysRevB.88.241103}{\bibinfo {journal} {Phys. Rev. B}}\ }%
  \textbf{\bibinfo {volume} {88}},\ \bibinfo {pages} {241103} (\bibinfo {month}
  {Dec}\ \bibinfo {year} {2013})%
  \bibAnnoteFile{NoStop}{barkeshli-PhysRevB.88.241103}%
\bibitem{barkeshli-PhysRevB.88.235103}%
  \BibitemOpen
  \bibfield{author}{%
  \bibinfo {author} {\bibfnamefont{M.}~\bibnamefont{Barkeshli}}, \bibinfo
  {author} {\bibfnamefont{C.-M.}\ \bibnamefont{Jian}},\ and\ \bibinfo {author}
  {\bibfnamefont{X.-L.}\ \bibnamefont{Qi}},\ }%
  \bibfield{journal}{%
  \Doi{10.1103/PhysRevB.88.235103}{\bibinfo {journal} {Phys. Rev. B}}\ }%
  \textbf{\bibinfo {volume} {88}},\ \bibinfo {pages} {235103} (\bibinfo {month}
  {Dec}\ \bibinfo {year} {2013})%
  \bibAnnoteFile{NoStop}{barkeshli-PhysRevB.88.235103}%
\bibitem{barkeshli-2014arXiv1410.4540B}%
  \BibitemOpen
  \bibfield{author}{%
  \bibinfo {author} {\bibfnamefont{M.}~\bibnamefont{{Barkeshli}}}, \bibinfo
  {author} {\bibfnamefont{P.}~\bibnamefont{{Bonderson}}}, \bibinfo {author}
  {\bibfnamefont{M.}~\bibnamefont{{Cheng}}},\ and\ \bibinfo {author}
  {\bibfnamefont{Z.}~\bibnamefont{{Wang}}},\ }%
  \bibfield{journal}{%
  \bibinfo {journal} {ArXiv e-prints}}%
   (\bibinfo {month} {Oct.}\ \bibinfo {year} {2014}),\
  \Eprint{http://arxiv.org/abs/1410.4540}{arXiv:1410.4540 [cond-mat.str-el]}%
  \bibAnnoteFile{NoStop}{barkeshli-2014arXiv1410.4540B}%
\bibitem{Teo-2015arXiv150306812T}%
  \BibitemOpen
  \bibfield{author}{%
  \bibinfo {author} {\bibfnamefont{J.~C.~Y.}\ \bibnamefont{{Teo}}}, \bibinfo
  {author} {\bibfnamefont{T.~L.}\ \bibnamefont{{Hughes}}},\ and\ \bibinfo
  {author} {\bibfnamefont{E.}~\bibnamefont{{Fradkin}}},\ }%
  \bibfield{journal}{%
  \bibinfo {journal} {ArXiv e-prints}}%
   (\bibinfo {month} {Mar.}\ \bibinfo {year} {2015}),\
  \Eprint{http://arxiv.org/abs/1503.06812}{arXiv:1503.06812 [cond-mat.str-el]}%
  \bibAnnoteFile{NoStop}{Teo-2015arXiv150306812T}%
\bibitem{repellin-PhysRevB.90.045114}%
  \BibitemOpen
  \bibfield{author}{%
  \bibinfo {author} {\bibfnamefont{C.}~\bibnamefont{Repellin}}, \bibinfo
  {author} {\bibfnamefont{T.}~\bibnamefont{Neupert}}, \bibinfo {author}
  {\bibfnamefont{Z.}~\bibnamefont{Papi\ifmmode~\acute{c}\else \'{c}\fi{}}},\
  and\ \bibinfo {author} {\bibfnamefont{N.}~\bibnamefont{Regnault}},\ }%
  \bibfield{journal}{%
  \Doi{10.1103/PhysRevB.90.045114}{\bibinfo {journal} {Phys. Rev. B}}\ }%
  \textbf{\bibinfo {volume} {90}},\ \bibinfo {pages} {045114} (\bibinfo {month}
  {Jul}\ \bibinfo {year} {2014})%
  \bibAnnoteFile{NoStop}{repellin-PhysRevB.90.045114}%
\bibitem{moller-PhysRevLett.107.036803}%
  \BibitemOpen
  \bibfield{author}{%
  \bibinfo {author} {\bibfnamefont{G.}~\bibnamefont{M\"oller}}, \bibinfo
  {author} {\bibfnamefont{A.}~\bibnamefont{W\'ojs}},\ and\ \bibinfo {author}
  {\bibfnamefont{N.~R.}\ \bibnamefont{Cooper}},\ }%
  \bibfield{journal}{%
  \Doi{10.1103/PhysRevLett.107.036803}{\bibinfo {journal} {Phys. Rev. Lett.}}\
  }%
  \textbf{\bibinfo {volume} {107}},\ \bibinfo {pages} {036803} (\bibinfo
  {month} {Jul}\ \bibinfo {year} {2011})%
  \bibAnnoteFile{NoStop}{moller-PhysRevLett.107.036803}%
\bibitem{bonderson-PhysRevLett.106.186802}%
  \BibitemOpen
  \bibfield{author}{%
  \bibinfo {author} {\bibfnamefont{P.}~\bibnamefont{Bonderson}}, \bibinfo
  {author} {\bibfnamefont{A.~E.}\ \bibnamefont{Feiguin}},\ and\ \bibinfo
  {author} {\bibfnamefont{C.}~\bibnamefont{Nayak}},\ }%
  \bibfield{journal}{%
  \Doi{10.1103/PhysRevLett.106.186802}{\bibinfo {journal} {Phys. Rev. Lett.}}\
  }%
  \textbf{\bibinfo {volume} {106}},\ \bibinfo {pages} {186802} (\bibinfo
  {month} {May}\ \bibinfo {year} {2011})%
  \bibAnnoteFile{NoStop}{bonderson-PhysRevLett.106.186802}%
\bibitem{Yang-PhysRevLett.108.256807}%
  \BibitemOpen
  \bibfield{author}{%
  \bibinfo {author} {\bibfnamefont{B.}~\bibnamefont{Yang}}, \bibinfo {author}
  {\bibfnamefont{Z.-X.}\ \bibnamefont{Hu}}, \bibinfo {author}
  {\bibfnamefont{Z.}~\bibnamefont{Papi\ifmmode~\acute{c}\else \'{c}\fi{}}},\
  and\ \bibinfo {author} {\bibfnamefont{F.~D.~M.}\ \bibnamefont{Haldane}},\ }%
  \bibfield{journal}{%
  \Doi{10.1103/PhysRevLett.108.256807}{\bibinfo {journal} {Phys. Rev. Lett.}}\
  }%
  \textbf{\bibinfo {volume} {108}},\ \bibinfo {pages} {256807} (\bibinfo
  {month} {Jun}\ \bibinfo {year} {2012})%
  \bibAnnoteFile{NoStop}{Yang-PhysRevLett.108.256807}%
\bibitem{Stanley198976}%
  \BibitemOpen
  \bibfield{author}{%
  \bibinfo {author} {\bibfnamefont{R.~P.}\ \bibnamefont{Stanley}},\ }%
  \bibfield{journal}{%
  \Doi{http://dx.doi.org/10.1016/0001-8708(89)90015-7}{\bibinfo {journal}
  {Advances in Mathematics}}\ }%
  \textbf{\bibinfo {volume} {77}},\ \bibinfo {pages} {76 } (\bibinfo {year}
  {1989}),\ ISSN \bibinfo {issn} {0001-8708}%
  \bibAnnoteFile{NoStop}{Stanley198976}%
\bibitem{PhysRevLett.103.016801}%
  \BibitemOpen
  \bibfield{author}{%
  \bibinfo {author} {\bibfnamefont{N.}~\bibnamefont{Regnault}}, \bibinfo
  {author} {\bibfnamefont{B.~A.}\ \bibnamefont{Bernevig}},\ and\ \bibinfo
  {author} {\bibfnamefont{F.~D.~M.}\ \bibnamefont{Haldane}},\ }%
  \bibfield{journal}{%
  \Doi{10.1103/PhysRevLett.103.016801}{\bibinfo {journal} {Phys. Rev. Lett.}}\
  }%
  \textbf{\bibinfo {volume} {103}},\ \bibinfo {pages} {016801} (\bibinfo
  {month} {Jun}\ \bibinfo {year} {2009})%
  \bibAnnoteFile{NoStop}{PhysRevLett.103.016801}%
\bibitem{Haldane85-PhysRevLett.55.2095}%
  \BibitemOpen
  \bibfield{author}{%
  \bibinfo {author} {\bibfnamefont{F.~D.~M.}\ \bibnamefont{Haldane}},\ }%
  \bibfield{journal}{%
  \Doi{10.1103/PhysRevLett.55.2095}{\bibinfo {journal} {Phys. Rev. Lett.}}\ }%
  \textbf{\bibinfo {volume} {55}},\ \bibinfo {pages} {2095} (\bibinfo {month}
  {Nov}\ \bibinfo {year} {1985})%
  \bibAnnoteFile{NoStop}{Haldane85-PhysRevLett.55.2095}%
\bibitem{Greiter93}%
  \BibitemOpen
  \bibfield{author}{%
  \bibinfo {author} {\bibfnamefont{M.}~\bibnamefont{Greiter}},\ }%
  \bibinfo {note} {bull. Am. Phys. Soc. \textbf{38}, 137 (1993).}%
  \bibAnnoteFile{Stop}{Greiter93}%
\bibitem{Haldane-PhysRevLett.67.937}%
  \BibitemOpen
  \bibfield{author}{%
  \bibinfo {author} {\bibfnamefont{F.}~\bibnamefont{Haldane}},\ }%
  \bibfield{journal}{%
  \Doi{10.1103/PhysRevLett.67.937}{\bibinfo {journal} {Phys. Rev. Lett.}}\ }%
  \textbf{\bibinfo {volume} {67}},\ \bibinfo {pages} {937} (\bibinfo {month}
  {Aug}\ \bibinfo {year} {1991})%
  \bibAnnoteFile{NoStop}{Haldane-PhysRevLett.67.937}%
\bibitem{Ardonne-2008JSMTE..04..016A}%
  \BibitemOpen
  \bibfield{author}{%
  \bibinfo {author} {\bibfnamefont{E.}~\bibnamefont{{Ardonne}}}, \bibinfo
  {author} {\bibfnamefont{E.~J.}\ \bibnamefont{{Bergholtz}}}, \bibinfo {author}
  {\bibfnamefont{J.}~\bibnamefont{{Kailasvuori}}},\ and\ \bibinfo {author}
  {\bibfnamefont{E.}~\bibnamefont{{Wikberg}}},\ }%
  \bibfield{journal}{%
  \Doi{10.1088/1742-5468/2008/04/P04016}{\bibinfo {journal} {Journal of
  Statistical Mechanics: Theory and Experiment}}\ }%
  \textbf{\bibinfo {volume} {4}},\ \bibinfo {pages} {16} (\bibinfo {month}
  {Apr.}\ \bibinfo {year} {2008})%
  \bibAnnoteFile{NoStop}{Ardonne-2008JSMTE..04..016A}%
\bibitem{Bergholtz-PhysRevB.77.155308}%
  \BibitemOpen
  \bibfield{author}{%
  \bibinfo {author} {\bibfnamefont{E.}~\bibnamefont{Bergholtz}}\ and\ \bibinfo
  {author} {\bibfnamefont{A.}~\bibnamefont{Karlhede}},\ }%
  \bibfield{journal}{%
  \Doi{10.1103/PhysRevB.77.155308}{\bibinfo {journal} {Phys. Rev. B}}\ }%
  \textbf{\bibinfo {volume} {77}},\ \bibinfo {pages} {155308} (\bibinfo {month}
  {Apr}\ \bibinfo {year} {2008})%
  \bibAnnoteFile{NoStop}{Bergholtz-PhysRevB.77.155308}%
\bibitem{Bergholtz-PhysRevLett.99.256803}%
  \BibitemOpen
  \bibfield{author}{%
  \bibinfo {author} {\bibfnamefont{E.}~\bibnamefont{Bergholtz}}, \bibinfo
  {author} {\bibfnamefont{T.}~\bibnamefont{Hansson}}, \bibinfo {author}
  {\bibfnamefont{M.}~\bibnamefont{Hermanns}},\ and\ \bibinfo {author}
  {\bibfnamefont{A.}~\bibnamefont{Karlhede}},\ }%
  \bibfield{journal}{%
  \Doi{10.1103/PhysRevLett.99.256803}{\bibinfo {journal} {Phys. Rev. Lett.}}\
  }%
  \textbf{\bibinfo {volume} {99}},\ \bibinfo {pages} {256803} (\bibinfo {month}
  {Dec}\ \bibinfo {year} {2007})%
  \bibAnnoteFile{NoStop}{Bergholtz-PhysRevLett.99.256803}%
\bibitem{Seidel-PhysRevLett.95.266405}%
  \BibitemOpen
  \bibfield{author}{%
  \bibinfo {author} {\bibfnamefont{A.}~\bibnamefont{Seidel}}, \bibinfo {author}
  {\bibfnamefont{H.}~\bibnamefont{Fu}}, \bibinfo {author}
  {\bibfnamefont{D.-H.}\ \bibnamefont{Lee}}, \bibinfo {author}
  {\bibfnamefont{J.}~\bibnamefont{Leinaas}},\ and\ \bibinfo {author}
  {\bibfnamefont{J.}~\bibnamefont{Moore}},\ }%
  \bibfield{journal}{%
  \Doi{10.1103/PhysRevLett.95.266405}{\bibinfo {journal} {Phys. Rev. Lett.}}\
  }%
  \textbf{\bibinfo {volume} {95}},\ \bibinfo {pages} {266405} (\bibinfo {month}
  {Dec}\ \bibinfo {year} {2005})%
  \bibAnnoteFile{NoStop}{Seidel-PhysRevLett.95.266405}%
\end{thebibliography}%
\end{document}